%% file: main.tex
\documentclass[10pt,journal,compsoc]{IEEEtran}
\usepackage{amsmath}
\usepackage{amssymb}
\usepackage{bm}
\usepackage{algorithm}
\usepackage{algpseudocode}
\usepackage{amsmath}
\usepackage{multirow}
\usepackage{indentfirst}
\usepackage{subfigure}
\usepackage{epsfig}
\usepackage{epstopdf}
\usepackage{url}
\usepackage{xspace}
\usepackage{booktabs}
\usepackage{subeqnarray}
\usepackage{color}

\newcommand{\paratitle}[1]{\vspace{1.5ex}\noindent\textbf{#1}}
\newcommand{\ie}{\emph{i.e.,}\xspace}
\newcommand{\aka}{\emph{a.k.a.,}\xspace}
\newcommand{\eg}{\emph{e.g.,}\xspace}

\newcommand{\ignore}[1]{}

\usepackage{makecell}

\usepackage{cite}

\begin{document}

\title{Dense Text Retrieval based on Pretrained Language Models: A Survey}

\author{Wayne Xin Zhao,
        Jing Liu,
        Ruiyang Ren
        and Ji-Rong Wen
\IEEEcompsocitemizethanks{\IEEEcompsocthanksitem Wayne Xin Zhao, Ruiyang Ren and Ji-Rong Wen are with  Renmin University of China; Jing Liu is with Baidu Inc., Beijing, China. \protect\\
E-mail: batmafly@gmail.com, reyon.ren@ruc.edu.cn, liujing46@baidu.com.}
}

\IEEEtitleabstractindextext{
\begin{abstract}
Text retrieval is a long-standing research topic on information seeking, where a system is required to return relevant information resources to user's  queries in natural language. 
From classic  retrieval methods to learning-based ranking functions, the underlying retrieval models have been continually evolved with the ever-lasting technical innovation. 
To design effective retrieval models, a key point lies in how to learn the text representation and model the relevance matching.  
The recent success of pretrained language models~(PLMs) sheds light on developing more capable text retrieval approaches by leveraging the excellent modeling capacity of PLMs.
With powerful PLMs, we can effectively learn the representations of queries and texts in the latent representation space, and further  construct the semantic matching function between the dense vectors for relevance modeling. Such a retrieval approach is referred to as \emph{dense retrieval}, since it employs dense vectors (\aka embeddings) to represent the texts.
Considering the rapid progress on dense retrieval,  
in this survey, we systematically review the recent advances on PLM-based dense retrieval. Different from previous surveys on dense retrieval, we take a new perspective to organize the related work  by four major aspects, including architecture, training, indexing and integration, and summarize  the  mainstream   techniques  for  each  aspect. We thoroughly survey the literature, and include 300+ related reference papers on dense retrieval. 
To support our survey, we 
create a website for providing useful resources, and release a code repertory and toolkit for implementing dense retrieval models.  
This survey aims to provide a comprehensive, practical reference focused on the major progress for dense text retrieval.   
\end{abstract}

}

\maketitle

\IEEEdisplaynontitleabstractindextext

\IEEEpeerreviewmaketitle

\input{sec/sec-intro}

\input{sec/sec-formulation}

\input{sec/sec-dataset}

\input{sec/sec-network}
\input{sec/sec-training}
\input{sec/sec-index}
\input{sec/sec-rerank}

\input{sec/sec-advancedtopics}
\input{sec/sec-application}

\section{Conclusion}
In this survey, we thoroughly review the recent progress of dense retrieval based on pretrained language models~(PLM). 
As an important evolution of language intelligence techniques,  PLMs empower dense retrieval models with excellent modeling capacities to capture and represent text semantics for relevance matching.  
Our survey has extensively discussed the key issues and the mainstream solutions in four major aspects to develop dense retrieval systems, including architecture, training,  indexing and integration. 
Next, we briefly summarize the discussions of this survey and introduce some remaining issues for dense retrieval.

\paratitle{PLM-based architecture}. 
In dense retrieval, two kinds of PLM architectures have been extensively used, namely bi-encoder (\emph{efficient} yet \emph{less effective}) and cross-encoder (\emph{effective} yet \emph{less efficient}). 
In particular, bi-encoder is usually employed to implement the retriever and cross-encoder is usually employed to implement the reranker. Since the bi-encoder has limited capacity in capturing the semantic interaction between query and text, the multi-representation technique has been widely explored in enhancing the fine-grained interaction modeling. It proposes to compute the relevance  score according to multiple contextual representations of a text. 
Although dense retrieval models have achieved decent retrieval performance on several benchmark datasets, it has been shown that their capacities in some specific settings are rather limited, 
especially the capability on zero-shot retrieval~\cite{thakur2021beir}. 
Compared to sparse retrievers (\eg BM25), dense retrievers perform better in terms of semantic matching, but less well in terms of  lexical matching~\cite{dpr2020,Sciavolino2021SimpleEQ}. Thus, 
there has been increasing attention in combining the merits of both  dense retrievers and sparse retrievers in a unified retrieval system~\cite{chen2021salient,Lin2022AggretrieverAS}.  
Moreover,  dense retrieval models rely on latent semantic  representations for relevance matching. It is not fully clear that how dense retrieval models behave and perform in response to different types of queries, \eg how they mimic  ``\emph{bag-of-words}" retrieval~\cite{MEBERT}. More theoretical discussions are needed in order to better understand and enhance the retrieval capacity of dense retrieval models. 
Despite the rapid improvement, the problem  whether dense retrieval can fully replace sparse retrieval is still being under explored by the research community, and there are also some studies that devise new framework consisting of 
multiple effectiveness measurements for examining this problem~\cite{Hofstatter2022AreWT}.

\paratitle{Training approach}. 
Compared with previous neural IR methods, PLM-based dense retrieval models are  more difficult to be optimized given the huge number of parameters. 
Even with the ever-increasing labeled dataset, it still requires specific training techniques to effectively optimize large retrieval models. 
We summarize three major issues to address for improving dense retrieval systems, \ie large-scale candidate space, limited relevance judgements, and pretraining discrepancy. 
Focused on the three issues,
the researchers have made extensive efforts by designing various effective training strategies, including negative selection, data augmentation and specific pretraining techniques. These studies have significantly improved the retrieval performance of dense retrievers on benchmark datasets~\cite{dpr2020,ANCE,rocketqa,gao2021condenser}. In particular, negative selection is a key step in the training approach, and using \emph{more} negatives (\eg in- and cross-batch negatives) or \emph{more high-quality} negatives (\eg sampling static or dynamic hard negatives) typically lead  to a better retrieval performance.
Besides, in order to enhance the capacity of bi-encoder, it is very useful to distill a more capable  teacher model (\eg cross-encoder) and design suitable pretraining tasks (\eg representation-enhanced pretraining). 
More recently, auto-encoder and contrastive learning based pretraining methods have been widely adopted to enhance the semantic representations of texts.  As an improvement  direction, it is important to design  data- and parameter-efficient training approaches for enhancing the retrieval capacity under low-resourced setting (\eg zero-shot retrieval).

\paratitle{Dense vector index}. 
Although the idea of representing texts by dense  vectors is intuitive, it is not easy to construct large-scale dense vector index for efficient search. Without the support of a suitable index structure, the retrieval process (\ie finding the most close text vectors \emph{w.r.t.} query vectors) would be extremely slow. For this issue, a popular solution is the use of  \emph{Approximate Nearest Neighbor Search~(ANNS)} algorithms~\cite{gionis1999similarity,ALSH,faiss}, which  can achieve highly efficient retrieval performance (\eg sublinear time cost~\cite{ALSH}). 
Compared with term-based index, it is more difficult to maintain the ANNS index for   dense retrieval, especially with the regular  operations of adding, deleting and updating the embeddings of texts. For example, when the optimization involves in  the update of indexed embeddings, it usually suffers from the  \emph{index staleness} issue~\cite{REALM}.  
Therefore, it is necessary to develop efficient online update algorithms for enhancing the flexibility of ANNS index\footnote{{There is commercial software  (\eg \url{https://pinecone.io}) for maintaining the embedding index instantly (\ie add, edit or delete data instantly), while academic solutions are still lacking.} }.
In addition, dense vector index is typically used in memory, taking  a higher cost compared to a term-based inverted index that can be stored on disk. Besides the effectiveness, the cost for maintenance and use is also  important to consider for the practical deployment of dense retrieval systems~\cite{Hofstatter2022AreWT}. 
Although quantization techniques can alleviate the cost issue to some extent, it is not easy to optimize quantization-based vector index due to the involved non-differential operation 
(with some joint optimization approaches proposed~\cite{zhan2021jointly,zhang2021joint}). Overall, the dense vector index should be well suited to the architecture and optimization of the PLM-based retrievers, which still requires more research in this line.

\paratitle{Retrieval pipeline}. 
An information retrieval system typically adopts 
a pipeline  way (consisting of first-stage retrieval and reranking stages) to recall and rank the texts, which gradually reduces the search space and refines the top ranked texts.
Generally, it is more  complicated to optimize the entire  retrieval pipeline than a single-stage component in it (\eg first-stage retriever). 
A preferred approach is to jointly optimize the retrieval pipeline, considering different components as a whole. 
As the major  difficulty for joint training, these stages are usually learned according to different optimization goals, based on varied input and output.
In early studies~\cite{multistage}, the multiple stages in  a retrieval pipeline are separately optimized.  Recent studies propose a series of improved optimization approaches for  jointly  training the retrieval pipeline~\cite{REALM,rocketqav2}.
The basic idea is to let the retriever and the reranker adjust according to the learned relevance information from each other. 
However, it is still challenging to optimize a multi-stage retrieval pipeline, especially when it is built in a  sparse-dense hybrid way. 
Besides, it is also meaningful to study how to automatically construct the pipeline  (\eg what component to use for each of the stages) by learning from labeled dataset. 

Besides the above four aspects, we have also extensively discussed 
several important advanced topics that have received increasing attention.
In order to support this survey, 
we have created a referencing website (at the link \url{ https://github.com/RUCAIBox/DenseRetrieval}) to provide useful resources on paper collection and  code repertory for dense retrieval. 
This survey is expected to  provide a systemic literature review that summarizes the major progress for PLM-based dense retrieval. 

\textbf{\emph{Coda}}: This survey was planned in June, 2021, when we drafted a preliminary outline to organize the content. We started the writing in September, 2021, and finished the initial version in December, 2021. Considering the rapid research progress of dense retrieval, we did not release the initial version online. In this year, we have largely revised this survey by extensively including recent papers and finally made it online, which we expect to be helpful for both researchers and engineers on the research of  dense retrieval. It is not easy to write a long survey, and we will keep updating  this survey and welcome any constructive comments to improve our survey.   

\ifCLASSOPTIONcaptionsoff
  \newpage
\fi

\bibliographystyle{IEEEtran}

\bibliography{reference}

\end{document}

%% file: sec/sec-intro.tex
\IEEEraisesectionheading{\section{Introduction}\label{sec:introduction}}

\IEEEPARstart{T}{ext} retrieval  aims at finding relevant information resources (\eg documents or passages) in response to user’s queries. 
It refers to a specific information seeking scenario where the queries and resources are present in the form of natural language text. As one of the most essential techniques to overcome  information overload, text retrieval systems have been widely employed to support many downstream applications,  including question answering~\cite{dpr2020,mrr}, dialog system~\cite{conversation,survey-DS}, entity linking~\cite{entitylinking,wu2020scalable} and Web search~\cite{mitra2017learning}.

The idea of developing text retrieval systems has  a long history in the research literature.
Early in the 1950s, pioneering researchers have started to study how to index the texts by selecting representative terms for information retrieval~\cite{joyce1958thesaurus}. 
Among the early efforts,  a significant achievement is the \emph{vector space model}~\cite{Salton-ref1,Salton-ref2} based on the ``bag-of-words'' assumption, representing both documents and queries as sparse term-based vectors\footnote{In this survey, sparse vectors mainly refer to term-based vector representations.}.  
To construct sparse vector representations, various term weighting methods have been designed and implemented, including the classic \emph{tf-idf} method~\cite{Salton1988TermWeightingAI,tf-idf-ref1,tf-idf-ref2}. Based on this scheme,  the relevance can be estimated according to the lexical similarity between sparse query and text vectors.  Such a representation scheme is further supported by the well-known data structure of \emph{inverted index}~\cite{Zobel2006InvertedFF,inverted-index}, which organizes the text content as term-oriented posting lists, for efficient text retrieval. 
In order to better understand the underlying retrieval mechanism, probabilistic relevance frameworks have been proposed for relevance modeling, exemplified by  the classic BM25 model~\cite{robertson1995okapi,robertson2009probabilistic}. Furthermore, statistical language modeling approaches~\cite{SLM4IR} have been widely explored for text ranking. 
These early contributions lay the foundation of modern information retrieval systems, while the proposed retrieval methods are usually based on heuristic  strategies or simplified probabilistic principles.

With the development of machine learning discipline, \emph{learning to rank}~\cite{Liu2009LearningTR,Li2011LearningTR} 
introduces \emph{supervised learning} for text ranking. 
The basic idea is to design feature-based ranking functions taking as input hand-crafted features (not only limited to lexical features) and then train the ranking function with relevance judgements (binary or graded relevance annotations over documents).  Despite the flexibility, learning to rank methods still rely on human efforts for feature engineering. 

Further, the re-surge of neural networks  sheds lights on  developing more capable text retrieval systems, which no longer require 
hand-crafted text features. 
As an important progress in information retrieval, deep learning approaches~\cite{huang2013learning} can learn  query and document representations from labeled data in an automatic way, where both queries and documents are mapped into low-dimensional vectors (called \emph{dense vectors} or \emph{embeddings})  in the latent representation space. In this manner, the relevance can be measured according to the semantic similarity between the dense vectors. 
In contrast to sparse vectors in the classic vector space model, embeddings do not correspond to explicit term  dimensions, but instead aim at capturing latent semantic characteristics for matching.  Such a retrieval paradigm is  called \emph{Neural Information Retrieval~(Neural IR)}~\cite{guo2016deep,mitra2017neural,guo2020deep}, 
which can be considered as initial explorations for dense retrieval techniques. Following the convention in \cite{lin2021pretrained,fan2021pre}, we refer to these neural IR methods (before the use of pretrained language models) as \emph{pre-BERT models}.

Recently, based on the powerful  Transformer architecture~\cite{transformer}, pretrained language models~(PLM)~\cite{bert2019naacl} have significantly pushed forward the progress of language intelligence.
Pretrained on large-scale general text data, PLMs can encode large amounts of semantic knowledge,  thus having an improved capacity to  understand and represent the semantics of  text content.
Following the ``\emph{pretrain then fine-tune}''  paradigm, PLMs can be further fine-tuned according to specific downstream tasks.
Inspired by the success of PLMs in natural language processing, {since 2019}, researchers have started to develop text retrieval models based on PLMs, leading to the new generation of text retrieval approaches, \ie \emph{PLM-based dense retrieval models}.

In the recent {four years}, a large number of studies on PLM-based dense retrieval have been proposed~\cite{lin2021pretrained,cai2021semantic,fan2021pre}, which have largely raised the performance bar on existing benchmark datasets.   It has been reported that PLM-based approaches have dominated the top results for both document and passage retrieval tasks  at the 2021 Deep Learning Track~\cite{craswell2022overview}. 
The reason for the success of PLM-based dense retrieval models can be given in two major aspects. 
First, the excellent text representation capability of PLMs enables the text retrieval systems to answer difficult queries that cannot be solved via simple lexical match\footnote{An example query that is difficult to answer by lexical match can be given as:  ``\emph{average salary for dental hygienist in nebraska}''. See more query examples in the TREC deep learning track report~\cite{craswell2021overview}.}.  Second, 
the availability of large-scale labeled retrieval datasets (\eg MS MARCO~\cite{msmarco} and Natural Questions~\cite{nq} datasets) makes it feasible to train (or fine-tune) capable  PLMs for text retrieval.
For example, TREC 2019 and 2020 Deep Learning Track~\cite{craswell2019overview,craswell2021overview} release a training set of 0.367 million queries  (derived from MS MARCO~\cite{msmarco}) for document retrieval task, which is significantly larger than those in previous retrieval tasks in TREC.

Considering the important progress on dense retrieval, this survey aims to provide a systematic review of existing text retrieval approaches.
In particular, we  focus on the PLM-based dense retrieval approaches (short as \emph{dense retrieval models} throughout this survey) instead of previous neural IR models (\ie the pre-BERT methods~\cite{huang2013learning,guo2016deep}). 
This survey takes \emph{first-stage retrieval} as the core, and extensively discusses four related aspects to build a dense retrieval system, including \emph{architecture} (how to design the network architectures for dense retrievers), \emph{training} (how to optimize the dense retriever with special training strategies), \emph{indexing} (how to design efficient data structure for indexing and retrieving dense vectors) and \emph{integration} (how to integrate and optimize a complete retrieval pipeline). 
Our survey extensively discusses various useful topics or techniques for building a dense retrieval system, which aims to provide a \emph{comprehensive}, \emph{practical} reference to this research direction for researchers and engineers.

We are aware of several closely related surveys or books on dense retrieval~\cite{guo2020deep, cai2021semantic, lin2021pretrained,fan2021pre,note}. For example, Guo et al.~\cite{guo2020deep}  review early contributions in neural IR, Cai et al.~\cite{cai2021semantic} summarize the major progress for the first-stage retrieval in three different paradigms, Lin et al.~\cite{lin2021pretrained}  present a review of text ranking methods mainly based on pretrained Transformers, Fan et al.~\cite{fan2021pre} review the major progress for information retrieval based on pretraining methods, and 
Nicola Tonellotto~\cite{note} present a lecture note of recent progress on neural information retrieval. 
Different from these literature surveys, we highlight three new features of this survey as follows:

$\bullet$~Firstly, we concentrate on the research of \emph{PLM-based dense retrieval}, and organize the related studies by a new categorization in four major aspects, \ie architecture, training, indexing and integration.

$\bullet$~Secondly, we take a special focus on \emph{practical techniques} for dense retrieval, extensively discussing the approaches to train the retrieval models, to build the dense index, and optimize 
the retrieval pipeline.

$\bullet$~Thirdly, we  cover the recent advances on dense retrieval, and  discuss several  emerging research topics (\eg model based retrieval and representation enhanced pretraining) in detail.

To support this survey, we create a referencing website for including related resources (\eg papers, dataset, and library) on dense retrieval research, at the link: \url{https://github.com/RUCAIBox/DenseRetrieval}. 
Furthermore, we implement and release a code repertory for a number of dense retrieval models based on PaddlePaddle\footnote{\url{https://github.com/PaddlePaddle/RocketQA}} (including RocketQA~\cite{rocketqa}, PAIR~\cite{pair} and RocketQAv2~\cite{rocketqav2}), and provide flexible interfaces to use or retrain dense retrievers. 

The remainder of the survey is organized as follows: Section 2 provides the overall introduction to the terminology, notations and task settings for dense retrieval, and Section 3 presents the   existing datasets and evaluation metrics for dense retrieval. 
As the core content, Section 4, 5, 6 and 7  review  the mainstream architecture, training approach, index mechanism and retrieval pipeline for dense retrieval, where we will thoroughly discuss the recent progress in the four aspects. 
Then, we continue the discussion on the advanced topics (Section 8) and the applications (Section 9). Finally, we conclude this survey by summarizing the major findings and remaining issues in Section 10.

%% file: sec/sec-formulation.tex
\section{Overview}\label{sec:overview}
This section first introduces the background about dense text retrieval,  and then discusses the key aspects for designing the dense retrieval models.

\begin{figure*}
    \centering
    \includegraphics[width=0.94\textwidth]{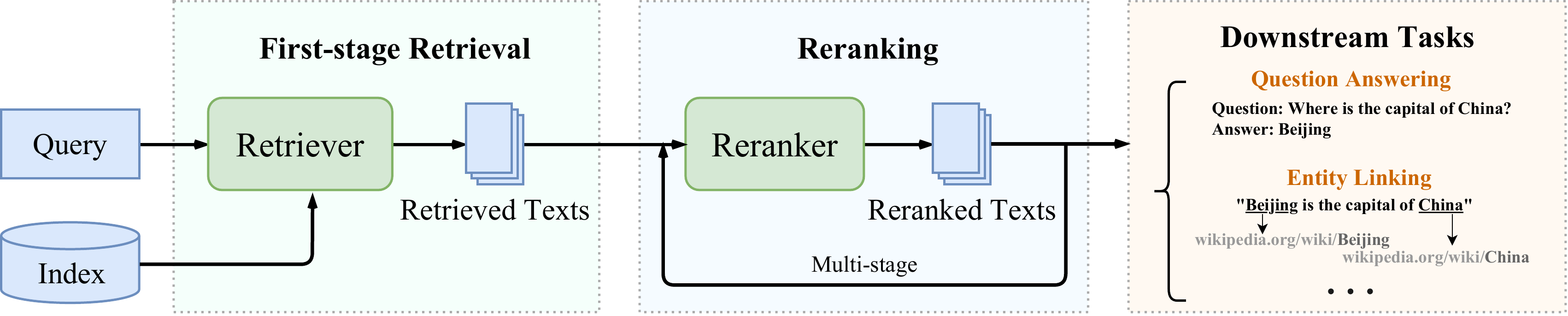}
    \caption{The illustration for the overall pipeline of an information retrieval system.}
    \label{fig:pipeline}
\end{figure*}

\subsection{Task Setting and Terminology} 
In this survey, we focus on the task of finding relevant texts from a large text collection in response to a natural language query issued by a user.
Specially, both  query and text (\eg a document) are presented in the form of a sequence of word tokens from a vocabulary.
In particular, \emph{texts} can be in different semantic granularities (\eg document, passage or sentence), leading to different types of retrieval tasks such as  document retrieval and passage retrieval. Since we aim to introduce general methodologies  for various retrieval tasks,  we refer to these tasks as \emph{text retrieval} in a unified way.

Typically, a complete information retrieval system consists of multiple procedures or stages arranged in a processing pipeline~\cite{fan2021pre,lin2021pretrained}\footnote{For ease of discussion, we omit other parts such as data collection and preprocessing.}.
In this pipeline, the first-stage retrieval aims to reduce the candidate space by retrieving relevant candidates, and we refer to the component that implements the first-stage retrieval as \emph{retriever}~\cite{cai2021semantic,lin2021pretrained}. 
Correspondingly, the subsequent stages mainly focus on 
\emph{reranking} (or \emph{ranking}) the candidate texts, called \emph{reranking stages}, which are supported by \emph{rerankers}~\cite{lin2021pretrained}.
Based on the retrieval results from the first-stage retriever, a retrieval system usually sets up one or multiple reranking stages to refine the initial results and derive the final search results. For other fine-grained retrieval tasks, \eg question answering, another component called \emph{reader}~\cite{chen17openqa} may be further integrated to analyze the returned texts by the retriever (or reranker) and locate the answers for the query. 

\subsection{Formulation for Dense Retrieval} 
\label{sec:formulation}
Formally, let $q$ denote a natural language query   and $d_i$ denote a text from a large text collection  $\mathcal{D}=\{ d_i \}_{i=1}^m$ consisting of $m$ documents. Given a query, text retrieval aims to return a ranked list of $n$ most relevant texts $\mathcal{L}=[d_1, d_2, \cdots d_n]$ according to the relevance scores of a retrieval model. As the technical  approach,  either sparse retrieval models (relying on  lexical matching) or dense retrieval models (relying on semantic  matching) can be used to implement the retriever. 

The key of dense retrieval lies in the fact that queries and texts are represented by  dense vectors, such that the relevance score can be 
computed according to some similarity function between these dense vectors~\cite{framework,cai2021semantic,lin2021pretrained,fan2021pre}, denoted by 
\begin{equation}
\label{equation:sim_definition}
\text{Rel}(q, d)=f_{\text{sim}}\big(\phi(q), \psi(d)\big),
\end{equation}
where $\phi(\cdot) \in \mathbb{R}^{l}$ and $\psi(\cdot) \in \mathbb{R}^{l}$ are functions mapping queries and texts into $l$-dimensional vectors, respectively\footnote{{This formulation can be also considered  as a generalized  representational approach~\cite{framework} to information retrieval, for both dense and sparse retrieval models.  }}. 
For dense retrieval, $\phi(\cdot)$ and $\psi(\cdot)$ are developed based on neural network encoders, and  similarity measurement functions (\eg inner product) can be used to implement $f_{\text{sim}}(\cdot)$.
At the pre-BERT time\footnote{Following the naming conventions in \cite{nogueira2020document}, we refer to the neural information retrieval methods as \emph{pre-BERT approaches}, which are before the use of Transformer and BERT family. }, the encoders are usually implemented by multi-layered perceptron networks (or other neural networks), while in this survey we focus on the text encoder based on pretrained language models~(PLMs), called \emph{dense retrievers}.

To learn the dense retrieval models, we also assume that a set of relevance judgements are provided for training or fine-tuning the PLMs.
Specially, we only consider the setting with binary (positive only) relevance judgements:  for a query $q$, a list of positive texts $\{ d^{+}_i \}$ are given as training data. 
Usually, negative texts (called \emph{negatives}) are not directly available, and we need to  obtain negatives through sampling or other strategies (detailed in Section~\ref{sec:negative-selection}). 
Note that in some cases, we can also obtain graded relevance judgements~\cite{NDCG,soboroff2019trec}, where a label indicates  the relevance degree of a text (\eg relevant, partially relevant and non-relevant). However, such fine-grained relevance labels are difficult to obtain in real-world retrieval systems. Thus, we mainly consider the binary relevance judgements in this survey.

\subsection{Key Aspects}

This survey takes the \emph{first-stage dense retrieval} as the core, and extensively discusses four key aspects for building a capable retrieval system, which are detailed as follows:

\begin{itemize}
\item 
\emph{Architecture} (Section 4). It is important  to design suitable network architectures  for building the relevance model. Currently, a number of general-purpose PLMs have been developed, and we need to adapt existing PLMs to  text retrieval  according to the task requirement.  

\item\emph{Training} (Section 5). Different from traditional retrieval models, it is more difficult to train PLM-based retrievers, consisting of a huge  number of parameters.   
We need to develop effective training approaches to achieving desirable retrieval performance.  

\item\emph{Indexing} (Section 6). Traditionally, inverted index can efficiently support the sparse retrieval models. However, dense vectors do not explicitly correspond to lexical terms. We need to design suitable indexing mechanism for dense  retrieval. 

\item\emph{Integration} (Section 7). As mentioned before, retrieval systems are usually implemented via a pipeline consisting of retrievers and rerankers. It is necessary to study how to integrate and optimize the dense retriever in a whole retrieval pipeline. 
\end{itemize}

%% file: sec/sec-dataset.tex
\section{Datasets, Evaluation and Resources}\label{sec:dataset}

In this section, we introduce the  available  datasets,  
 evaluation metrics, and supporting libraries  for dense retrieval.

\begin{table*}[htbp]
    \centering
    \small
    \caption{Detailed statistics of  available retrieval datasets. Here, ``q'' is the abbreviation of queries, and ``instance'' denotes a candidate text in the collection. }
    \begin{tabular}{c|c|c|c|c|c|c|c}
    \toprule
     \textbf{Categorization} & \textbf{Domain} & \textbf{Dataset}   &  \textbf{\#q in train} &  \textbf{\#label in train} & \textbf{\#q in dev} & \textbf{\#q in test} & \textbf{\#instance}  \\
    \midrule
        \multirow{12}*{\makecell[c]{Information \\ retrieval}} & 
        Web & MS MARCO~\cite{msmarco}& 502,939 & 532,761 & 6,980 & 6,837 & 8,841,823 \\
        & Web & mMARCO~\cite{bonifacio2021mmarco}& 808,731 & -- & 101,093  & -- & 8,841,823 \\
        & News & TREC-NEWS~\cite{soboroff2019trec} & -- & -- & -- & 57 & 594,977 \\
        & Biomedical & TREC-COVID~\cite{roberts2020trec} & -- & -- & -- & 50 & 171,332 \\
        & Biomedical &NFCorpus~\cite{boteva2016full}   & 5,922 & 110,575 & 324 & 323 & 3,633 \\
        & Twitter &Signal-1M~\cite{suarez2018data} & -- & -- & -- & 97 & 2,866,316 \\
        & Argument &Touché-2020~\cite{bondarenko2020overview}   & -- & -- & -- & 249 & 528,155 \\
        & Argument & ArguAna~\cite{wachsmuth2018retrieval} & -- & -- & -- & 1,406 & 8,674 \\
        & Wikipedia &DBPedia~\cite{hasibi2017dbpedia} & -- & -- & 67 & 400 & 4,635,922 \\
        & Web &ORCAS~\cite{craswell2020orcas} & 10.4M & 18.8M & -- & -- & 3,213,835  \\
        & Wikipedia &EntityQuestions~\cite{Sciavolino2021SimpleEQ} & 176,560 & 186,367 & 22,068 & 22,075 & -- \\
        & Web & MS MARCO v2~\cite{msmarcov2} & 277,144 & 284,212 & 8,184 & -- & 138,364,198 \\
        & Web & DuReader$_\text{retrieval}$~\cite{qiu2022dureader_retrieval} &97,343 & 86,395 & 2,000 & 8,948 & 8,096,668\\
        \midrule
        \multirow{12}*{\makecell[c]{Question \\ answering}} &
        Wikipedia & Natural Questions~\cite{nq} & 152,148 & 152,148 & 6,515 & 3,610 & 2,681,468 \\
        & Wikipedia & SQuAD~\cite{rajpurkar2016squad}  & 78,713 & 78,713 & 8,886 & 10,570 & 23,215 \\
        & Wikipedia & TriviaQA~\cite{joshi2017triviaqa} & 78,785 & 78,785 & 8,837 & 11,313 & 740K \\
        & Wikipedia & HotpotQA~\cite{yang2018hotpotqa}  & 85,000 & 170,000 & 5,447 & 7,405 & 5,233,329 \\
        & Web &WebQuestions~\cite{berant2013semantic}  & 3,417 & 3,417 & 361 & 2,032 & -- \\
        & Web & CuratedTREC~\cite{baudivs2015modeling} & 1,353 & 1,353 & 133 & 694 & -- \\
        & Finance &FiQA-2018~\cite{maia201818}  & 5,500 & 14,166 & 500 & 648 & 57,638 \\
        & Biomedical &BioASQ~\cite{tsatsaronis2015overview} & 3,743 & 35,285 & -- & 497 & 15,559,157 \\
        & StackEx. &CQADupStack~\cite{hoogeveen2015cqadupstack} & -- & -- & -- & 13,145 & 457,199 \\
        & Quora &Quora~\cite{quora}  & -- & -- & 5,000 & 10,000 & 522,931 \\
        & News &ArchivalQA~\cite{wang2021archivalqa} & 853,644 & 853,644 & 106,706 & 106,706 & 483,604 \\
        & Web &CCQA~\cite{huber2021ccqa}  & 55M & 130M & -- & -- & -- \\
        \midrule
        \multirow{4}*{\makecell[c]{Other tasks}} 
        & Wikipedia &FEVER~\cite{thorne2018fever}  & -- & 140,085 & 6,666 & 6,666 & 5,416,568 \\
        & Wikipedia &Climate-FEVER~\cite{diggelmann2020climate}  & -- & -- & -- & 1,535 & 5,416,593 \\
        & Scitific &SciFact~\cite{wadden2020fact} & 809 & 920 & -- & 300 & 5,183 \\
        & Scitific &SciDocs~\cite{cohan2020specter} & -- & -- & -- & 1,000 & 25,657 \\
    \bottomrule
    \end{tabular}
    \label{tab:datasets}
\end{table*}

\subsection{Datasets}
Compared with traditional retrieval models, dense retrieval models are more data hungry, requiring large-scale labeled datasets to learn the parameters of the PLMs.
In recent years, there are a number of retrieval datasets with relevance judgements released publicly, which significantly advances the research on dense retrieval.  We categorize the available retrieval datasets into three major categories according to their original tasks, namely information retrieval, question answering and other tasks.   
The statistics of the available datasets are shown in Table~\ref{tab:datasets}.

As we can see from Table~\ref{tab:datasets}, 
Wikipedia and Web are  two major resources for creating these datasets. Among these datasets,  \emph{MS MARCO} dataset~\cite{msmarco}  contains a large amount of queries with annotated relevant passages in Web documents, and   \emph{Natural Questions (NQ)} dataset~\cite{nq} contains Google search queries with annotations (paragraphs and answer spans) from the top ranked Wikipedia pages. Among these datasets, MS MARCO and NQ datasets have been widely used for evaluating dense retrieval models. 
A recent study~\cite{Craswell2021MSMB} has summarized the promotion effect of MS MARCO on the progress of dense retrieval: ``\emph{the MS MARCO datasets have enabled large-data exploration of neural models}''.
Besides, based on MS MARCO, several variants have been created in order to enrich the evaluation characteristics on some specific aspect, \eg  
the multilingual version  mMARCO~\cite{bonifacio2021mmarco}
and the MS MARCO Chameleons dataset~\cite{arabzadeh2021ms} (consisting of obstinate queries that are difficult to answer by neural retrievers despite having the similar query length and distribution of relevance judgements as queries that are easier to answer). 
Besides, Sciavolino et al.~\cite{Sciavolino2021SimpleEQ} create EntityQuestions dataset as a challenging test set for the models trained on NQ, which contains simple factoid questions about entities from Wikipedia. 
Another interesting observation is that there are increasingly more domain-specific retrieval datasets, including COVID-19 pandemic dataset~\cite{roberts2020trec}, financial dataset~\cite{maia201818}, biomedical dataset~\cite{tsatsaronis2015overview}, climate-specific dataset~\cite{diggelmann2020climate} and scientific dataset~\cite{wadden2020fact}.
 
Besides the presented datasets in Table~\ref{tab:datasets}, several more comprehensive benchmark datasets are released to evaluate the overall retrieval capability of the retrieval models by aggregating representative datasets and conducting diverse evaluation tasks, such as  \emph{BEIR}~\cite{thakur2021beir} and \emph{KILT}~\cite{petroni2020kilt}.

Although existing datasets largely improve the training of dense retrievers,
in these datasets,  a  query  typically corresponds to  very few relevance judgements.
For example, Nogueira et al. observe that most of queries in MS MARCO dataset contains only one labeled positive\cite{arabzadeh2021shallow}, which is likely to be smaller than the actual number of relevant ones in the collection. It is mainly because it is time-consuming to construct complete relevance judgements for a large dataset. 
Incomplete relevance annotations lead to potential training issues such as false negative, which would harm the retrieval performance.
 
Besides, to date, most of the released datasets are created in English, and it is more difficult to obtain sufficient labeled data for training a non-English dense retriever. 
More recently, DuReader-retrieval~\cite{qiu2022dureader_retrieval} releases a large-scale Chinese dataset consisting of 90K queries from Baidu search and  over 8M passages for passage retrieval. In order to enhance the evaluation quality, DuReader-retrieval tries to reduce the false negatives in development and testing sets, and also  
removes the training queries that are semantically similar to the development and testing queries. 
Besides, it also provides human-translated queries (in English) for cross-lingual retrieval.

\subsection{Evaluation Metrics}

To evaluate the retrieval capacity of an information retrieval system, a number of factors need to be considered~\cite{harman2011information,harman2011information,vargas2014novelty}: effectiveness, efficiency, diversity, novelty and so on.
This survey mainly focuses on the effectiveness for the retrieval  system\footnote{Note that these metrics can be used in both first-stage retrieval and  reranking. However, we only describe the evaluation metrics from a general ranking perspective without considering specific task scenarios.}. 
We next introduce the commonly used evaluation metrics for ranking, including Recall, Precision,  MAP, MRR and NDCG.
For dense retrieval tasks, top ranked texts are more important for evaluation, and therefore cut-off metrics are often adopted to examine the quality of  texts at top positions. Next, we introduce several commonly used metrics for dense retrieval  in detail. 

In the traditional retrieval benchmarks, 
Recall is the fraction of relevant texts that are actually retrieved by a retrieval model among all the relevant ones, and Recall@$k$~\cite{R-P-AP} calculates a truncated Recall value at the $k$-th position  of a retrieved list:
\begin{equation}
    \text{Recall}@k = \frac{1}{\left|\mathcal{Q}\right|} \sum_{q=1}^{\left|\mathcal{Q}\right|} \frac{\text{\#retr}_{q,k}}{\text{\#rel}_q},
\end{equation}
where $\text{\#retr}_{q,k}$ denotes the number of relevant texts \emph{w.r.t.} query $q$ retrieved at top $k$ positions by a retrieval method, and $\text{\#rel}_q$ denotes the total number of relevant texts for query $q$. Here, we average the $\text{Recall}@k$ values over the queries from the query set $\mathcal{Q}$.

In dense retrieval, there is a commonly used metric, \emph{Top-$k$ Accuracy}, for computing the proportion of queries for which the top-$k$ retrieved texts contain the answers~\cite{dpr2020}, defined as: 
\begin{equation}
    \text{Accuracy}@k = \frac{1}{\left|\mathcal{Q}\right|} \sum_{q=1}^{\left|\mathcal{Q}\right|} \mathbb{I}(\text{\#retr}_{q,k}>0),
\end{equation}
where $\mathbb{I}(\cdot)$ is an binary indicator function that only returns 1 when the case is true. 
In contrast to traditional TREC benchmarks, mainstream dense retrieval benchmarks, such as NQ, aim to find answers to queries instead of retrieving all relevant texts.
According to \cite{dpr2020,rocketqa}, a retrieved list is considered to \emph{accurately} solve a query when it contains the answer, not necessarily retrieving all the relevant texts
\footnote{Note that Accuracy@$k$ is also known as Recall@$k$ in some dense retrieval studies~\cite{rocketqa, gao2021unsupervised}, which is somehow different from the definition used in traditional retrieval benchmarks. }.

Besides, Precision@$k$~\cite{R-P-AP} calculates the average proportion of relevant texts among the top $k$ positions over the query set $\mathcal{Q}$:
\begin{equation}
    \text{Precision}@k = \frac{1}{\left|\mathcal{Q}\right|} \sum_{q=1}^{\left|\mathcal{Q}\right|} \frac{\text{\#retr}_{q,k}}{k}.
\end{equation}

Based on Precision@$k$, the Average Precision~(AP) further averages the precision values at the positions of each positive text for a query:
\begin{equation}
    \text{AP}_q = \frac{1}{\text{\#rel}_q} \sum_{k=1}^L \text{Precision}@k \times \mathbb{I}(q,k),
\end{equation}
where $\text{Precision}@k$ is the per-query version of the precision value at the $k$-th position,  $L$ is the length of a retrieved list and $\mathbb{I}(q,k)$ is an indicator function returning 1 only when the $k$-th position corresponds to a relevant text for query $q$.
Furthermore, 
Mean Average Precision~(MAP)~\cite{R-P-AP} calculates the mean of  Precision scores over a set of queries $Q$:
\begin{equation}
    \text{MAP} = \frac{1}{\left|\mathcal{Q}\right|} \sum_{q=1}^{\left|\mathcal{Q}\right|} \text{AP}_q.
\end{equation}

Not only counting the occurrence  of positive texts, DCG~(Normalized Discounted Cumulative Gain)~\cite{NDCG}  further incorporates the position of a relevant text into consideration, and it prefers a ranking that places a relevant text at a higher position:
\begin{equation}
    \label{equation:dcg}
    \text{DCG}_q@k = \sum_{i=1}^{k}\frac{2^{g_i}-1}{\log_2(i+1)},
\end{equation}
where $g_i$ is the graded relevance score for the $i$-th retrieved text. Based on the above definition of DCG, 
NDCG is the sum of the normalized DCG values at a particular rank position:
\begin{equation}
    \label{equation:ndcg}
    \text{NDCG}@k = \frac{1}{\left|\mathcal{Q}\right|} \sum_{q=1}^{\left|\mathcal{Q}\right|} \frac{\text{DCG}_q@k}{\text{IDCG}_q@k},
\end{equation}
where DCG@$k$ and IDCG@$k$ denote discounted cumulative gain and ideal discounted cumulative gain at a particular rank position $k$, respectively.

Furthermore, MRR~\cite{mrr} 
averages the reciprocal of the rank of the first retrieved positive text over a set of queries $\mathcal{Q}$: 
\begin{equation}
\label{equation:mrr}
\text{MRR} = \frac{1}{\left|\mathcal{Q}\right|} \sum_{q=1}^{\left|\mathcal{Q}\right|} \frac{1}{\text{rank}_q},
\end{equation}
where $\text{rank}_q$ is the position of the first retrieved positive text \emph{w.r.t.} query $q$.

For first-stage retrieval, Recall@$k$ and Accuracy@$k$ are the most commonly used metrics, since its main focus is to recall as many relevant texts or answers as possible at a truncation length; while for ranking, MRR, NDCG and MAP are more commonly used in practice. 
For a comprehensive discussion about IR evaluation, 
the readers are suggested to read more focused references \cite{IR-Book,evaluation-ref1,harman2011information,vargas2014novelty}. 

\subsection{Code Library Resource}

Recently,  several open-sourced dense retrieval libraries have been released for research purpose. As a representative library, Tevatron~\cite{Gao2022TevatronAE} has developed a modularized framework for building dense retrieval models based on PLMs via command-line interfaces. 
It supports a number of important procedures involved in a complete retrieval pipeline including text processing, model training, text encoding, and text retrieval.
It can be accessed at the link: \url{https://github.com/TextTron/Tevatron}.

Besides, Pyserini~\cite{Lin_etal_SIGIR2021_Pyserini}  is a toolkit that is designed to facilitate reproducible research for information retrieval. 
Specifically, it supports both sparse retrieval and dense retrieval with Anserini IR toolkit~\cite{Yang2017AnseriniET} and FAISS~\cite{faiss}. 
It also provides the evaluation scripts for the standard IR test collections. It can be accessed at the link:~\url{http://pyserini.io/}.

In order to enhance the validation of dense retriever checkpoints, 
Asyncval~\cite{Zhuang2022AsyncvalAT} is released to ease and accelerate the  checkpoint validation for dense retrieval models. An important merit is that the training can be decoupled from checkpoint validation with Asyncval. It can be accessed at the link: \url{https://github.com/ielab/asyncval}.

OpenMatch~\cite{openmatch} has been  originally proposed for pre-BERT neural IR models (v1), and extended  (v2) to support dense retrieval models on commonly used benchmarks such as MS MARCO and NQ.  It can be accessed at the link: \url{https://github.com/thunlp/OpenMatch}.

MatchZoo~\cite{MatchZoo} is a text matching library that supports a number of neural text matching models, and allow users to develop new models by providing rich interfaces. It can be accessed at the link: \url{https://github.com/NTMC-Community/MatchZoo}.

PyTerrier~\cite{PyTerrier} is a retrieval framework that supports the declarative use of Python operators for building retrieval pipelines and evaluating retrieval models, covering representative learning-to-rank, neural reranking and dense retrieval models.   It can be accessed at the link: \url{https://github.com/terrier-org/pyterrier}.

SentenceTransformers~\cite{reimers2019sentence} is another library that provides an easy way to compute dense embeddings for sentences and paragraphs based on Transformer  networks. 
Specifically, it integrates the implementation of Sentence-BERT~\cite{reimers2019sentence} and Transformer-based Sequential Denoising Auto-Encoder (TSDAE)~\cite{Wang2021TSDAEUT}. 
It can be accessed at the link:~\url{https://www.sbert.net/}. 

As the supporting resource, we also release an open-sourced implementation of  dense retrievers based on our previous work RocketQA~\cite{rocketqa}, at the link: \url{https://github.com/PaddlePaddle/RocketQA}, including RocketQA~\cite{rocketqa}, RocketQA$_{PAIR}$~\cite{pair} and RocketQAv2~\cite{rocketqav2}.
This software also provides an easy-to-use toolkit with pre-built models (including both English and Chinese models) for direct use after  installation.  We also aggregate other open-sourced codes for related dense retrieval papers in our survey website, which can be found in the link: \url{https://github.com/RUCAIBox/DenseRetrieval}.

%% file: sec/sec-network.tex
\begin{table*}[htbp]
    \centering
    \caption{A detailed list of different dense retrieval methods in the literature with detailed configurations (abbreviations are from Table~\ref{tb:abbr}).}
    \footnotesize
    \renewcommand\tabcolsep{2.5pt}
    \begin{tabular}{llcccccccc}
    \toprule
    \multirow{2}{*}{\textbf{Years}} & \multirow{2}{*}{\textbf{Methods}} &  \multirow{2}{*}{\textbf{PLMs}} & \multirow{2}{*}{\textbf{Arch.}} & \multicolumn{4}{c}{\textbf{Training}} & \multirow{2}{*}{\textbf{Other Tricks}} \\
    &&&& \textbf{Loss} & \textbf{Negative} & \textbf{Data Aug.} & \textbf{Pretrain} \\
    \midrule
    2020 & Poly-encoders~\cite{humeau2019poly} & BERT$_\text{base}$  & MR & CE & && TAP & Context-candidate attention\\ 
    2019 & ICT~\cite{latent2019acl} & BERT$_\text{base}$  & SR & IOC &&& TAP & Joint training\\
    \midrule
    2020 & ICT+BFS+WLP~\cite{pretrain2020iclr} & BERT$_\text{base}$ & SR & IOC&&& TAP \\ 
    2020 & REALM~\cite{REALM} & BERT$_\text{base}$ & SR & IOC & & & RAP & Joint training + Async. index\\ 
    2020 & DPR~\cite{dpr2020} & BERT$_\text{base}$ & SR & NLL & IB+SHN & ALD &  \\ 
    2020 & ColBERT~\cite{colbert2020sigir} & BERT$_\text{base}$  & MR & CE & &&& Multi-core pre-processing  \\ 
    2020 & ME-BERT~\cite{MEBERT} & BERT$_\text{large}$ & MR & CE & IB+DyHN &&& Sparse-dense hybrid\\ 
    2020 & RepBERT~\cite{zhan2020repbert} & BERT$_\text{base}$ & SR & IOC & IB &  \\ 
    2020 & ANCE~\cite{ANCE} & RoBERTa$_\text{base}$ & SR & NLL & DyHN & & & Async. index\\ 
    2020 & EZRQG~\cite{liang2020embedding} & BERT$_\text{base}$ & SR & IOC & &&& Query generation\\ 
    2020 & AugmentedBERT~\cite{google2020augmentation} & BERT$_\text{base}$ & SR & IOC+KL & SHN & KD & & Kernel density estimation \\ 
    2020 & In-batch KD~\cite{hofstatter2020improving} & BERT$_\text{base}$ & SR & MSE & IB & KD && Cross-architecture KD \\ 
    2020 & RocketQA~\cite{rocketqa} & ERNIE$_\text{base}$ & SR & NLL & CB+DeHN & KD & & Iter. training \\ 
    2020 & AugmentSBERT~\cite{thakur2020augmented} &  BERT$_\text{base}$ & SR & NLL & SHN+DeHN & KD && Kernel density estimation \\ 
    2020 & TCT-ColBERT~\cite{lin2020distilling} & BERT$_\text{base}$ & SR & KL & IB+SHN & KD && Sparse-dense hybrid \\ 
    2020 & Multi-stage~\cite{lu2021multi} & BERT$_\text{base}$  & SR & NLL & IB+SHN & KD & TAP & Embedding fusion \\ 
    2020 & DKRR\cite{izacard2021distilling} & BERT$_\text{base}$ & SR & KL & SHN & KD && Iter. training\\ 
    2020 & DensePhrases~\cite{lee2021learning} & SpanBERT$_\text{base}$ & PR & IOC+KL & IB & KD && Pre-batch negatives\\ 
    2020 & UnifiedQA~\cite{oguz2020unified} & BERT$_\text{base}$ & SR & NLL & IB & ALD \\ 
    \midrule 
    2021 & Individual Top-k~\cite{sachan2021end} & BERT$_\text{large}$ & SR & IOC+NLL & IB+SHN & & TAP & Joint training \\
    2021 & Grad. Cache~\cite{gao2021scaling} & BERT$_\text{base}$  & SR & NLL & CB &&& Gradient caching \\ 
    2021 & TAS-Balanced~\cite{tasbalanced2021} & BERT$_\text{base}$  & SR & MSE & IB+SHN & KD && Balanced topic aware sampling\\ 
    2021 & ADORE+STAR~\cite{zhan2021optimizing} & RoBERTa$_\text{base}$ & SR & NLL & SHN+DyHN &&& Async. index  \\ 
    2021 & Condenser~\cite{gao2021condenser} & Condenser & SR & CE & IB &  & REP &   \\ 
    {2021} & DRPQ~\cite{tang2021improving} & BERT$_\text{base}$ & MR & NLL & IB+DyHN & && Multiple pseudo query embeddings \\ 
    2021 & RANCE~\cite{prakash2021learning} & RoBERTa$_\text{base}$ & SR & NLL & DeHN\\ 
    2021 & DANCE~\cite{li2021more} & RoBERTa$_\text{base}$ & SR & IOC & SHN &&& Contrastive dual learning\\ 
    2021 & DPR-PAQ~\cite{ouguz2021domain} & RoBERTa$_\text{large}$ & SR & NLL & IB+SHN & ALD & GAP\\ 
    2021 & JPQ~\cite{zhan2021jointly} & BERT$_\text{base}$ & SR & IOC & DyHN &&& Quantization\\ 
    2021 & coCodenser~\cite{gao2021unsupervised} & Condenser & SR & NLL & SHN & & REP & Corpus-aware pretraining  \\ 
    2021 & PAIR~\cite{pair} & ERNIE$_\text{base}$ & SR & IOC+NLL & IB+SHN & KD & REP & Passage-centric constrain\\ 
    2021 & ST5~\cite{Ni2021SentenceT5SS} & T5$_\text{11B}$ & SR & NLL & IB & &  \\ 
    2021 & ANCE-PRF~\cite{yu2021improving} & RoBERTa$_\text{base}$ & SR & NLL & DyHN & && Pseudo relevance feedback \\ 
    2021 & Trans-Encoder~\cite{liu2021trans} & RoBERTa$_\text{base}$ & SR & MSE & DyHN & KD && Iter. training + Mutual distillation \\ 
    2021 & AR2-G~\cite{zhang2021adversarial} & ERNIE$_\text{base}$ & SR & IOC & DyHN & && Adversarial training \\ 
    2021 & RepCONC~\cite{zhan2021learning} & BERT$_\text{base}$ & SR & IOC & DyHN &&& Quantization \\ 
    2021 & RocketQAv2~\cite{rocketqav2} & ERNIE$_\text{base}$ & SR & KL & SHN+DeHN & KD & & Joint training + Mutual distillation \\ 
    2021 & SPAR~\cite{chen2021salient} & BERT$_\text{base}$ & SR & NLL & SHN & KD & & \\
    2021 & SEED~\cite{lu2021less} & BERT$_\text{base}$ & SR & NLL & DyHN & & TAP & Autoencoder pretraining \\ 
    2021 & ColBERTv2~\cite{santhanam2021colbertv2} & BERT$_\text{base}$ & SR & NLL & DeHN & KD && Residual compression + Quantization \\ 
    2021 & GPL~\cite{wang2021gpl} & DistilBERT$_\text{base}$ & SR & MSE & IB+SHN & KD & TAP & Query generation \\ 
    2021 & Spider~\cite{Ram2021LearningTR} & BERT$_\text{base}$ & SR & NLL & IB+SHN & & TAP & Sparse-dense hybrid \\ 
    2021 & DrBoost~\cite{Lewis2021BoostedDR} & BERT$_\text{base}$ & SR & NLL & SHN &&& Boosting \\ 
    2021 & GTR~\cite{Ni2021LargeDE} & T5$_\text{11B}$ & SR & NLL & IB+DeHN & & TAP & \\ 
    2021 & Contriever~\cite{Izacard2021TowardsUD} & BERT$_\text{base}$ &SR& NLL & IB & & REP & Contrastive learning\\ 
    \midrule
    2022 & Uni-Retriever~\cite{Zhang2022UniRetrieverTL} & BERT$_\text{base}$ & SR & IOC+NLL & CB + SHN & &&    \\ 
    2022 & LaPraDoR~\cite{xu2022laprador} & DistilBERT$_\text{base}$ & SR & IOC+NLL & IB & & REP & Sparse-dense hybrid \\ 
    {2022} & HLP~\cite{zhou2022hyperlink} & BERT$_\text{base}$  & SR & NLL & IB && TAP &\\ 
    {2022} & DAR~\cite{jeong2022augmenting} & RoBERTa$_\text{large}$ & SR & NLL & IB+SHN &&& Interpolation and perturbation \\ 
    2022 & MVR~\cite{zhang2022multi} & BERT$_\text{base}$ & MR & IOC+NLL & SHN & & & Multi-view representation \\ 
    2022 & ColBERTer~\cite{Hofsttter2022IntroducingNB} & DistilBERT$_\text{base}$ & MR & IOC+MSE & SHN & & &  \\ 
    2022 & CharacterBERT~\cite{zhuang2022characterbert} & CharacterBERT & SR & NLL+KL & IB+SHN & & & Self-teaching  \\ 
    {2022} & GNN-encoder~\cite{liu2022gnn} & ERNIE$_\text{base}$ & SR & IOC+NLL & IB+DeHN & KD &  & Query-passage interaction \\ 
    {2022} & COSTA~\cite{ma2022pre} & BERT$_\text{base}$ & SR & NLL & SHN & & TAP & Contrastive span prediction\\ 
    2022 & CL-DRD~\cite{CL-DRD} & DistilBERT$_\text{base}$ & SR & KL & KD && & Curriculum Learning \\ 
    2022 & ERNIE-Search~\cite{ERNIE-Search} & ERNIE$_\text{large}$ & SR & NLL+KL & IB & KD & & Joint training + Mutual distillation \\ 
    2022 & RetroMAE~\cite{RetroMAE} & BERT$_\text{base}$ & SR & IOC+NLL & DyHN & KD & REP & MAE pretraining \\ 
    {2022} & DCSR~\cite{hong2022sentence} & BERT$_\text{base}$ & MR & NLL & IB+SHN &  \\ 
    {2022} & ART~\cite{sachan2022questions} & BERT$_\text{large}$ & SR & KL & &&& Query reconstruction \\ 
    2022 & SimLM~\cite{SimLM} & BERT$_\text{base}$  & SR & NLL & IB+DeHN & KD & REP & Iter. training\\ 
    2022 & RoDR~\cite{LocalRanking} & BERT$_\text{base}$ & SR & NLL+KL & IB+SHN & & & Passage-centric constraint\\ 
    2022 & Aggretriever~\cite{Lin2022AggretrieverAS} & Condenser & SR & NLL & IB+SHN & & &Text aggregation  \\ 
    {2022} & CoT-MAE~\cite{wu2022contextual} & BERT$_\text{base}$  & SR & NLL & SHN & & GAP  \\ 
    {2022} & CPDAE~\cite{ma2022contrastive} & BERT$_\text{base}$ & SR & IOC+NLL & SHN & & TAP & Autoencoder pretraining \\ 
    2022 & DPTDR~\cite{tang2022dptdr} & RoBERTa$_\text{large}$ & SR & NLL & IB+DeHN & KD & TAP & Deep prompt tuning \\ 
    2022 & LED~\cite{Zhang2022LEDLD} & Condenser &SR&IOC+NLL & IB+SHN & KD & & Rank-consistent regularization  \\ 
    2022 & LexMAE~\cite{LexMAE} & BERT$_\text{base}$ & SR & IOC+NLL+KL & SHN & KD & REP & MAE pretraining \\ 
    2022 & Promptagator~\cite{Dai2022PromptagatorFD} & FLAN & SR & NLL & IB+SHN &&&Prompt-base query generation\\ 
    2022 & PROD~\cite{PROD} & ERNIE$_\text{base}$ & SR & NLL+KL & IB+SHN & KD & & Curriculum learning  \\ 
    {2022} & TASER~\cite{cheng2022task} & Condenser & SR & NLL & IB+SHN &&& Mixture-of-experts \\
    \bottomrule
    \end{tabular}
    \label{tab:methods}
\end{table*}

\begin{table*}[]
    \renewcommand\tabcolsep{16pt}
    \caption{Abbreviations for different techniques or strategies.}\label{tb:abbr}
    \centering
    \begin{tabular}{c|c|l|l|c}
    \toprule
    \multicolumn{4}{c|}{\textbf{Type}} & \textbf{Abbr.} \\
    \midrule
    \multirow{3}{*}{Architecture} & \multicolumn{3}{c|}{Single-representation~(Sec.~\ref{sec:bi-encoder})} & SR \\
    & \multicolumn{3}{c|}{Multi-representation~(Sec.~\ref{sec:bi-encoder})} & MR \\
     & \multicolumn{3}{c|}{Phrase-level representation~(Sec.~\ref{sec:bi-encoder})} & PR \\
    \midrule
    \multirow{17}{*}{Training} & \multirow{6}{*}{Loss} & \multicolumn{2}{c|}{Negative log-likelihood loss~(Sec.~\ref{sec:loss})}  & NLL \\
     &  & \multicolumn{2}{c|}{Cross-entropy loss~(Sec.~\ref{sec:loss})} & CE \\
     &  & \multicolumn{2}{c|}{Triplet ranking loss~(Sec.~\ref{sec:loss})} & TR \\
     &  & \multicolumn{2}{c|}{MSE loss~(Sec.~\ref{sec:knowledge-distillation})} & MSE \\
     &  & \multicolumn{2}{c|}{KL-divergence loss~(Sec.~\ref{sec:knowledge-distillation})} & KL \\
     &  & \multicolumn{2}{c|}{Incorporating optimization constraints~(Sec.~\ref{sec:knowledge-distillation})} & IOC \\
    \cmidrule{2-5}
     & \multirow{5}{*}{Negative selection} & \multicolumn{2}{c|}{In-batch negatives~(Sec.~\ref{sec:in-batch})} & IB \\
     &  & \multicolumn{2}{c|}{Cross-batch negatives~(Sec.~\ref{sec:cross-batch})} & CB \\
    \cmidrule{3-5}
     &  & \multicolumn{1}{l|}{\multirow{3}{*}{Hard negatives}} & Static hard negatives~(Sec.~\ref{sec:hard-negatives}) & SHN \\
     &  &  & Dynamic hard negatives~(Sec.~\ref{sec:hard-negatives}) & DyHN \\
     &  &  & Denoised hard negatives~(Sec.~\ref{sec:hard-negatives}) & DeHN \\
    \cmidrule{2-5}
     & \multirow{2}{*}{Data augmentation} & \multicolumn{2}{c|}{Auxiliary labeled datasets~(Sec.~\ref{sec:auxiliary})} & ALD \\
     &  & \multicolumn{2}{c|}{Knowledge  distillation~(Sec.~\ref{sec:knowledge-distillation})} & KD \\
    \cmidrule{2-5}
     & \multirow{4}{*}{Pretraining} & \multicolumn{2}{c|}{Task adaptive pretraining~(Sec.~\ref{sec:task-adaptive})} & TAP \\
     &  & \multicolumn{2}{c|}{Generation-augmented pretraining~(Sec.~\ref{sec:generation-augmented-pretraining})} & GAP \\
     &  & \multicolumn{2}{c|}{Retrieval-augmented pretraining~(Sec.~\ref{sec:retrieval-augmented-pretraining})} & RAP \\
     &  & \multicolumn{2}{c|}{Representation enhanced pretraining~(Sec.~\ref{sec:representation-enhanced-pretraining})} & REP \\
    \bottomrule
    \end{tabular}
\end{table*}

\section{Architecture}

This section describes the major network architectures for dense text retrieval. We start with the  background of Transformer and PLMs, then introduce two typical neural architectures for building dense retrieval models, and finally compare sparse and dense retrieval models. 

\subsection{Background}

\subsubsection{Transformer and Pretrained Language Models}

Transformer~\cite{transformer} has become the mainstream backbone for language model pretraining, which was originally proposed for efficiently modeling sequence data. Different from traditional sequence neural networks (\ie RNN and its variants~\cite{hochreiter1997long,cho2014properties}),  Transformer no longer processes the data in order,  but instead introduces a new \emph{self-attention} mechanism:  a token attends to all the positions from the input. 
Such an architecture is particularly suited to be parallelized with the support of GPU or TPU.
As a remarkable contribution, Transformer makes it more flexible and easier to train very large neural networks.  
There are a large number of variants that improve the basic Transformer architecture, and the readers can refer to the related surveys~\cite{transformersurvey-1,transformersurvey-2} for a comprehensive discussion.

Based on Transformer and its variants, pretrained language models~(PLM) are proposed~\cite{bert2019naacl,liu2019roberta,brown2020language} in the field of natural langauge processing. PLMs are pretrained over a large-scale general text corpus with specially designed self-supervised loss functions, and can be 
further fine-tuned  according to various downstream tasks. Such a learning approach is called \emph{pretraining then fine-tuning paradigm}~\cite{bert2019naacl,Liu2022PretrainPA}, which is one of the most striking achievements  on language intelligence in recent years. 
As one of the most representative PLMs, BERT~\cite{bert2019naacl}  pretrains deep bidirectional architectures for learning general text representations (with the trick of word masking), which has largely raised the performance bar in a variety of natural language processing tasks. 
The success of BERT inspires a series of follow-up studies, including the improved pretraining approach~\cite{liu2019roberta},  the refined bidirectional representations~\cite{ma2019universal}, and model compression~\cite{sanh2019distilbert}. Furthermore, several studies attempt to train extremely large PLMs in order to explore the performance limit as a universal learner~\cite{brown2020language,Raffel2020ExploringTL}, achieving very impressive results on downstream tasks. 

Beyond language modeling, PLMs have so far become a general technical approach to model, represent and manipulate large-scale unlabeled data, which are also called \emph{foundation models}~\cite{FoundationModels}. 
A detailed introduction of recent progress on foundation models is beyond the scope of this survey, which is omitted in this survey. 

\subsubsection{PLMs for Sparse Retrieval}
\label{sec:sparse-retrieval-PLM}
Before introducing the use of PLMs for dense retrieval, this part briefly reviews how to utilize PLMs for improving sparse retrieval models that rely on lexical match. 
Basically speaking, this line of research work can be roughly divided into two major categories, namely term weighting and term expansion.

The first category of work aims to improve term weighting based on per-token contextualized representations. 
As a representative study, DeepCT~\cite{Dai2020ContextAwareTW} utilizes the learned BERT token representations for estimating context-specific importance of an occurring  term in each passage. 
The basic idea is to regress the token representations into real-valued term weights. 
HDCT~\cite{dai2020context} extends it to long documents by splitting the documents into passages, and  aggregates the estimated term weights in each passage. 
Specifically, it leverages three kinds of data resources (\ie  titles, Web inlinks, and pseudo-relevance feedback) to generate weak supervision signals for learning term weights. 
It can better learn the importance of a term by 
modeling its text context. 
Further, COIL~\cite{gao2021coil} utilizes the contextualized token representations of \emph{exact matching terms} for estimating the relevance. 
It computes the dot product between the token representations from the query encoder and text encoder (only considering the overlapping terms), and then sums the term-specific similarity scores as the relevance score\footnote{In order to alleviate the term mismatching problem, COIL also incorporates semantic matching based on ``\textsc{[CLS]}'' embeddings.}. 
To better understand the above models,   Lin et al.~\cite{Lin2021AFB} propose a conceptual framework to unify these approaches:  
DeepCT aims to assign \emph{scalar weights} to query terms, while COIL aims to assign \emph{vector weights} to query terms. Furthermore, uniCOIL~\cite{Lin2021AFB} is also proposed by reducing the weight vector in COIL to one dimension, and the experiments in \cite{Lin2021AFB} show that uniCOIL can retain the effectiveness while increasing the efficiency.

The second category of work expands queries or documents by using PLMs to mitigate the vocabulary mismatching problem.  
For example, docTTTTTquery~\cite{doctttttquery}  predicts a set of queries that a document will be relevant to, such that the predicted queries can be used to enrich the document content. 
In addition, SPLADE~\cite{formal2021splade} and SPLADEv2~\cite{formal2021spladev2} project each term in queries and texts to a vocabulary-sized weight vector, in which each dimension represents the weight of a term in the vocabulary of PLMs. 
The weights are estimated by using the logits of masked language models. 
Then, the final representation of a whole text (or query) is  obtained by combining  (\eg sum or max pooling) the  weight vectors estimated by all the  tokens of the text. 
Such a representation can be viewed as an expansion of a query (or a text), since it contains the terms that  do not occur in the query (or the text). 
A sparsity regularization is further applied to obtain the sparse representation, so as to efficiently use the inverted index.

Despite the use of PLMs, these approaches are still based on  lexical matching. 
They can reuse the traditional index structure by incorporating additional payloads (\eg contextualized embeddings~\cite{gao2021coil}). 
For a more thorough discussion of PLM-enhanced sparse retrieval models, the readers can refer to \cite{fan2021pre, cai2021semantic,lin2021pretrained}. 
 
\subsection{Neural Architecture for Dense Retrieval}
The essence of dense retrieval is to model the \emph{semantic interaction} between queries and texts based on the representations learned in latent semantic space. 
Based on different interaction modeling ways,  there are two mainstream architectures for dense retrieval, namely the cross-encoder and bi-encoder. 

\begin{figure}[t]
	\centering
	\subfigure[Dual-encoder architecture]{\label{fig:dual-encoder}
		\centering
		\includegraphics[width=0.44\textwidth]{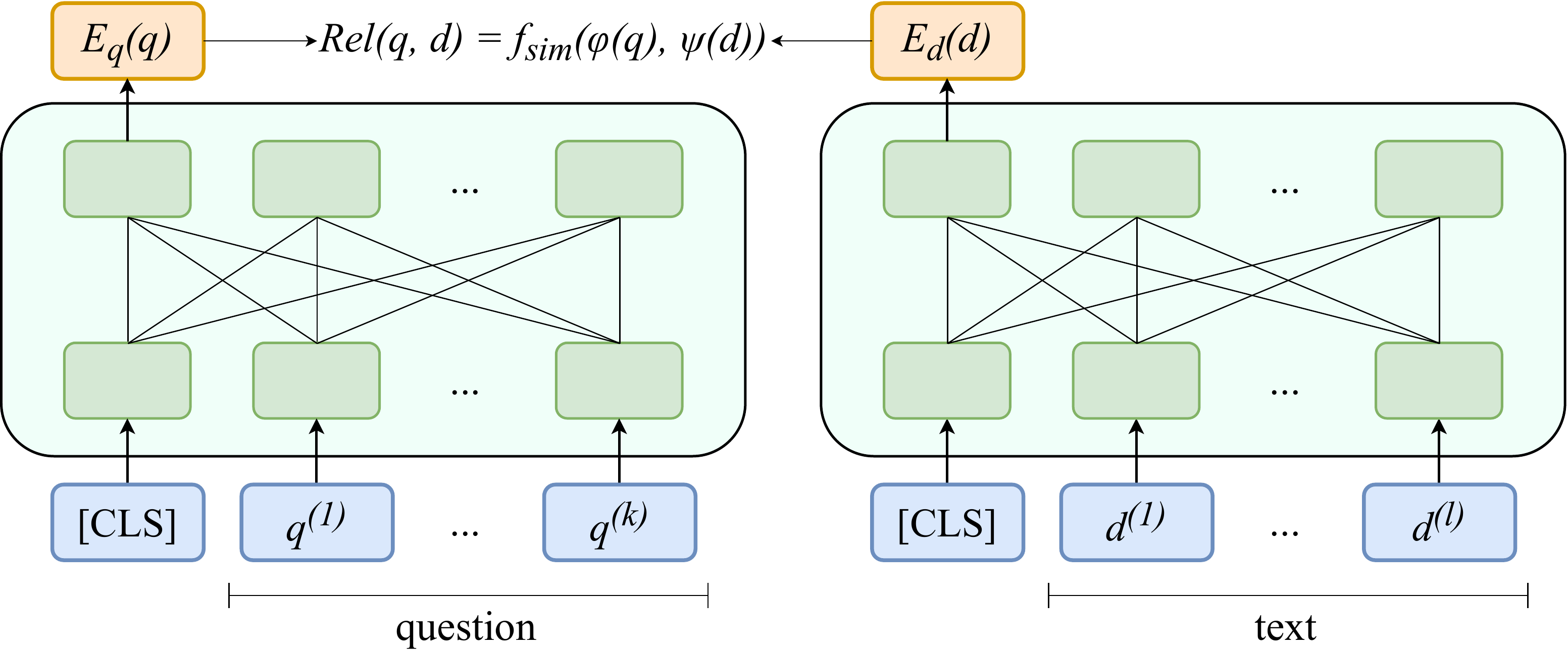}
	}
	\subfigure[Cross-encoder architecture]{\label{fig:cross-encoder}
		\centering
		\includegraphics[width=0.44\textwidth]{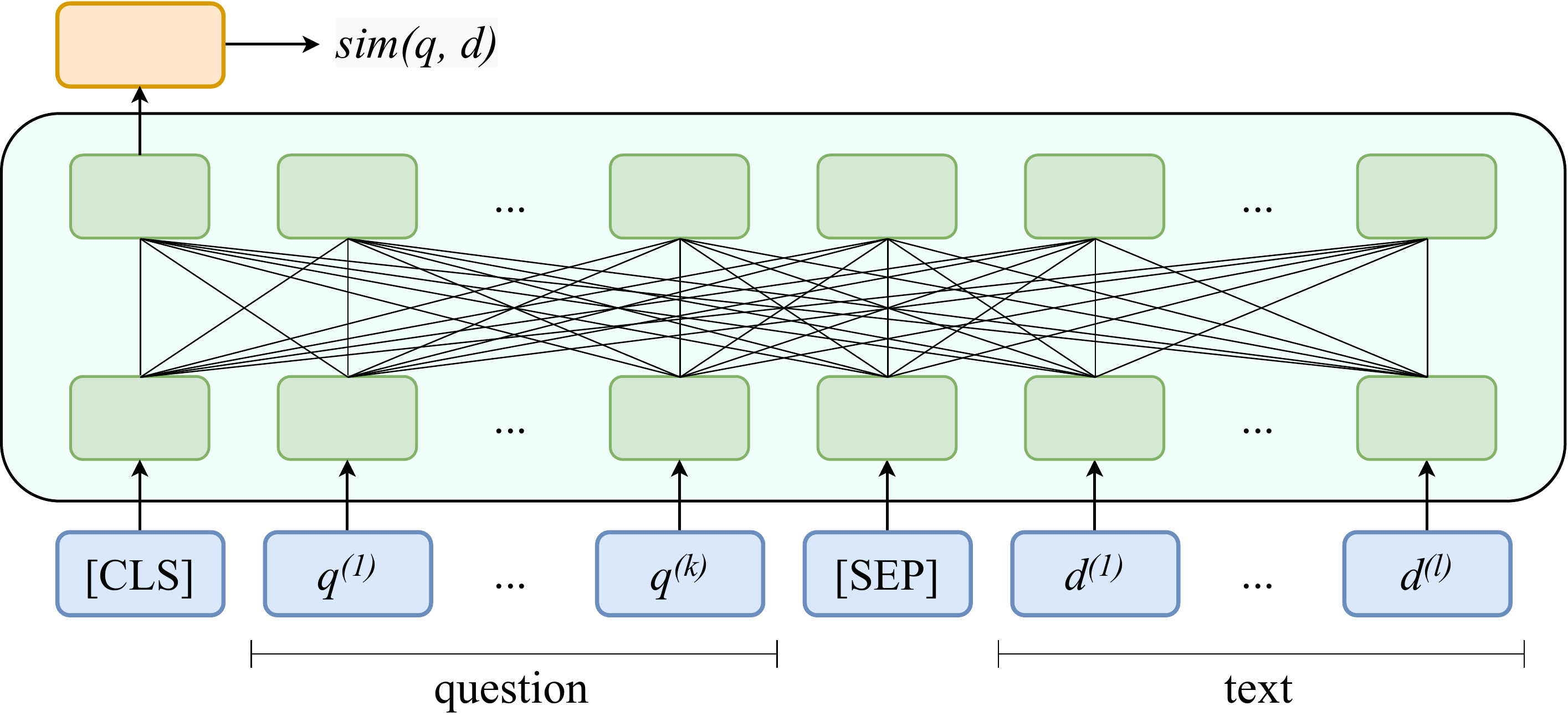}
	}
	\caption{Visual illustration of dual-encoder and cross-encoder architectures.}
	\label{fig:encoders}
\end{figure}

\subsubsection{The Cross-Encoder Architecture}
As a direct application of PLMs,  cross-encoder considers a query-text pair as an entire ``sentence''.
To be specific, the input of cross-encoder is the concatenation of a query and a text, separated by a special symbol ``\textsc{[SEP]}''.
In order to obtain the overall representation, another symbol  ``\textsc{[CLS]}'' is inserted at the beginning of the concatenated sentence.  
Then, the query-text sequence is fed into a PLM for modeling the semantic interaction between any two tokens of the input sequence.  
After the representations for each token in the sequence have been learned, we can obtain the match representation for this query-text pair. 
A commonly used way is to take the ``\textsc{[CLS]}'' representation as the semantic matching representation~\cite{Qiao2019UnderstandingTB}.
Other variants can be also used, \eg averaging token embeddings~\cite{reimers2019sentence}. 
Such an architecture is similar to  \emph{interaction-based architecture} in the pre-BERT studies~\cite{guo2016deep, xiong2017end}, since it allows the tokens to interact across queries and texts. 

\subsubsection{The Bi-Encoder Architecture}
\label{sec:bi-encoder}
The bi-encoder (\aka dual-encoder) adopts the two-tower architecture, which is similar to  
representation-based approaches (\eg DSSM~\cite{huang2013learning}) in pre-BERT studies.
The major difference is that it replaces the previously used multi-layered perceptions (or other neural networks) with PLM-based encoders. 
Next, we will discuss the major progress of bi-encoder architecture for dense retrieval.

\paratitle{Single-representation bi-encoder}. 
As the basic form of bi-encoder architecture, it first learns latent semantic representations for both query and text with two separate encoders, called \emph{query embedding} and \emph{text embedding}, respectively. Then,   
the relevance score can be computed via some similarity function (\eg cosine similarity and inner product) between the query embedding and text embedding. As mentioned before, we can directly encode the query and text by placing a special symbol (\eg ``\textsc{[CLS]}'' in BERT) at the beginning of a text sequence, 
so that the learned representation of the special symbol can be used to represent the semantics of a text (or query). 
Most of the single-representation dense retrievers~\cite{dpr2020,ANCE,rocketqa}  learn the query and text representations with encoder-only PLMs (\eg BERT~\cite{bert2019naacl}, RoBERTa~\cite{bert2019naacl}, and ERNIE~\cite{ernie20aaai}). 
More recently, text-to-text Transformer (T5)~\cite{Raffel2020ExploringTL}, which is an encoder-decoder based PLM,  has  been explored to learn text representations for dense retrieval~\cite{Ni2021SentenceT5SS,Ni2021LargeDE}. It has been shown that a T5-based sentence embedding model outperforms  Sentence-BERT~\cite{Reimers2019SentenceBERTSE} on SentEval~\cite{Conneau2018SentEvalAE} and SentGLUE~\cite{Raffel2020ExploringTL}  tasks.

\paratitle{Multi-representation bi-encoder}. A major limitation of single-representation bi-encoder is that it cannot well model  fine-grained semantic interaction between query and text. To address  this issue, several researchers propose to explore multiple-representation bi-encoder for enhancing the text representation and semantic interaction. 
The poly-encoder model~\cite{humeau2019poly} learns $m$ different representations for modeling the semantics of a text in multiple views, called \emph{context codes} in the original paper. 
During query time, by attending to the query vector, these $m$ representations are then combined into a single vector.
Finally, the inner product between the query vector and aggregated vector is computed as the relevance score.
As another related study, ME-BERT~\cite{MEBERT} also generates $m$ representations for a candidate text, by directly taking 
the contextualized representations of the first $m$ tokens. The text relevance is computed using 
the maximum inner product between the query vector and the $m$ contextual representations.
Furthermore, ColBERT~\cite{colbert2020sigir} designs an extreme multi-representation semantic matching model, where 
per-token contextualized representations are kept for query-text interaction.
Different from the representation scheme of cross-encoder (Section 4.2.1), these contextualized representations do not directly interact across queries and texts during encoding.  
Instead, they introduce a \emph{query-time} mechanism for per-token representation interaction, called \emph{late interaction}~\cite{colbert2020sigir}.
As the extension of ColBERT, 
ColBERTer~\cite{Hofsttter2022IntroducingNB} designs an approach by combining single-representation (the ``\textsc{[CLS]}'' embedding) and multi-representation (per-token embeddings) mechanisms for matching, achieving performance improvement over ColBERT.
More recently, MVR~\cite{zhang2022multi} proposes to insert multiple  
``\textsc{[VIEW]}'' tokens (similar to ``\textsc{[CLS]}'') at the beginning of a text, and aims to learn the text representations in multiple different views. Further, they design a special local loss to distinguish the best matched view from the rest views of a text for a query.
 {MADRM~\cite{multi-aspect} proposes to learn multiple aspect embeddings for both query and text, and employ explicit aspect annotations to supervise the aspect learning. 
Instead of using standard dense retrieval benchmarks, this work conducts the evaluation experiments based on the e-commerce datasets.} 

These multi-representation bi-encoders share the idea of using multiple contextual embeddings for representing queries or texts, such that the similarity can be measured from different semantic views. These contextualized representations are learned and stored during 
training and indexing; while at query time, we can model fine-grained semantic interaction between query embeddings and pre-computed text embeddings. Such an approach is effective to improve  the retrieval performance, but causes a significantly high cost to  maintain the multi-representation implementation (\eg increased index size), which is not practical in real-world applications.
Considering this issue, some specific  strategies proposed in ColBERTer~\cite{Hofsttter2022IntroducingNB} can be used to reduce the  multi-representation costs, such as embedding dimension reduction, bag of unique whole-word representations and  contextualized stopword removal. The experiments in \cite{Hofsttter2022IntroducingNB} demonstrate that ColBERTer can significantly reduce the space cost while retaining the effectiveness.

\paratitle{Phrase-level representation}. Generally, dense retrieval models focus on solving document- or paragraph-level text retrieval tasks. It is natural to extend existing retrieval models by considering more fine-grained retrieval units (\eg phrases or contiguous text segments) for specific tasks, such as open-domain question answering and slot filling~\cite{lee2021learning}.
Recently, several studies propose to learn phrase-level representations for directly retrieving phrases as the answers to  queries~\cite{seo2019real, lee2021learning, lee2021phrase, seo2018phrase}. 
The basic idea is to preprocess all the documents in a query-agnostic way 
and generate phrase-level represenations (called \emph{phrase embeddings}) for contiguous text segments in the documents. 
Then, the answer finding task is cast into the problem of retrieving the nearest phrase embeddings to the query embedding.
PIQA (phrase-indexed question answering)~\cite{seo2018phrase} presents the first approach that leverages dense phrase index for question answering. 
DenSPI~\cite{seo2019real} further combines dense and  sparse vectors of phrases to capture both semantic and lexical information, so as to improve the performance of phrase-indexed question answering.
While, such an  approach heavily relies  on sparse vectors for achieving good performance.  
Considering this issue,  
Seo et al.~\cite{lee2021learning}  propose DensePhrases, an approach for improving the learning of dense phrase encoder without using sparse representations. DensePhrases learns dense phrase representations 
by data augmentation and knowledge distillation, and further employs enhanced negative training (considering both in-batch and pre-batch negatives) and query-side fune-tuning. 
An interesting observation found in \cite{lee2021phrase} is that a dense phrase retrieval system can achieve a better passage retrieval accuracy than DPR on NQ and TriviaQA, without any re-training.
It shows that phrase-level information is  useful for relevance matching by capturing fine-grained semantic characteristics. 
They further propose that dense  phrase retriever can be considered as a dynamic version of multi-representation retriever, since it dynamically generates a set of phrase embeddings for each text.

\paratitle{Other improvement variants.} Besides the above major representation ways, there are also several improvement variants in different aspects.

$\bullet$ \emph{Composite architectures}. The above bi-encoders and variants can be applied for either document retrieval or passage retrieval, depending on specific tasks.
We can also devise composite architectures by integrating text encoders in different  semantic  granularities.
In a recent work~\cite{liu2021dense}, a hierarchical retrieval approach is proposed by combining a document retriever and a passage retriever, considering both document-level and passage-specific semantics. 
Besides, a boosting approach has been proposed in \cite{Lewis2021BoostedDR} that ensembles multiple ``weak'' dense retrievers for achieving a good retrieval performance.  { Further, 
Li et al.~\cite{uncertaintyfusion}  propose to combine multiple dense retrievers trained with different tasks in an  uncertainty-weighted way. }

$\bullet$ \emph{Lightweight architectures}. Besides retrieval performance,  several  approaches  aim to learn  lightweight representations, including the query embedding pruning methods~\cite{tonellotto2021query} according to the estimated term importance, lightweight late-interaction variant~\cite{santhanam2021colbertv2} (\ie \emph{ColBERTv2}) that utilizes cross-encoder distillation with hard negatives and  residual representation compression, and product quantization approaches~\cite{zhan2021jointly} that uses discrete  quantization representations  (detailed in Section~\ref{sec:quantization}). 

$\bullet$ \emph{Incorporating relevance feedback}.  As a classic strategy, pseudo relevance feedback has been re-visited in the setting of dense retrieval to enhance the query representation~\cite{li2021pseudo,yu2021improving,wang2021pseudo,Li2021ImprovingQR}. The basic idea is to utilize the initially retrieved documents to derive the 
enhanced query embedding. 
Furthermore, Zhuang et al.~\cite{Zhuang2022ImplicitFF} explore click-through data as feedback signal by using simulation based on click models, and then propose a counterfactual approach to dealing with the bias issue in click data.  {Wang et al.~\cite{feed-multi} employ relevance feedback to enhance the multi-representation dense retrieval models, showing a large performance improvement over the base model ColBERT. }

$\bullet$ 
\emph{Parameter-efficient tuning}. Since it is costly to fine-tune large-scale PLMs, parameter-efficient tuning~\cite{Liu2022PretrainPA}  has been proposed for adapting PLM to the retrieval task~\cite{tang2022dptdr,Tam2022ParameterEfficientPT}. Generally, parameter-efficient tuning tends to underperform fine-tuning methods when applied directly to dense retrieval. 
Considering this issue, DPTDR introduces retrieval-oriented  pretraining and negative mining~\cite{tang2022dptdr} for enhancing the parameter-efficient tuning, which achieves a comparable performance as  fine-tuned retrievers.  Besides, the experiments in~\cite{Tam2022ParameterEfficientPT} have shown that parameter-efficient tuning (\eg P-tuning v2~\cite{Liu2021PTuningVP}) can improve the zero-shot retrieval performance of DPR and ColBERT.

\subsubsection{Comparison between Cross- and Bi-Encoder}
Ever since the pre-BERT age, \emph{interaction-based} or \emph{representation-based} approaches have been proposed as the two major architectures for neural IR~\cite{guo2016deep}, according to whether fined-grained tokens can interact across queries and texts. 
Following this convention, cross-encoder and bi-encoder can be understood as PLM-based instantiations for the two kinds of architectures, respectively.   
In order to have a deeper understanding, we next compare cross-encoder and bi-encoder from the perspective of dense text retrieval.

Firstly,  cross-encoder is more capable of learning fine-grained semantic interaction information for the query-text pair. 
It is widely recognized that cross-encoder is more effective in relevance matching~\cite{rocketqa, nogueira2019passage, google2020augmentation}.
In contrast, bi-encoder (in an original implementation) cannot  capture the fine-grained interaction between query and text representations. 
As discussed above, a number of studies aim  to improve the representation capacity of bi-encoder by using multiple context embeddings~\cite{humeau2019poly, colbert2020sigir}. 
According to the reported results on MS MARCO~\cite{rocketqav2, santhanam2021colbertv2}, the multi-representation bi-encoder performs better than the single-representation bi-encoder, but still worse than the cross-encoder, since it attempts   to mimic the cross-encoder in modeling fine-grained semantic interaction. 

Second, bi-encoder is more flexible and efficient in architecture. For flexibility, it can employ different encoding networks for encoding queries and texts, which allows more flexible architecture design for dense retrieval, \eg phrase index~\cite{seo2019real}.  
For efficiency, it can be accelerated via approximate nearest neighbor  search for large-scale vector recall~\cite{faiss}, which is particularly important for practical use of dense retrieval approaches.

Considering the pros and cons of the two architectures, in practice, they are often jointly used in retrieval systems for  a trade-off between effectiveness and efficiency. 
Typically,  bi-encoder is used for large-scale candidate recall (\ie first-stage retrieval), and cross-encoder is adopted to implement the reranker or reader (Section 7.1.2). Besides, cross-encoder is often utilized to improve the bi-encoder, \eg knowledge distillation~\cite{thakur2020augmented, rocketqav2} (Section 5.3.2) and pseudo labeled data generation~\cite{pair, rocketqa} (Section 5.4.2).

\subsection{Sparse Retrieval \emph{v.s.} Dense Retrieval}

In this part, we first discuss the difference/connection between sparse and dense retrieval, and then introduce how to combine both kinds of approaches for retrieval.

\subsubsection{Discussions}
To understand how dense retrieval models behave and their connections with sparse retrieval models, we present some discussions by focusing on the following questions.

\begin{itemize}
\item[$A$] 
 Do dense retrieval models always outperform sparse retrieval models? What are their respective strengths and weaknesses? 
\end{itemize}

On the one hand, in the early literature of dense retrieval~\cite{dpr2020, ANCE, rocketqa}, a number of experimental studies have reported  that dense retrieval models often outperform classic sparse models (\eg BM25), especially on the benchmarks (\eg MS MARCO, Natural Questions and TriviaQA) that were originally designed for question answering.
Since these benchmarks focus on complex query solving, deep semantic matching capabilities are required to resolve queries containing  few overlapping terms with the answers. 
Thus, sparse retrievers are unable to well resolve these queries due to the incapability in handling the term mismatch issue. 
By contrast, dense retrievers are more effective to answer complex queries by measuring the latent semantic similarity. 
Take an example query that cannot be resolved by a sparse retriever from the  DPR paper~\cite{dpr2020}: given the question ``\emph{Who is the bad guy in lord of the rings?}'',  a dense retriever can retrieve a relevant text  span ``\emph{Sala Baker is best known for portraying the villain Sauron in the Lord of the Rings trilogy}" by capturing the semantic relatedness between ``\emph{bad guy}" and ``\emph{villain}"~\cite{dpr2020}.

On the other hand, a major limitation of dense retrievers is that they rely on labeled data for learning the model parameters. It has been  reported that dense retrieval models 
might perform worse than sparse retrieval models under the \emph{zero-shot} setting (without relevance judgement data from the target domain) on the test datasets from the BEIR benchmark~\cite{thakur2021beir}.  
The conducted experiments on BEIR show that the zero-shot retrieval capacity of dense retrieval models is highly limited compared to classic retrieval models such as BM25~\cite{thakur2021beir}. Hence, a number of studies  are developed to improve the zero-shot retrieval capabilities of dense retrievers (see Section~\ref{sec:advanced-topics-zero} for a detailed discussion). 
Besides, sparse models are more capable in solving the queries that require exact match (\eg keyword or entity retrieval), while dense retrievers seem to perform worse in these cases, especially when the entities or compositions are rare or do not appear in the training set~\cite{Liu2021ChallengesIG,Sciavolino2021SimpleEQ}. 
For example, the DPR paper~\cite{dpr2020} shows that it fails to capture the entity phrase ``\emph{Thoros of Myr}" for the query ``\emph{Who plays Thoros of Myr in Game of Thrones?}", while BM25 succeeds in this case. 
Such a finding is indeed consistent with the underlying mechanism of the two kinds of technical approaches:  sparse retrievers are based on lexical matching,  while dense retrievers employ latent  embeddings for semantic matching, which may result in the loss of salient information due to the compressed semantic representations. Section~\ref{sec:combine-sparse-dense} will discuss  how to enhance the exact match capacity of dense retrieval models by combining sparse  and dense retrievers.

\begin{itemize}
\item[$B$] 
{
Is there any theoretical analysis on the capacity of dense representations on  relevance matching, especially how they mimic and relate to sparse retrieval? Is it possible to apply the traditional IR axioms to dense retrieval? 
}
\end{itemize}
 
Dense retrieval  employs latent  semantic representations for relevance matching, showing different retrieval performance compared with sparse retrieval models: they perform better on complicated queries but worse on exact match queries. 
It is important to understand the retrieval behavior of dense retrieval models and their connection with sparse retrieval models (\eg how they perform exact match). 
For this purpose,  Luan et al.~\cite{MEBERT}  investigate the effect of the embedding size on 
the ability to mimic sparse retrieval (\eg bag-of-words models).
They demonstrate that the embedding size of dense retrievers should be increased to achieve comparable retrieval performance of bag-of-words models, when the document length increases. 
Besides, another related study shows that  corpus scale has a larger effect on dense retrieval~\cite{reimers2020curse}: ``the performance for dense representations can decrease quicker for increasing index sizes than for sparse representations''. They explain this  finding based on a proof that the probability for false positives becomes larger when the index size increases, especially with a decreasing dimensionality.

Another possible approach to understanding the  behavior of dense retrievers is \emph{axiomatic analysis} from classic IR literature~\cite{Fang2004AFS,Fang2005AnEO,Fang2006SemanticTM}.
Specifically,  IR axioms are formalized as a set of relevance constraints that reasonable IR models are desired to satisfy or at least partially satisfy~\cite{Fang2005AnEO,axiomatic-NIR}, \eg a document containing more occurrences of a query term should receive a higher score, which can provide both interpretable evidence and learning guidance for traditional IR models. 
However, it has been found that existing IR axioms may  only partially explain~\cite{volske2021towards} or even not be suitable to analyze~\cite{camara2020diagnoising} the behavior of PLM-based IR models. 
Some further analysis~\cite{macavaney2020abnirml} also shows that  neural ranking models have different characteristics compared with sparse ranking models. For example,  dense models are easier to be affected by adding non-relevant contents, and different ranking architectures based on the same language model may lead to varied retrieval  behaviors.
Although  BERT-based ranking models are not well aligned with traditional IR axioms, recent research~\cite{Formal2021AWB} also shows that some dense retrieval model (\eg  ColBERT~\cite{colbert2020sigir}) is aware of term importance  in text representations, and  can also implicitly mimic exact matching for important terms. 

Overall, the theoretical exploration of  dense retrieval models still remains to be an open research direction.  More efforts in this line are required to understand the relevance matching behaviors of dense retrieval models.

\subsubsection{Combining Sparse and Dense Retrieval models}
\label{sec:combine-sparse-dense}
In practice, the two kinds of retrieval  models can benefit each other by leveraging complementary relevance information. For example, sparse retrieval models can be utilized to 
provide initial search results for training dense retrieval models~\cite{negative2020google, dpr2020}, 
while PLMs can be also used to learn term weights to improve sparse retrieval~\cite{dai2020context,Dai2020ContextAwareTW} (as discussed in Section~\ref{sec:sparse-retrieval-PLM}). In particular, it also shows that sparse retrieval models can be utilized to enhance the zero-shot retrieval capacity of dense retrieval models (see more discussions in Section~\ref{sec:advanced-topics-zero}).

Considering different relevance characteristics captured by both approaches, there are increasing studies that develop a hybrid of sparse and dense retrieval models. A straightforward approach is to aggregate the retrieval scores of the two approaches~\cite{MEBERT,dpr2020}. 
Further, a classifier is employed to select  among  sparse, dense, or hybrid retrieval strategies specially for each individual query\cite{arabzadeh2021predicting}, which aims to balance the cost and utility of retrievers.

However, these hybrid approaches have to maintain two different indexing systems (inverted index and dense vector index), which is too complicated to be deployed in real practice.
To address this issue, Lin et al.~\cite{Lin2021DensifyingSR,Lin2022AggretrieverAS} propose to
learn low-dimensional dense lexical representations (DLRs) by densifying high-dimensional sparse lexical representations, making it feasible to end-to-end learn lexical-semantic combined  retrieval systems. 
The basic idea of densification is to first divide the high-dimensional lexical representations into slices and reduce each sliced representation by specifically designed pooling approaches. Then, these  representations are concatenated  as the dense lexical representation. 
For text ranking, the dense lexical representations and the dense semantic representations (\ie the representation of ``\textsc{[CLS]}'') are combined for computing similarity scores. 
The experiment results in  \cite{Lin2021DensifyingSR,Lin2022AggretrieverAS} demonstrate that this approach can largely improve the retrieval effectiveness by combining the benefits of both kinds of representations, showing a superior performance on zero-shot evaluation.
Furthermore, Chen et al.~\cite{chen2021salient} propose to learn a dense lexical retriever from the weakly supervised data constructed by BM25, so that the learned dense lexical retriever can imitate the retrieval behavior of the  sparse retrievers. In this approach, the  representations from the dense lexical retriever and dense semantic retriever (\eg DPR and RocketQA) are combined  for relevance computation.
It has  been shown that the knowledge learned from BM25 can help the dense retriever on queries containing rare entities, so as to improve the robustness~\cite{chen2021salient}. 

Besides, LED~\cite{Zhang2022LEDLD}  proposes to employ PLM-based lexical-aware models (\ie SPLADE~\cite{formal2021splade}) to enhance the lexical matching capacity of dense retrievers.
It introduces lexicon-augmented contrastive learning and rank-consistent regularization
for developing the lexicon-enhanced dense retrieval models. The experiment results show that the proposed approach outperforms a number of competitive baselines. 

%% file: sec/sec-training.tex
\section{Training}
\label{sec:training}
After introducing the network architecture, we next discuss how to effectively train PLM-based dense retrievers, which is the 
key to achieve good retrieval performance. Focused on the training of bi-encoder for first-stage retrieval, we will first formulate the loss functions and introduce three major issues for training (\ie large-scale candidate space, limited relevance judgements and pretraining discrepancy), and then discuss how to address the three issues, respectively. 

\subsection{Formulation and Training Issues}
This section presents the formulation and issues for training the PLM-based dense retrievers. 

\subsubsection{Loss Function}
\label{sec:loss}
In this part, we first formulate the retrieval task as a learning task via the \emph{negative log-likelihood loss}, and then discuss several variants to implement the loss functions.

\paratitle{Negative log-likelihood loss}. Following the notations introduced in Section~\ref{sec:formulation}, 
we assume that a set of binary positive relevance judgements (\eg like or click)\footnote{As we introduced in Section~\ref{sec:formulation}, only binary relevance judgements are considered here, which are the most common form of user feedback data in real-world search systems. } are  given as supervision signals, denoted by $\mathcal{R}=\{\langle q_i, d^{+}_i \rangle\}$, where $d^{+}_i$ denotes a relevant document for query $q_i$. To optimize the dense retrievers, the core idea is to maximize the probabilities of the relevant texts \emph{w.r.t.} queries.  For simplicity, we introduce the negative log-likelihood loss for a query-positive pair as follows:
\begin{eqnarray}\label{eq-L}
			&&\mathcal{L}(\langle q_i, d^{+}_i \rangle) \\ 
			&=&-\log \frac{e^{f(\phi(q_i), \psi(d^{+}_i))}}{e^{f(\phi(q_i), \psi(d^{+}_i))}+\sum_{d' \in \mathcal{D}^{-}}e^{f(\phi(q_i), \psi(d'))}},\nonumber
\end{eqnarray}
where $f(\cdot)$ is a similarity function (often set to the dot product) that measures the similarity between query embedding $\phi(q_i)$ and text embedding $\psi(d^{+}_i)$, and $\phi(\cdot)$ and $\psi(\cdot)$ are the query encoder and text encoder (Equation~\eqref{equation:sim_definition}), respectively. Here, we enumerate all the text candidates in the collection $\mathcal{D}$ except the positives for computing the exact likelihood. 
Since the normalization term is  time-consuming to compute, a negative sampling trick is usually adopted to reduce the computational cost~\cite{dpr2020,rocketqa}:  

\begin{eqnarray}\label{eq-L-Neg}
			&&\mathcal{L}(\langle q_i, d^{+}_i \rangle) \\ 
			&=&-\log \frac{e^{f(\phi(q_i), \psi(d^{+}_i))}}{e^{f(\phi(q_i), \psi(d^{+}_i))}+\sum_{d' \in \mathcal{N}_{q_i}} e^{f(\phi(q_i), \psi(d'))}},\nonumber
\end{eqnarray}
where we sample or select a small set of negative samples for query $q_i$, denoted by  $\mathcal{N}_{q_i}$.
The above 
training objective advocates to increase the likelihood of positive texts and decrease the likelihood of sampled negative ones.
Actually, such an optimization objective is similar to the InfoNCE loss~\cite{InfoNCE} from contrastive learning, where the loss is constructed by contrasting a positive pair of examples against a set of random example pairs.

\paratitle{Other loss functions}. Instead of optimizing over a set of negatives, 
triplet ranking loss~\cite{dpr2020,HEARTS} directly optimizes the difference between a positive text and a negative text given a query, which is defined as:
\begin{eqnarray}\label{eq-L-Triplet}
			&&\mathcal{L}(\langle q_i, d^{+}_i,  d^{-}_i \rangle) \\ 
			&=& \max\bigg(0, 1-(f(\phi(q_i), \psi(d^{+}_i)) - f(\phi(q_i), \psi(d^{-}_i))) \bigg). \nonumber
\end{eqnarray}
It has been reported that negative log-likelihood loss is better than the triplet ranking loss~\cite{dpr2020} based on the retrieval accuracy on NQ dataset.
Different from the above loss functions,  binary cross-entropy~(BCE) loss is more commonly used to optimize the reranker model~\cite{nogueira2019passage} (detailed in Section 7). It firstly utilizes the dense representations of query and text to derive a match vector, and then predicts the relevance probability of a query-text pair based on the match vector:
\begin{equation}\label{eq-prob-rel}
\text{Pr}(\text{rel}=1| q_i, d_i) = \sigma\big(g(\phi(q_i) \odot \psi(d_i))\big),
\end{equation}
where $\sigma(\cdot)$ is the sigmoid function, $g(\cdot)$ is a linear function and $\odot$ is vector combination operation (\eg concatenation). Then, the BCE loss can be constructed as follows:
\begin{eqnarray}\label{eq-L-BCE}
			&&\mathcal{L}(\langle q_i, d_i \rangle) \\ 
			&=& - y_{q_i, d_i} \cdot \text{Pr}_{q_i, d_i} - (1-y_{q_i, d_i} ) \cdot (1-\text{Pr}_{q_i, d_i}), \nonumber
\end{eqnarray}
where $d_i$ can be either positive or negative, and $\text{Pr}_{q_i, d_i}=\text{Pr}(\text{rel}=1| q_i, d_i)$ which is defined in Equation~\eqref{eq-prob-rel}.

\paratitle{Similarity measurement}. To instantiate the function $f(\cdot)$, 
various vector similarity (or distance) measurements can be used to compute the query-text similarity, including inner product, cosine similarity and Euclidean distance. Several studies have examined the effect of different similarity functions on the retrieval performance. Thakur et al.~\cite{thakur2021beir} 
conduct an analysis experiment by training two BERT-based models (an identical parameter configuration on MS MARCO dataset) with  cosine similarity and inner product, respectively. They observe that the variants with cosine similarity and inner product prefer retrieving shorter and longer documents, respectively. 
Karpukhin et al.~\cite{dpr2020} also examine the effect of Euclidean distance as the distance function, however, no significant differences were observed in retrieval performance. So far, in 
the literature, inner product has been widely adopted as the similarity measurement.

\subsubsection{Incorporating Optimization Constraints}

Besides the above formulation, there are several variants that aim to improve the optimization objective by incorporating more constraints.

\paratitle{Text-oriented optimization}. 
In the above, the negative log-likelihood loss (Equation~\eqref{eq-L-Neg}) is modeled in a query-oriented way, which 
optimizes the likelihood of texts conditioned on the query and the text set. Similarly, we can model the relevance in a text-oriented way as follows: 
\begin{eqnarray}\label{eq-L-qlikelihood}
			&&\mathcal{L}_T(\langle q_i,  d^{+}_i\rangle) \\ 
			&=& -\log \frac{e^{f(\psi(d^{+}_i),\phi(q_i))}}{e^{f( \psi(d^{+}_i),\phi(q_i))}+\sum_{q^- \in \mathcal{Q}^-} e^{f(\psi(d^{+}_i), \phi(q^-))}}, \nonumber
\end{eqnarray}
where a set of sampled negative queries denoted by $\mathcal{Q}^-$ is incorporated to compute the negative query likelihood. 
Given both query-oriented and text-oriented loss functions, it has become an important 
optimization trick to  train the dense retrievers by jointly optimizing  Equation~\eqref{eq-L-Neg} and Equation~\eqref{eq-L-qlikelihood}.
Similar symmetric optimization tricks have been used in a number of follow-up studies~\cite{yang2021xmoco, li2022learning, li2021more, xu2022laprador},
showing an improved retrieval performance.

\paratitle{Text-text similarity constraints}. Besides directly optimizing query-text similarities, Ren et al.~\cite{pair} find that it is useful to incorporate \emph{text-text} similarity constraints in the optimization objective. 
They argue that previous optimization goals are mainly query-centric, which is difficult to discriminate between positive texts and semantically-similar negative texts. 
Therefore,  a text similarity constraint has been proposed, assuming that the similarity between  positive passage $d^+$ and query $q$ should be larger than the similarity between positive passage $d^+$ and negative passage $d^-$, \ie $f_{\text{sim}}(d^+, q) > f_{\text{sim}}(d^+, d^-)$.
Such a loss is formulated as follows:
\begin{eqnarray}\label{eq-L-pcentric}
			&&\mathcal{L}_{TT}(\langle q_i, d^{+}_i \rangle) \\ 
			&=&-\log \frac{e^{f(\psi(d^{+}_i),\phi(q_i) )}}{e^{f(\psi(d^{+}_i),\phi(q_i) )}+\underline{\sum_{d' \in \mathcal{N}_{q_i}} e^{f(\psi(d^{+}_i),\psi(d'))}}}.\nonumber
\end{eqnarray}
The major difference between Equation~\eqref{eq-L-Neg} and Equation~\eqref{eq-L-pcentric} lies in the underlined part, where it incorporates text-text  similarity as the normalization term.  
Not only optimizing the similarity between query and positive text, this approach tries to increase the distance between positive and semantically-similar negative texts.

\subsubsection{Major  Training Issues}
In this part, we summarize the major training issues of  dense retrieval.  

$\bullet$ Issue 1: \emph{Large-scale candidate space}. 
In  retrieval tasks, there is typically a large number of candidates  in a text collection, while only a few texts from the collection are actually relevant to a query. 
Consequently, it is challenging to train capable dense retrievers that perform well on large-scale text collections.  
Specifically, due to the computational space and time limits, we can   sample only a small number of negative samples for computing the loss function~(Equation~\eqref{eq-L-Neg}),  resulting  in a shift in the candidate space between training (a small population of texts) and testing (the entire collection). 
It has been found the sampled negatives have a significant effect on the retrieval performance~\cite{dpr2020,ANCE,rocketqa}.

$\bullet$ Issue 2: \emph{Limited relevance judgements}. 
In practice, it is difficult to construct large-scale labeled data with complete relevance judgements for dense retrieval. 
Even though several benchmark datasets with large sizes have been released (\eg MS MARCO), it is still limited to training  very large PLM-based dense retrievers. 
Additionally, the relevance annotations in these datasets are far from complete. 
Therefore, ``\emph{false negatives}'' (actually relevant but not annotated) are likely to occur, which has become a major challenge for training dense retrievers.

$\bullet$ Issue 3: \emph{Pretraining discrepancy}. PLMs are typically trained using pre-designed self-supervised loss functions, such as masked word prediction and next sentence prediction~\cite{bert2019naacl}. These pretraining tasks are not specially optimized for  the retrieval tasks~\cite{pretrain2020iclr,ma2021prop,Ram2021LearningTR}, thus likely leading to suboptimal retrieval performance. 
Besides, to represent a text, existing work usually adopts the ``\textsc{[CLS]}'' embedding of PLMs as the text representation.  
while the ``\textsc{[CLS]}'' embedding is not explicitly designed to capture the semantics of the whole text.  
Considering this issue, it needs to design specific   pretraining tasks that are more suited to dense retrieval for the underlying PLMs.

The above three  issues have become the major technical bottlenecks for dense retrieval, attracting much research attention. 
Next, we will review the recent progress on addressing the above issues, and discuss three major techniques for dense retrieval, detailed in the following three subsections: Section~5.2 (negative selection), Section~5.3 (data augmentation) and Section~5.4 (pretraining). 

\subsection{Negative Selection}
\label{sec:negative-selection}
To optimize the dense retriever (Equation~\eqref{eq-L-Neg}),  a certain number of sampled negatives are needed for computing the negative log-likelihood loss. Thus, 
how to select high-quality negatives  has become an important issue for improving the retrieval performance. 
Next, we will review three major negative selection methods, and then present the theoretical discussions.

\subsubsection{In-batch Negatives}
\label{sec:in-batch}
A straightforward approach for negative selection is random sampling, \ie each positive text is paired with several random negatives. 
However, most PLMs are optimized in a batch mode on GPU with limited memory, which makes it infeasible to use a large number of negatives during training. 
Considering this problem, in-batch negatives are used for optimizing the dense retriever:
given a query,  the positive texts paired with the rest queries from the same batch are considered as negatives. 
Assume that there are $b$ queries ($b> 1$) in a batch and each query is associated with one relevant text.
With in-batch negative technique, we can obtain $b-1$ negatives for each query in the same batch, which largely increases the number of available negatives per query under the memory limit. The in-batch technique was firstly proposed in \cite{henderson2017efficient}  for the response selection task of \emph{Smart Reply}, while it has been explicitly utilized for dense retrieval by  DPR~\cite{dpr2020}. In-batch negatives are shown to be   effective  to improve the learning of bi-encoder by increasing the number of negatives~\cite{dpr2020,rocketqa}.

\subsubsection{Cross-batch Negatives}
\label{sec:cross-batch}
By reusing the examples from a batch,
in-batch negative training can increase the number of negatives for each query in a memory-efficient way. 
In order to further optimize the training process with more negatives, another improvement strategy called \emph{cross-batch negatives}~\cite{rocketqa, gao2021scaling} are proposed under the multi-GPU setting.
The basic idea is to reuse examples across different GPUs. Assume there are $a$ GPUs for training the dense retriever. We first compute the text embeddings at each single GPU, and then communicate them across all the GPUs. In this way, the text embeddings from other GPUs can be also used as negatives.  
For a given query, we obtain $a \times b-1$ negatives from $a$ GPUs, 
which is approximately $a$ times as in-batch negatives.  
In this way, more negatives can be used during training for improving the retrieval performance. 
The  idea of cross-batch negatives can be also extended to a single-GPU setting with the technique of gradient caching~\cite{gao2021scaling}, where one can accumulate multiple mini-batches for increasing the number of negatives (while taking a larger computational cost). 

\subsubsection{Hard Negatives}
\label{sec:hard-negatives}
Although in-batch and cross-batch negatives can increase the number of available negatives, they cannot guarantee to generate \emph{hard negatives}, which refer to the irrelevant texts but having a high semantic similarity with the query.  
It is particularly important to incorporate hard negatives to improve the capacity in discriminating between relevant and irrelevant texts~\cite{dpr2020,ANCE,rocketqa}.
A number of studies have designed different hard negative selection strategies for improving the retrieval performance. According to whether the negative selector (\ie a sampler  over the text collection based on some relevance model) is fixed or updated, hard negatives can be roughly divided into \emph{static hard negatives} and \emph{dynamic hard negatives}~\cite{ANCE,zhan2021optimizing}. Besides, the selected hard negatives might contain noisy data (\eg false negatives), and thus \emph{denoised hard negatives} are also used to train dense retrievers~\cite{rocketqa}. Next, we present the detailed discussions for each kind of hard negatives. 

\paratitle{Static hard negatives}. For static hard negatives,  the negative selector is fixed during the training of the dense retriever. 
The goal is to select negatives that are difficult to be discriminated by the dense retriever. 
To achieve this, a straightforward way is to sample negatives from top retrieval results from some other retriever, either sparse or dense. 
Since BM25~\cite{maron1960relevance}  usually gives very competitive retrieval performance, several studies select hard negatives based on BM25, \ie sampling lexically similar texts (but without containing the answers) returned by BM25~\cite{dpr2020}.  In \cite{negative2020google}, multiple kinds of negatives are also mixed for training, including retrieved negatives (based on BM25, coarse and fine semantic similarity) and  heuristics-based context negatives. 
This study shows that an ensemble approach combining models trained with mixed hard negatives is able to improve the retrieval performance. As a follow-up study, the authors~\cite{lu2021multi} further design three fusion strategies to combine different kinds of negatives, namely mixing fusion, embedding fusion and rank fusion.
Besides, Hofstatter et al.~\cite{tasbalanced2021} propose a  topic-aware sampling method to compose  training batches for dense retrieval, which first clusters the queries before training and then samples queries out of one cluster per batch. In this way, the examples in a batch are highly similar, which implicitly derives hard negatives with the in-batch sampling technique. 

\paratitle{Dynamic (or periodically updated) hard negatives}. As discussed above, static hard negatives are obtained from a \emph{fixed} negative selector. Since the training of dense retriever is an iterative process, it is better to use adaptively updated negatives, called \emph{dynamic hard negatives}, for model training. The ANCE approach~\cite{ANCE} proposes to sample from the top retrieved texts by the optimized retriever itself as negatives, which they call \emph{global hard negatives}. 
It has been shown in \cite{ANCE} that globally selected negatives can
lead to faster learning convergence.
For retrieving global negatives, it needs to refresh the indexed text embeddings after updating the model parameters, which is very time-consuming.  
To reduce the time cost, ANCE uses an asynchronous index refresh strategy during the training, \ie it performs a periodic update for each $m$ batches. 
Furthermore, Zhan et al.~\cite{zhan2021optimizing} propose a new approach \emph{ADORE} for selecting dynamic hard negatives, which samples negatives from dynamic retrieval results according to the being-updated query encoder. 
Unlike ANCE~\cite{ANCE}, ADORE fixes the text encoder and the text embedding index, and utilizes an adaptive query encoder to retrieve top ranked texts as hard negatives. 
At each iteration, 
 since the query encoder for negative selection is optimized during training, it can generate adaptive negatives for the same queries. 
A note is that before training with dynamic hard negatives, the model should be warmed up (\ie BM25 and the STAR training approach in \cite{ANCE}).  {In order to better understand the effect of  static and dynamic negatives, we can roughly take an adversarial perspective to illustrate their difference: for a given query, static approaches use fixed negatives during training (\ie fixed generator), while dynamic approaches generates adaptive negatives according to the being-optimized retriever (\ie adaptive generator). Intuitively, dynamic hard negatives are more informative to train the dense retrievers (\ie the discriminator). Besides, considering the large-scale text corpus, dynamic negatives can   potentially alleviate the training-test discrepancy in the candidate space, since it can ``\emph{see}'' more  informative  negatives during training.   
}

\paratitle{Denoised hard negatives}.
Negatives play a key role in the training of dense retriever, and it has been found that dense retrievers are actually sensitive to the quality of negatives~\cite{rocketqa, prakash2021learning}.
When the sampled texts  contain noisy negatives, they tend to affect the retrieval performance. This issue becomes more severe for hard negatives, which are more likely to be false negatives~\cite{arabzadeh2021shallow, cai2022hard, debiasedsent} (\ie unlabeled positives), because they are top-ranking retrieval results of high relevance scores to queries. To resolve this issue, RocketQA~\cite{rocketqa} proposes a denoised negative selection approach, and it utilizes a well-trained cross-encoder to filter top ranked texts that are likely to be false negatives. Since the cross-encoder architecture is more powerful in capturing rich semantic interactions, it can be utilized to refine the selected hard negatives from the bi-encoder.  
Given the top ranked texts retrieved by a bi-encoder,
only the predicted negatives with confident scores by the cross-encoder are retained as final hard negatives. In this way, the originally selected negatives are denoised, which are more reliable to be used for training. 
More recently, SimANS~\cite{Zhou2022SimANS} propose to sample \emph{ambiguous negatives} that  are ranked near the positives, 
with a moderate similarity (neither too hard nor too easy) to the query. They empirically show that such negatives are more informative, and less likely to be false negatives. 

\subsubsection{Discussions on the Effect of Negative Sampling}
\label{sec:negative-discussion}

As discussed in previous subsections,  a number of negative sampling approaches have been developed to enhance the retrieval performance. Here, we present some discussions about the effect of negative sampling for dense retrieval. 

It has been shown that in-batch sampling (Section 5.2.1) cannot generate sufficiently informative negatives for dense retrieval~\cite{negative2020google,lu2021multi,ANCE}. In particular, Lu et al.~\cite{lu2021multi} analyze why in-batch negatives may not include informative negatives. They cast the  negative log-likelihood objective with in-batch negatives as a special case of the ranking-based Noise Contrastive Estimation~(NCE). Instead of sampling a negative from the whole collection, in-batch sampling considers a rather small collection (reflecting a different distribution) for sampling, \ie the annotated relevant texts to  the queries in the query set. 
A similar discussion about the informativeness of in-batch negatives is also presented in \cite{ANCE}: since the batch size and the number of informative negatives are significantly smaller than the collection size, the probability of sampling informative negatives from a random batch (or mini-batch) tends to be close to zero.
Besides, the study in \cite{ANCE}  shows that the informativeness of negatives is key to the training of dense retrieval models, from the perspective of convergence rate \emph{w.r.t.} gradient norms. 

Furthermore, Zhan et al.~\cite{zhan2021optimizing} show that random negative sampling and hard negative sampling indeed optimize  different retrieval objectives: random negative sampling mainly minimizes the total pairwise errors, which might over-emphasize the optimization of difficult queries; while in contrast, hard negative sampling minimizes top  pairwise errors, which is more suited  for optimizing top rankings.

In machine learning, negative sampling is a commonly adopted learning strategy for optimizing over the large candidate space~\cite{goldberg2014word2vec, yang2020understanding}, which has been extensively studied in the literature. Here, we limit the scope of our discussion to dense retrieval.

\subsection{Data Augmentation}
To train PLM-based dense retrievers, the amount of available  relevance judgements is usually limited \emph{w.r.t.} the huge number of model parameters. Therefore, it is important to increase the availability of (pseudo) relevance judgement data. In this section, we discuss two major approaches for data augmentation: the former incorporates additional labeled datasets, while the latter generates  pseudo relevance labels by knowledge distillation.

\subsubsection{Auxiliary Labeled Datasets}
\label{sec:auxiliary}
Several studies propose to incorporate auxiliary labeled datasets for enriching the relevance judgements. 
Karpukhin et al.~\cite{dpr2020} collect five datasets of NQ~\cite{nq}, TriviaQA~\cite{joshi2017triviaqa}, WebQuestions~\cite{berant2013semantic}, TREC~\cite{baudivs2015modeling} and SQuAD~\cite{rajpurkar2016squad},  then train a multi-dataset encoder by leveraging all the training data (excluding the SQuAD dataset), and test the unified encoder on each of the five datasets. As shown in \cite{dpr2020}, the performance on most of the datasets benefits from more training examples, especially the smallest dataset in these five datasets. 
They further conduct an analysis experiment by training DPR on NQ dataset only and testing it on the WebQuestions and CuratedTREC datasets. The results show that DPR transfers well across different datasets (focusing on the QA task), with some slight performance loss.
Furthermore, Maillard et al.~\cite{maillard2021multi} propose to train a universal dense retriever across multiple retrieval tasks. Specifically, they devise two simple variants (variant 1: separate query encoders and shared passage encoders for multiple tasks; variant 2: shared query and passage encoders for multiple tasks) of DPR, and examine the universal retrieval capacity with multi-dataset training. 
They show that such a multi-task trained model can yield comparable performance with task-specific models and achieves a better performance in a few-shot setting. 

Besides leveraging multiple QA datasets, one can also utilize  different types of data sources.  Oguz et al.~\cite{oguz2020unified}  propose a unified open-domain question answering approach by utilizing multiple resources of text, tables, lists, and knowledge bases. The basic idea is to flatten the structured data into plain texts, so that 
we can process these data in a unified text form. 
As shown in \cite{oguz2020unified}, overall, 
it is useful to combine multiple sources in the   experiments on five QA datasets, in both per-dataset and multi-dataset settings. 

When there is no training data for the target task, it becomes the  \emph{zero-shot retrieval}. In this setting, though we can leverage 
auxiliary datasets for allievating the data scarcity, it should be noted that the final performance is highly affected by these auxiliary datasets~\cite{ren-zero,Extrapolation,survey-lowresource}  (detailed in Section~\ref{sec:advanced-topics-zero}). 

\subsubsection{Knowledge Distillation}
\label{sec:knowledge-distillation}
Considering that human-generated relevance judgement is limited, knowledge distillation becomes an important approach to improving the capacity of the bi-encoder. In machine learning, knowledge distillation refers to the process of transferring knowledge from a more capable model (called \emph{teacher}) to a less capable model (called \emph{student})~\cite{Hinton2015distilling}. Following this convention, our goal is to improve the standard bi-encoder (the student) with a more powerful  model (the teacher) on a given labeled dataset. 

\paratitle{Training the teacher network}.
To implement the teacher network, we can adopt a 
well-trained cross-encoder for  knowledge distillation, since it is more capable in  modeling fined-grained semantic interaction across queries and texts. 
Typically, the training of the teacher network is independent from the training of the student network. As an improved approach, RocketQA~\cite{rocketqa} introduces an improved strategy by incorporating the information interaction between the teacher and student when training the teacher network. The basic idea is to randomly sample the top ranked texts from the student network as negatives for training the teacher. This strategy enables the cross-encoder to adjust to the retrieval results by the bi-encoder. 
Besides, dual teachers respectively trained with pairwise and in-batch negatives  
are adopted to improve the retrieval performance of the student network~\cite{tasbalanced2021}.
Further, an empirical study~\cite{hofstatter2020improving} is conducted to analyze the  effectiveness of knowledge distillation with a single teacher and a teacher ensemble, and it has been found that the performance improves with  increasing effectiveness of a single teacher or the ensemble of teachers. 
While, it should be noted that the teacher network is not necessary to be the cross-encoder. In principle, any retriever that is more capable than the basic student network can serve as the teacher network.  For example, Lin et al.~\cite{lin2020distilling} explore the possibility of using an enhanced bi-encoder (\ie ColBERT~\cite{colbert2020sigir} that uses late interaction) as the teacher network. 
As another interesting work, Yu et al.~\cite{ConDR} utilize a well-trained ad-hoc dense retriever to improve the query encoder in a conversational dense retriever, in order to mimic the corresponding ad-hoc query representation.

\paratitle{Distillation for the student network}. After training the teacher network, we utilize it to improve  the student network.  The basic idea is to run the well-trained teacher network on the unlabeled (or even labeled) data to produce pseudo relevance labels for training the student network. According to the types of the derived labels, we can categorize the distillation approaches into  two major categories, namely  hard-label and soft-label distillation. 

\emph{Hard-label distillation}.
Given the unlabeled texts, the first approach directly sets the binary relevance labels for unlabeled texts according to the  relevance scores of the teacher network.   Since the predictions of the teacher network might contain noise or errors, thresholding methods are often adopted to remove texts with low confidence scores. 
For example, RocketQA~\cite{rocketqa}  generates pseudo relevance labels for top ranked passages according to their ranking scores: positive (higher than 0.9) and negative (lower than 0.1), and discards the rest with unconfident  predictions.   
They also manually examine a small number of denoised texts, and find that this method is generally effective to remove false negatives or positives.

\emph{Soft-label distillation}. Instead of using hard labels, we can also approximate the outputs of the teacher network by tuning the student network. Formally, let $r^{(t)}_{q,d}$ and $r^{(s)}_{q,d}$ denote the relevance scores of text $d$ \emph{w.r.t.} to query $q$ assigned by the teacher network and the student network, respectively. Next, we introduce several distillation functions. 

\begin{itemize}
\item MSE loss. This function directly minimizes the difference between the relevance scores between the teacher and student using mean squared error loss: 
\begin{equation}
\mathcal{L}^{KD}_{MSE} = \frac{1}{2}\sum_{q \in \mathcal{Q}}\sum_{d \in \mathcal{D}} (r^{(t)}_{q,d}-r^{(s)}_{q,d})^2,
\end{equation}
where $\mathcal{Q}$ and $\mathcal{D}$ denote the sets of queries and texts, respectively. 
\item KL-divergence loss. This function first normalizes the relevance scores of candidate documents into probability distributions by queries, denoted by $\tilde{r}^{(t)}_{q,d}$ and $\tilde{r}^{(s)}_{q,d}$, respectively, and then reduces their KL-divergence loss:
\begin{equation}\label{eq-KL}
\mathcal{L}^{KD}_{KL} = - \sum_{q \in \mathcal{Q}, d \in \mathcal{D}}  \tilde{r}^{(s)}_{q,d} \cdot \big( \log \tilde{r}^{(s)}_{q,d} - \log \tilde{r}^{(t)}_{q,d} \big).
\end{equation}
\item Max-margin loss. This function adopts a max-margin loss for penalizing the inversions in the ranking generated by the retriever:
\begin{equation}
\mathcal{L}^{KD}_{MM} = \sum_{q, d_1, d_2} \max\bigg( 0, \gamma- \text{sign}(\Delta^{(t)}_{q,d_1,d_2} \times \Delta^{(s)}_{q,d_1,d_2})\bigg),
\end{equation}
where $q\in \mathcal{Q}$, $d_1, d_2 \in \mathcal{D}$, $\Delta^{(t)}_{q,d_1,d_2}=r^{(t)}_{q,d_1}-r^{(t)}_{q,d_2}$, $\Delta^{(s)}_{q,d_1,d_2}=r^{(s)}_{q,d_1}-r^{(s)}_{q,d_2}$, $\gamma$ is the margin and $\text{sign}(\cdot)$ is the sign function indicating whether the value is positive, negative or zero.
\item Margin-MSE  loss. This function reduces the margin difference for a positive-negative pair between the teacher and student, via the MSE loss:
\begin{equation}
\mathcal{L}^{KD}_{M-MSE} = \text{MSE}( r^{(t)}_{q,d^+}-r^{(t)}_{q,d^-} , r^{(s)}_{q,d^+}-r^{(s)}_{q,d^-} ).
\end{equation}
\end{itemize}
To examine the effectiveness of different distillation functions, Izacard et al.~\cite{izacard2021distilling} perform an empirical comparison of the above loss functions, and they find that  the KL-divergence loss leads to a better distillation performance than MSE loss and max-margin loss for the task of question answering. 
Further, the researchers empirically find that the Margin-MSE loss is more effective than the other options~\cite{hofstatter2020improving,tasbalanced2021}, \eg pointwise MSE loss. 

\paratitle{Advanced distillation methods}. 
Recent studies~\cite{Pro-KD,distillation} show that knowledge distillation might become less effective when there exists a large capacity gap between the teacher and student models. 
Thus, instead of using direct distillation, a progressive distillation approach~\cite{PROD,ERNIE-Search,CL-DRD} should be adopted when using a strong teacher model, which can be implemented in two ways. 
As the first way, we use   gradually enhanced  
teacher models  at different stages of distillation, in order to fit the learning of the student model. PROD~\cite{PROD} proposes to 
use a progressive chain of the teacher model with improved model architectures and increased network layers: 12-layer bi-encoder $\rightarrow$ 12-layer cross-encoder $\rightarrow$ 24-layer cross-encoder (given the 6-layer bi-encoder as the student model). ERNIE-Search~\cite{ERNIE-Search} introduces two distillation mechanisms for reducing the large capacity gap between the teacher and student models, namely (i) \emph{late-interaction models} (\eg ColBERT) to \emph{bi-encoder}, and (ii) \emph{cross-encoder} to \emph{late-interaction models} and then to \emph{bi-encoder}. As the second way, we fix the strong   teacher model, and gradually increase the difficulty of the distilled knowledge. CL-DRD~\cite{CL-DRD} proposes a curriculum learning approach, and schedules the distillation process with gradually increased difficulty levels: 
 more difficult samples (with a larger query-text similarity) are arranged  at a later stage. 
In the above, we assume that the teacher model is fixed during a distillation process. RocketQA-v2~\cite{rocketqav2} extends the distillation way by introducing a \emph{dynamic listwise distillation} mechanism: both the retriever (student) and the reranker (teacher) are mutually updated and improved. Based on the KL-divergence loss (Equation~\eqref{eq-KL}),  the teacher model is also able to be adjusted according to the student model, yielding a better distillation performance.

The above two approaches can directly produce pseudo supervision signals for \emph{fine-tuning} the underlying PLMs for dense retrieval. Besides, 
there are also other augmentation related methods for dense retrieval, such as synthetic data generation~\cite{Alberti2019SyntheticQC,lewis2021paq,ouguz2021domain} and contrastive learning~\cite{Gao2021SimCSESC,Yan2021ConSERTAC,xu2022laprador}, which are more related to \emph{pretraining}. We   leave the discussion of these topics in Section 5.4.

\subsection{Pretraining for Dense Retrieval Models}~\label{sec:training-pretrain}
The original purpose of PLMs is to learn universal semantic representations that can generalize across different tasks. However, such representations often 
 lead to suboptimal performance on downstream applications due to the lack of task-specific  optimization~\cite{latent2019acl, pretrain2020iclr,ma2021prop,Ram2021LearningTR}. 
Considering this issue, recent studies employ task-related pretraining strategies for dense retrieval. Besides reducing  the pretraining discrepancy (Issue 2), these approaches  can also alleviate the scarcity of labeled relevance data (Issue 3)\footnote{Note that some of the techniques presented in this part are also related to data augmentation. We discuss these techniques in this section,  because they are more focused on the pretraining stage instead of the fine-tuning stage. }.   
Next, we describe the pretraining approaches for dense retrieval in detail. 

\subsubsection{Task Adaptive Pretraining}
\label{sec:task-adaptive}

This line of pretraining tasks  essentially follow what have been used in BERT~\cite{bert2019naacl}, but try to mimic 
the retrieval task in a self-supervised way. 
Below, we list several representative pretraining tasks for dense retrieval.
\begin{itemize}
\item Inverse Cloze Task~(ICT)~\cite{latent2019acl}: ICT randomly selects a sentence of a given text as the query, and the rest sentences are considered as gold matching text. This task aims to capture the semantic context of a sentence and enhance the matching capacity between query and relevant contexts.
\item Body First Selection~(BFS)~\cite{pretrain2020iclr}: BFS utilizes the sentences from the first section of a Wikipedia article as anchors to the rest sections of texts. It considers a randomly sampled sentence from the first section as the query and a randomly sampled passage in the following sections as a matched text. 
\item Wiki Link Prediction~(WLP)~\cite{pretrain2020iclr}: WLP utilizes the hyperlink to associate the \emph{query} with \emph{text}, where a sampled sentence of the first section from a  Wikipedia page is considered as the query and a passage from another article containing the hyperlink link to the page of the query is considered text. 
\item Recurring Span Retrieval~(RSR)~\cite{Ram2021LearningTR}: RSR proposes to use  \emph{recurring spans} (\ie ngrams with multiple occurrences in the corpus) to associate related passages and conduct unsupervised pretraining for dense retrieval. Given a recurring span, it first collects a set of passages that contain the recurring span, and then transforms one of the passages into a query and treats the rest passages as positives. 
\item Representative wOrds Prediction~(ROP)~\cite{ma2021prop}: ROP utilizes a document language model to sample a pair of word sets  for a given text. Then, a word set with a higher likelihood is considered to be more “representative” for the document, and the PLM is pretrained to predict the pairwise preference between the two sets of words. Following ROP, another variant  called \emph{B-ROP}~\cite{ma2021bprop} replaces  the unigram language model by a BERT model, and samples the representative words from a \emph{contrastive term distribution} constructed based on document-specific and random term distributions.
\end{itemize}

\subsubsection{Generation-Augmented Pretraining}
\label{sec:generation-augmented-pretraining}
Although the above pretraining tasks can improve the retrieval performance to some extent, they still rely on self-supervised signals that are derived from original text. 
Considering this issue, several studies propose to directly generate pseudo question-text pairs for retrieval tasks. 
Specially, these data generation approaches can be divided into two major categories, either a pipeline or end-to-end way.

For the first category, Alberti et al.~\cite{Alberti2019SyntheticQC} propose a pipeline approach to constructing question-answer pairs from large text corpora. It  consists of three major stages, including answer extraction, question generation and roundtrip  filtering. 
Experimental results show that pretraining on the synthetic data significantly improves the performance of question answering on SQuAD 2.0 and NQ datasets. 
In a similar way, 
Lewis et al.~\cite{lewis2021paq} further  introduce passage selection and global filtering to construct a dataset called \emph{PAQ} for question answering, which contains 65 million synthetically generated question-answer pairs from Wikipedia.
Based on the PAQ dataset, a related study~\cite{ouguz2021domain} pretrains the bi-encoder retrievers, leading to consistent performance improvement 
over the variants pretrained with the tasks of ICT and BFS (Section 5.4.1). 

Instead of using a pipeline way, Shakeri et al.~\cite{shakeri2020end} propose an end-to-end approach to generating question-answer pairs based on machine reading comprehension by using a pretrained LM (\eg BART~\cite{Lewis2020BARTDS}). It aims to train a sequence-to-sequence network for generating a pair of question and answer conditioned on the input text. Reddy et al.~\cite{reddy2021towards} further extend this approach by incorporating an additional  selection step for enhancing the generated  question-text pairs for retrieval tasks.

Furthermore, several studies~\cite{liang2020embedding, ma2021zero, reddy2021towards, wang2021gpl} also  explore the generation-augmented pretraining approach in the \emph{zero-shot setting}, where there are no training datasets in the target domain. Their results show that generation-augmented pretraining is useful to improve the zero-shot retrieval capacity of dense retrievers. We will discuss more on zero-shot retrieval in Section~\ref{sec:advanced-topics-zero}.

\subsubsection{Retrieval-Augmented  Pretraining}
\label{sec:retrieval-augmented-pretraining}
To enhance the modeling capacity of PLMs, another line of pretraining tasks is 
to incorporate an external retriever by enriching the relevant contexts. 
The basic idea is to enhance the training of the masked language modeling~(MLM) task with more referring contexts from a knowledge retriever. 

As a representative work, REALM~\cite{REALM} utilizes a knowledge retriever for 
retrieving relevant texts from a large background corpus, and the retrieved texts are further encoded and attended to augment language model pretraining.
In REALM, the basic idea  is to reward or discourage the retriever according to whether the retrieved context is useful to improve the prediction of the masksed words. 
Without using explicit human annotation, the context retriever is further trained via  language modeling pretraining, in which the retrieved texts are modeled by a latent variable through marginal likelihood. 
By fine-tuning the model, REALM performs well on the task of open-domain question answering. It further proposes to use an asynchronous optimization approach based on the maximum inner product search.

As an extension work, Balachandran et al.~\cite{balachandran2021simple} perform a thorough tuning of REALM  on  a variety of QA tasks.  
 They conduct extensive experiments with  multiple training tricks, including using exact vector similarity search, 
training with a larger batch, retrieving more documents for the reader, and incorporating human annotations for evidence passages. 
Their experiments show that REALM was not sufficiently trained for fine-tuning, and it is important to design suitable training and supervision strategies for improving open-domain question answering systems.

This part is also related to the topic of retrieval-augmented language model, which will be discussed in Section~\ref{sec:application-retrieval-augmented-lm}.

\subsubsection{Representation Enhanced Pretraining}
\label{sec:representation-enhanced-pretraining}

For bi-encoder based dense retrievers, a typical approach is to utilize the inserted ``\textsc{[CLS]}'' token to obtain the representation of a query or text. 
However, the original ``\textsc{[CLS]}'' embedding is not explicitly designed to represent the meaning of a whole text. Hence, this may lead to suboptimal  retrieval performance. 
, and it needs more effective approaches to enhance the ``\textsc{[CLS]}'' embedding for dense retrieval.

\paratitle{Autoencoder enhanced pretraining}. 
Inspired by the success of autoencoders in data representation learning~\cite{Socher2011SemiSupervisedRA,Socher2011DynamicPA,Li2015AHN,Dai2015SemisupervisedSL}, several studies 
explore autoencoders to enhance the text representation for dense retrieval.
The basic idea is to compress the text information into the ``\textsc{[CLS]}'' embedding by using an encoder network, 
and then use a paired decoder to reconstruct the original text  based on the ``\textsc{[CLS]}'' embedding. In this way, the learned ``\textsc{[CLS]}'' embedding is enforced to capture more sufficient text semantics than the original attention mechanisms in PLMs. 
Gao et al.~\cite{gao2021condenser} propose the Condenser architecture consisting of three orderly parts: early encoder backbone layers,  late encoder backbone layers and Condenser head layers (only used during pretraining). 
The key point is that Condenser removes fully-connected attention  across late and Condenser head layers, while keeping a short circuit from early layers to Condenser head layers. 
Since Condenser head layers can only receive information of late backbone layers via the late ``\textsc{[CLS]}'', the late ``\textsc{[CLS]}'' embedding is therefore enforced to aggregate the information of the whole text as possible,  for recovering the masked tokens. Similar attempts have been made in TSDAE~\cite{Wang2021TSDAEUT}, which is  a Transformer-based sequential denoising autoencoder for enhancing the text representation.
Furthermore, co-Condenser~\cite{gao2021unsupervised} extends the Condenser architecture by incorporating a query-agnostic contrastive loss based on the retrieval corpus, which  pulls close the text segments from the same document while pushing away other unrelated segments. 
Following the information bottleneck principle~\cite{IB}\footnote{In \cite{IB}, the authors define the goal of information bottleneck as ``finding a maximally compressed mapping of the input variable that preserves as much as possible the information on the output variable''.}, recent studies~\cite{SimLM,LexMAE} refer to  the key representations (\eg the ``\textsc{[CLS]}'' embedding) that aims to capture all the important semantics in the above autoencoder approaches as \emph{representation bottlenecks}.

\paratitle{Unbalanced autoencoder based pretraining}. Recently, Lu et al.~\cite{lu2021less} have reported an interesting finding that a stronger decoder may lead to worse sequence representations for the autoencoder. 
It is explained as the \emph{bypass effect}:  when the decoder is strong, it may not refer to the information representation from the encoder, but instead perform the sequence generation conditioned  on previous tokens. 
They provide theoretical analysis on this finding, and 
implement a weak decoder based on a shallow Transformer with restricted access to previous context.
In this way, it explicitly enhances the dependency of the decoder on the encoder.   
This work has inspired several studies that use the   ``\emph{strong encoder, simple decoder}'' architecture for unsupervised text representation  pretraining. 
Based on such an unbalanced architecture, 
SimLM~\cite{SimLM}  pretrains the encoder and decoder with replaced language modeling, where it aims to recover the original tokens after replacement.  To optimize  the dense retriever, it further employs a multi-stage training  procedure, which uses hard negative training and cross-encoder distillation. 
Furthermore,  RetroMAE~\cite{RetroMAE} proposes to use a large masking ratio for the decoder (50$\sim$90\%) while a common masking ratio for the encoder (15\%). It also introduces an enhanced decoding mechanism with two-stream attention and position-specific attention mask. Based on RetroMAE, Liu et al.~\cite{RetroMAE++} present a rigorous two-stage pretraining approach (\ie general-corpus pretraining and domain-specific continual pretraining), which shows strong performance on a variety of benchmark datasets. 
Besides, LexMAE~\cite{LexMAE} applies  such a representation  enhanced pre-training strategy to learned sparse retrieval (\eg SPLADE~\cite{formal2021splade}) based on PLMs, where it incorporates lexicon-based representation bottleneck (\ie continuous bag-of-words representations with learned term  weights) for pretraining. 

\paratitle{Contrastive learning enhanced pretraining}. 
Another promising direction is to apply contrastive learning (either unsupervised or supervised) to enhance the text representations for dense retrieval.
Contrastive learning is originated from the field of computer vision (pioneered by the works of  SimCLR~\cite{Chen2020ASF} and MoCo~\cite{He2020MomentumCF}), where the key idea is to learn the representation of images by making similar ones close and vice versa. 
Typically, contrastive learning is conducted in two major steps. 
First, the augmented views are generated for each image using various transformation methods (\eg cropping and rotation). 
The augmented views are usually considered as \emph{positives} to the original images, while randomly sampled instances are considered  as \emph{negatives} to the original images. 
Such a data augmentation process is particularly important to the final performance of contrastive learning.  
Second, a discrimination task is designed  assuming that the positives should be closer to the original one than the negatives. 
Following the same idea, we can generate large-scale positive and negative text pairs for unsupervised pretraining of text embeddings. 
As a representative study, SimCSE~\cite{Gao2021SimCSESC} produces positives of one sample text by applying different dropout masks,  and uses in-batch negatives. 
ConSERT~\cite{Yan2021ConSERTAC} uses four ways to generate different views (\ie positives) for texts, including adversarial attacks, token shuffling, cutoff and dropout. 
Contriever~\cite{Izacard2021TowardsUD} proposes generating positives by using ICT and cropping (\ie sampling two independent spans from a text to form a positive pair) and generating negatives by using in-batch and cross-batch texts. 
Experimental results demonstrate that unsupervised contrastive pretraining leads to good performance in both zero-shot and few-shot settings~\cite{Izacard2021TowardsUD}. 
In a similar manner, LaPraDoR~\cite{xu2022laprador} generates positives by using ICT and dropout, and proposes an iterative contrastive learning approach that  trains the query and document encoders in an alternative way. 
In contrast to autoencoder enhanced pretraining, Ma et al.~\cite{ma2022pre} further propose to pretrain an encoder-only network with a new contrastive span prediction task, which aims to to fully reduce the bypass effect of the decoder. It designs a group-wise contrastive loss, considering the representations of a text and the spans that it contains as positive pairs and cross-text representations as negative pairs.
Besides, it has shown that unsupervised contrastive pretraining~\cite{Neelakantan2022TextAC} (pretraining with text pairs and text-code pairs) can also improve both text and code retrieval performance.

\paratitle{Discussion}.
Compared with the  pretraining approaches in previous parts (Section 5.4.1,  5.4.2 and  5.4.3),
representation enhanced pretraining does not explicitly use retrieval (or retrieval-like) tasks as optimization goals. 
Instead, it aims to enhance the informativeness of the text representation, \ie the ``\textsc{[CLS]}'' embedding. 
For dense retrieval, 
the text representations should capture the semantic meaning of the whole text.  
By designing effective pretraning strategies  (autoencoder or constrastive learning), these approaches can produce more informative text representations, thus improving the retrieval performance~\cite{gao2021unsupervised,Wang2021TSDAEUT,lu2021less,SimLM,xu2022laprador}.  
Besides, it has been shown that these approaches can improve the retrieval performance even in zero-shot or low resource  settings~\cite{gao2021condenser,xu2022laprador,Izacard2021TowardsUD}, which also alleviates the scarcity issue of  relevance judgements.

\subsection{Empirical Performance Analysis with RocketQA}

Previous sections have extensively discussed various optimization techniques to improve the training of dense retrievers. In this section, we take RocketQA~\cite{rocketqa} as the studied model  and examine how different optimization techniques improve its retrieval performance on benchmark datasets.

\subsubsection{Experimental Setup}

To prepare our experiments, we adopt the widely used MS MARCO passage retrieval dataset~\cite{msmarco} for evaluation, and the detailed statistics of this dataset are reported in  Table~\ref{tab:datasets}. 

For comparison, we consider two variants of RocketQA, namely RocketQA~\cite{rocketqa}  and RocketQAv2~\cite{rocketqav2}, and a related extension called RocketQA$_{PAIR}$~\cite{pair}\footnote{It was originally called \emph{PAIR}, and we call it RocketQA$_{PAIR}$ for the consistency of the naming. }. For RocketQA, it uses three major training tricks, namely cross-batch negatives (Section~\ref{sec:cross-batch}), denoised hard negatives (Section~\ref{sec:hard-negatives}) and distillation-based data augmentation  (Section~\ref{sec:knowledge-distillation}). For RocketQAv2, it incorporates a new soft-label distillation mechanism, called dynamic listwise distillation (Section~\ref{sec:knowledge-distillation}). For RocketQA$_{PAIR}$, it is built on RocketQA and incorporates the passage-centric pretraining technique (Section ~\ref{sec:loss}). 
Considering the tunable configuration (\eg the number of negatives and batch size) of these models, we include multiple comparisons with different settings. 

Although RocketQA variants do not include all the optimization techniques introduced in Section~\ref{sec:training}, they provide a unified base model by examining the effects of different optimization techniques in a relatively fair way. 
As a comparison, we incorporate the classic BM25~\cite{robertson1995okapi} and DPR~\cite{dpr2020} methods as baselines.

To reproduce the experiments in this part, we implement a code repertory for dense retrieval at the link \url{https://github.com/PaddlePaddle/RocketQA} and release the script or code to reproduce the results in Table~\ref{tab:empirical}. 
This code repertory is implemented based on the library PaddlePaddele, consisting of RocketQA, RocketQA$_{PAIR}$ and RocketQAv2.
 For a fair comparison,  we re-implement the DPR model with the library PaddlePaddele~\cite{ma2019paddlepaddle} and enhance it with the ERNIE model~\cite{ernie20aaai} as the base PLM\footnote{ERNIE~\cite{ernie20aaai}  is a knowledge enhanced PLM by incorporating knowledge bases into text-based PLMs for pretraining.}. 
For simplicity, we only show the key commands or code to reproduce these comparison results in Table~\ref{tb-command}.

\subsubsection{Results and Analysis}
Table~\ref{tab:empirical} presents the performance comparsion of different methods or variants on the MS MARCO dataset for the passage retrieval task. We have the following observations:

$\bullet$ Firstly, cross-batch technique is able to improve the retrieval performance.  
Compared with DPR$_{ERNIE}$, the RocketQA variants $2a\sim 2g$ have the same configuration and implementation, except the cross-batch technique. Such a technique can lead to significant  improvement,  \ie  0.9 percentage point in absolute $MRR@10$ performance (variant $1$ \emph{v.s.} $2a$). Besides, we can see that it is key to use as more negatives as possible during the cross-batch training, which means that  a large batch size should be used.

$\bullet$ Secondly, it is effective to use denoised hard negatives, which leads to a significant improvement of 3.1 percentage points in absolute $MRR@10$ performance (variant $2a$ \emph{v.s.} $2i$). As a comparison, when using non-denoised hard negatives, the performance decreases quickly. A possible reason is that  hard negatives are more likely to be false negatives that will harm the final performance.  

$\bullet$ Thirdly, data augmentation can further improve the retrieval performance with both cross-batch technique and denoised hard negatives, which leads to a substantial improvement of 0.6 percentage point in absolute $MRR@10$ performance (variant $2i$ \emph{v.s.} $2l$). Besides, as we can see, the performance improves with the increasing amount of pseudo labeled data. 

$\bullet$ Fourthly,  passage-centric pretraining can boost the performance, which leads to a substantial improvement of 0.7 percentage point in absolute $MRR@10$ performance (variant $3a$ \emph{v.s.} $3b$). This technique  characterizes a more comprehensive semantic relation among query, positive texts, and negative texts. 

$\bullet$ Finally, the variant with dynamic listwise distillation (variant $4e$) achieves the best performance among all the RocketQA variants. The  
dynamic listwise distillation technique provides a joint training approach for retriever and reranker based on soft-label distillation (detailed in Section 7.2.3).

To conclude the discussion of this part, we can see that effective optimization techniques are key to achieve good retrieval performance for dense retrievers. 

\begin{table}[ht]
    \renewcommand\tabcolsep{4pt}
    \caption{Empirical comparison of different RocketQA variants on MS MARCO dataset for the passage retrieval task (in percentage). CB=cross-batch, IB=in-batch, HLD=hard-label distillation, SLD=soft-label distillation, and DeHN = denoised hard negatives. }
    \begin{tabular}{l|ll|c}
    \toprule
    \textbf{Model} & \multicolumn{2}{l|}{\textbf{Technique}} & \textbf{MRR@10} \\
    \midrule
    BM25 & (0) & BM25 & 18.7 \\
    \midrule
    DPR$_{ERNIE}$ & (1) & in-batch ($batchsize=4096$) & 32.4 \\
    \midrule
    \multirow{11}{*}{RocketQA} & (2a) & cross-batch ($batchsize=4096$) & 33.3 \\
    & (2b) & cross-batch ($batchsize=2048$) & 32.9 \\
    & (2c) & cross-batch ($batchsize=1024$)  & 31.4 \\
    & (2d) & cross-batch ($batchsize=512$)  & 29.8 \\
    & (2e) & cross-batch ($batchsize=256$)  & 29.5 \\
    & (2f) & cross-batch ($batchsize=128$)  & 28.6 \\
    & (2g) & cross-batch ($batchsize=64$)  & 26.9 \\
    & (2h) & CB + hard neg w/o denoising  & 26.0 \\
    & (2i) & CB + hard neg w/ denoising  & 36.4 \\
    & (2j) & CB + DeHN + data aug. (208K)  & 36.5 \\
    & (2k) & CB + DeHN + data aug. (416K)  & 36.9 \\
    & (2l) & CB + DeHN + data aug. (832K)  & 37.0 \\
    \midrule
    \multirow{2}{*}{RocketQA$_{PAIR}$} & (3a) & IB + hard-label distillation  & 37.2 \\
    & (3b) & IB + HLD + p-centric pretraining & 37.9 \\
    \midrule
    \multirow{6}{*}{RocketQAv2} & (4a) &  soft-label distillation &  36.5 \\
    &  (4b) & SLD + joint training (7 HNs)  & 36.5 \\
    &  (4c) & SLD + joint training (31 HNs)  & 37.3 \\
    &  (4d) & SLD + joint training (63 HNs)  & 38.1 \\
    &  (4e) & SLD + joint training (127 HNs)  & 38.3 \\
    \bottomrule
    \end{tabular}
    \label{tab:empirical}
\end{table}

\begin{table*}[ht]
\centering
\caption{Important command parameters to implement different dense retrieval models in our toolkit.}\label{tb-command}
\begin{tabular}{l|l|l|l}
    \toprule
    \textbf{Parameter name} & \textbf{Value} & \textbf{Setting} & \textbf{Related model} \\
    \midrule
    \multirow{2}{*}{use\_cross\_batch} & false & In-batch negative & \multirow{2}{*}{RocketQA} \\
     & true & Cross-batch negative &  \\
    \midrule
    batch\_size & number & Batch size per GPU & RocketQA, PAIR, RocketQAv2 \\
    \midrule
    train\_set & file\_name & Training   data for different strategies (DeHN, data aug., etc.) & RocketQA, PAIR, RocketQAv2 \\
    \midrule
    \multirow{2}{*}{is\_pretrain} & true & Incorporating   passage-centric pretraining & \multirow{2}{*}{PAIR} \\
     & false & Query-centric   fine-tuning & \\
    \bottomrule
\end{tabular}
\end{table*}

%% file: sec/sec-index.tex
\section{Indexing for Dense Retrieval}

Previously, we have extensively  discussed how to design and train a dense retrieval model from an algorithmic perspective. 
In order to implement a dense retrieval system, it is important to develop a suitable index structure that can  support efficient search  in dense vector space. In this section, we discuss the data structures and algorithms for efficient dense retrieval. 

\subsection{Traditional Inverted Index for Sparse Retrieval}

To start with, we review 
the key data structure, \ie term-based inverted index, to support traditional sparse retrieval. 

In essence, inverted index maintains an efficient mapping from terms to their locations in documents~\cite{mcdonell1977inverted}. 
The basic idea is to construct term-based posting lists by aggregating the occurrences of a term. To be specific, each term in the vocabulary is associated with a unique posting list, and the posting list stores the identifiers of the documents (possibly with more detailed positional information) in which the term occurs. 
The basic idea of using the inverted index for relevance evaluation is to search the documents by query terms and evaluate the documents that at least contain a query term. 
For retrieval, given a query, it first fetches the postings lists associated with query terms and traversing the postings to compute the result set. 

As major sparse retrieval toolkits, Apache Lucene\footnote{~\url{https://lucene.apache.org}} and its extensions including Elastic Search\footnote{\url{https://elastic.co}} and Apache Solr\footnote{\url{https://solr.apache.org}} have become \textit{de facto} software for building inverted index based retrieval systems. 
Besides, Anserini~\cite{Yang2017AnseriniET},  also built on top of Lucene, allows researchers to easily reproduce sparse retrieval models on standard IR test collections.

\subsection{Approximate Nearest Neighbor Search for Dense Retrieval}\label{sec:ANNS}
Since dense retrieval does not rely on lexical match,  term-based inverted index is no longer suited for embedding-based retrieval. 
Dense retrieval represents both queries and texts as  dense vectors, which can be  cast  into the problem of \emph{nearest neighbor search}:
finding the most  {close} vector(s) from a collection of candidate vectors (\ie the texts in the collection) \emph{w.r.t.} a query vector based on some similarity or distance measurements. 

In existing literature of dense retrieval, most of previous studies adopt the  Faiss library~\cite{faiss} to implement nearest neighbor search, while lacking a detailed discussion about the underlying data structure and implementation algorithms. 
In practice, nearest neighbor search is a crucial technique in order to design efficient dense retrieval systems.  
Additionally, there exists some connection between the two research communities of  {\emph{information retrieval}} and the \emph{nearest neighbor search}. 
Thus, it is worthwhile to have a  discussion  of nearest neighbor search in this survey. 
In what follows, we first formulate the nearest neighbor search under the setting of dense retrieval,  then discuss two major directions to improve search efficiency, and finally introduce the implementation and software.

\subsubsection{Formulation and Overview}
In this part, we formulate the dense retrieval task as the nearest neighbor search problem in the vector space of embeddings. 
Given a collection of candidate text embeddings $\mathcal{P}$ ($\mathcal{P} \subset \mathbb{R}^l$ and $\lvert \mathcal{P} \rvert = m$) and a query embedding $\mathbf{q} \in \mathbb{R}^l$, the goal of nearest neighbor search~\cite{ALSH,ANNS-survey} is to efficiently search for $\mathbf{p} \in \mathcal{P}$ 
{that are most close to  $\mathbf{q}$}, \ie $\mathbf{p}^{*} = \arg\max_{\mathbf{p} \in \mathcal{P}} f_{\text{sim}}(\mathbf{q}, \mathbf{p})$, where $f_{\text{sim}}(\cdot)$ can be implemented by 
various similarity or distance functions, \eg inner product or cosine similarity. 
The most straightforward approach for nearest neighbor search is to enumerate all the candidate embeddings, \ie computing the similarity between a query embedding and each candidate embedding in the collection. 
However, it will be very time-consuming when the number of candidate embeddings is large.
To reduce the high computational costs of brute-force enumeration,  a number of \emph{Approximate Nearest Neighbor Search (ANNS)} algorithms~\cite{ALSH,ANNS-ref1,ANNS-ref2,ANNS-survey} are developed to retrieve the approximates of the exact nearest neighbors, possibly with some loss in search quality.

Generally speaking, the design of ANNS algorithms needs to make a trade-off between search efficiency and  quality. In order to evaluate different ANNS aglorithms, ANN-Benchmarks~\cite{Aumller2017ANNBenchmarksAB}\footnote{~\url{http://ann-benchmarks.com/}} has been released with a  million-scale benchmark, and an enlarged version called Big-ANN-Benchmarks~\cite{Simhadri2022ResultsOT}\footnote{~\url{https://big-ann-benchmarks.com/}} further extends the data size to a billion scale. These two benchmarks maintain public performance rankings of 
different ANNS algorithms under different settings, which provides a guidance to select suitable algorithms according to dataset size and efficiency/accuracy requirement.

In general, there are two feasible directions to improve the search efficiency of ANNS:
(1) reducing the amount of similarity computes based on various index structures, and (2) reducing the overhead of each similarity compute by product quantization. 
In what follows, we will discuss  index structures for ANNS (Section~\ref{sec:anns-index}) and product quantization algorithms for ANNS (Section~\ref{sec:quantization}), and  also introduce publicly available software of ANNS for building efficient dense retrieval systems (Section~\ref{sec:ann-software}).

\subsubsection{Improving Search Efficiency by Index Structures}
\label{sec:anns-index}
As a major direction to improve search efficiency, we can design special index structures to reduce the amount of similarity computes, and there are several ways to develop efficient index structure for dense retrieval, including hashing-based approaches, 
clustering-based inverted index approaches, and graph-based approaches.

$\bullet$ \emph{Hashing-based approaches}{~\cite{Indyk1998ApproximateNN,ALSH,Wang2018ASO}}  assign vectors into different buckets according to a pre-designed hashing function. The idea is to allocate highly similar vectors (\ie similar texts) into the same buckets. Given a query vector, we employ a hash function to map it into some specific bucket, and then search the vectors only in these buckets. In this way, the search efficiency will be improved. 
Such an approach is more  suitable when the dimension of vectors is small.

$\bullet$ 
\emph{Clustering-based index approaches}~\cite{Sivic2003VideoGA} partition the search space by clustering. Given a query vector, we first compare it with the cluster centroids for finding the most similar clusters. 
Then, we can locate a limited number of clusters that are likely to contain the target vectors, and further perform in-cluster search for finding the target vectors. 
This approach is flexible to use, with high search quality and reasonable efficiency.
{
Further, this approach can be implemented in a memory-disk hybrid way for indexing large-scale data, where cluster centroids are stored in memory  while posting lists to centroids are stored in disk~\cite{Chen2021SPANNHB}. 
}

{
$\bullet$ 
\emph{Graph-based approaches}{~\cite{Wang2012QuerydrivenIN,Malkov2014ApproximateNN} (\eg SPTAG and HNSW)} work by navigating a graph that is built by associating the vertices with their nearest neighbors. 
Given a query, it searches on the graph by greedily selecting the similar neighbors. 
Such approaches are efficient due to the \emph{small world phenomenon}~\cite{small-wolrds}: the average path length of between two vertices is short. 
Graph-based approaches usually perform well on large high-dimensional datasets, while taking a high memory cost. 
}

\begin{figure*}[t]
    \centering
    \includegraphics[width=0.99\textwidth]{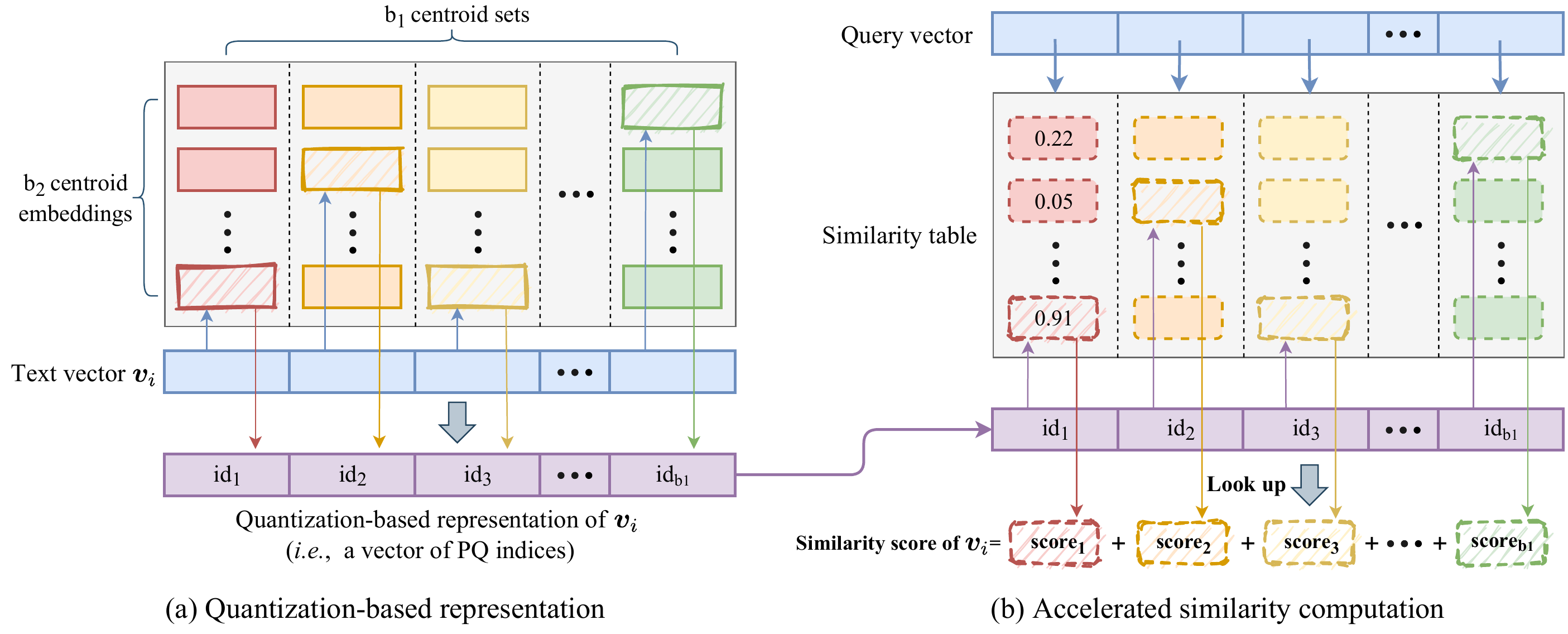}
    \caption{Illustration of text representation and similarity search based on product quantization~(PQ). Here, we assume that there are $b_1$ centroid sets, each with $b_2$ centroid embeddings. In this representation, an original text vector will be assigned with $b_1$ PQ indices, where each PQ index maps to a centroid embedding from the corresponding centroid set. The quantization-based  representation of a text is a vector of  $b_1$ PQ indices  (corresponding to the $b_1$ nearest centroid embeddings).
    When evaluating the similarity of a text, we can simply sum the entries from  the similarity table with its $b_1$ PQ indices. 
    }
    \label{fig:quantization}
\end{figure*}
\subsubsection{Improving Search Efficiency by Product Quantization}
\label{sec:quantization}

Besides reducing the amount of similarity computes, another possible way is to reduce the overhead of each similarity compute. 
Different from sparse representations (mostly composed of integer identifiers), dense representations (\ie text embeddings) are composed of real-value numbers, and it takes significantly more time for embedding similarity computation,  also  leading to an increase in memory cost.  

In this part, we describe a \emph{product quantization}~(PQ) approach~\cite{PQ-ref1,PQ-ref2} to compressing the text embeddings and accelerating similarity compute, reducing both time and memory costs. In practice, product quantization can be also jointly used with efficient index structures, \eg     clustering-based index and graph-based index. 
In the following, we will first describe how to compress the text embeddings as quantization-based representations, and then introduce how such representations reduce the cost of similarity computation when searching for a query. 
To introduce product quantization, we focus on discussing the case of a single embedding vector $\bm{v}_i \in \mathbb{R}^l$, while it can be easily extended to a set of text embeddings.

\paratitle{Quantization-based representations}.
As shown in Figure~\ref{fig:quantization}(a), to compress the text embeddings, given a corpus of $m$ texts, product quantization~\cite{PQ-ref1,PQ-ref2} firstly constructs $b_1$ centroid sets of embeddings, with each consisting of $b_2$ centroid embeddings. 
To represent an $l$-dimensional vector $\bm{v}_i$, we first split it into $b_1$ equal-length subvectors $\{\bm{v}_i^{(j)}\}_{i=1}^{b_1}$. 
Then, the $j$-th subvector $\bm{v}_i^{(j)} \in \mathbb{R}^{l / b_1}$ from the vector $\bm{v}_i$ is mapped to the nearest centroid (recorded as a \emph{PQ index} ranging from 1 to $b_2$) in the corresponding $j$-th centroid set. 
Since the number of centroid embeddings in a centroid set is usually smaller than 256  (\ie $b_2 < 256$), we can use one byte to represent a PQ index, which is able to largely reduce the space cost. 
Finally, the original vector   is compressed as a vector of $b_1$ PQ indices (with $b_1$ bytes), called \emph{quantization-based representation}. 
The quantization-based representations of all the text embeddings in a corpus can be pre-computed and stored before search.

\paratitle{Accelerated similarity computation}. 
Figure~\ref{fig:quantization}(b) illustrates how similarity computation can be implemented in an efficient way by using quantization-based representations. 
First, a query embedding is split into $b_1$ subvectors in the same way as that text vectors are split. 
Then, we compute the similarity between each query subvector and the centroid embeddings in the corresponding centroid set. 
As a result, a similarity table is constructed with $b_1 \cdot b_2$ entries (each entry represents the similarity score between one query subvector and one centroid embedding), and the table entries can be accessed by the PQ indices. 
In order to evaluate the similarity of a text \emph{w.r.t.} the query, we employ the quantization-based representation  of a text (\ie a vector of $b_1$ PQ indices) to look up the similarity table, and then sum the corresponding $b_1$ entries as the similarity score of the text. Note that the similarity table will be shared for all the texts: it only requires a total of $b_1 \cdot b_2$ embedding similarity computes, independent of the  corpus size. 
With this similarity table, the similarity evaluation of a text can be efficiently implemented by table lookup. 

\paratitle{Optimizing  quantization based index for retrieval}.
The original purpose of product quantization is to solve general vector compression problems, and it cannot be directly end-to-end learned in neural network models, due to the involved  non-differentiable operations. Besides, dense retrieval also requires specific  optimization strategies (\eg negative sampling), which are directly applicable to product quantization techniques.
To address the above issues, Zhan et al.~\cite{zhan2021jointly} propose an end-to-end learning approach based on product quantization for dense retrieval, where three major optimization strategies are proposed, including ranking-oriented loss, centroid optimization, and end-to-end negative sampling. 
Based on this work~\cite{zhan2021jointly}, the authors further incorporate the technique of constrained clustering to learn a better centriod assignment for a given document~\cite{zhan2021learning}. 
Besides, Zhang et al.~\cite{zhang2021joint} also propose several improved optimization strategies (including gradient straight-through estimators, warm start centroids and Givens rotation) to jointly train deep retrieval models with  quantization based embedding index.
Yamada et al.~\cite{Yamada2021EfficientPR} incorporate a learned hash function to represent both queries and texts as binary codes in an end-to-end way.  The proposed model achieves comparable performance to DPR, and further  significantly reduces the  computational and memory costs.

Besides efficiency improvement, it has been found that discrete representations derived from dense embeddings are useful to interpret the retrieval results by analyzing different aspects of input that a dense retriever focuses on~\cite{zhan2021interpreting}.

\subsection{Implementation for ANNS Algorithms}
\label{sec:ann-software}

In dense retrieval systems,  efficient similarity search is used at multiple stages. The primary use is to recall the most relevant text embeddings given the query vector.  
Besides, it can be also applied to select the  (static or dynamic) hard negatives~\cite{ANCE} or retrieve supporting context~\cite{REALM}.   

As aforementioned, existing dense retrieval studies mainly adopt Faiss~\cite{faiss} for implementing  nearest neighbor search. Faiss is a publicly released library for efficiently searching and clustering large-scale dense vectors. 
Basically, it provides the exact nearest neighbor search function, and further supports the implementations of several ANNS approaches discussed in Section~\ref{sec:ANNS} (including clustering-based approaches, graph based approaches, and quantization approaches) in both CPU implementations and GPU implementations. 
It also supports different similarity functions for vector search, including dot product and cosine similarity. 
In order to meet the requirements of effectiveness and efficiency, users can set the appropriate configuration with Faiss according to different hardware profile and data sizes. 

Besides,   {HNSWlib~\cite{malkov2018efficient} and SPTAG~\cite{ChenW18}} are libraries focusing on the efficient implementation for graph-based approach. 
ScaNN~\cite{avq_2020} is developed with quantization-based approaches, which achieves excellent performance on  ANN-Benchmarks. 
Distributed-Faiss~\cite{distributed-faiss} provides the distributed solutions for ANNS, when the index is too large to fit into the memory of a single machine.

Besides the above open-sourced software, Pinecone\footnote{~\url{https://pinecone.io/}} and Google's Vertex AI Matching Engine\footnote{~\url{https://cloud.google.com/vertex-ai}} provide commercial service for ANNS algorithms, which can help developers easily conduct dense retrieval systems for applications. 
Note that this survey mainly limits the discussion to in-memory solutions. When the data size is extremely large (\eg 100 billion scale),  it is not feasible to use completely in-memory ANNS index for dense retrieval, and we need to develop multi-level index using a hybrid of  memory index and disk index, \eg DiskANN~\cite{Subramanya2019DiskANNFA}  and SPANN~\cite{Chen2021SPANNHB}.

%% file: sec/sec-rerank.tex
\begin{figure*}[t]
	\centering
	\subfigure[Separate training.]{\label{fig:separate-optimization}
		\centering
		\includegraphics[width=0.25\textwidth]{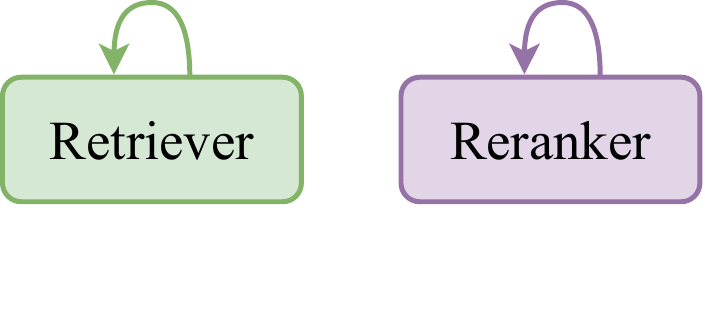}
	}\hspace{10mm}
	\subfigure[Adaptive training.]{\label{fig:coupled-optimization}
		\centering
		\includegraphics[width=0.25\textwidth]{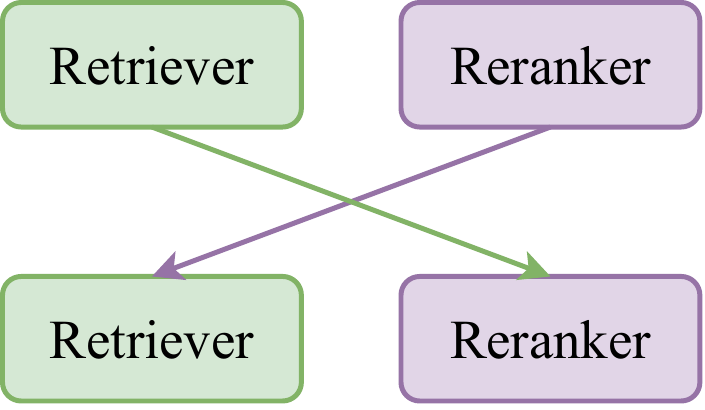}
	}\hspace{10mm}
	\subfigure[Joint training.]{\label{fig:joint-optimization}
		\centering
		\includegraphics[width=0.29\textwidth]{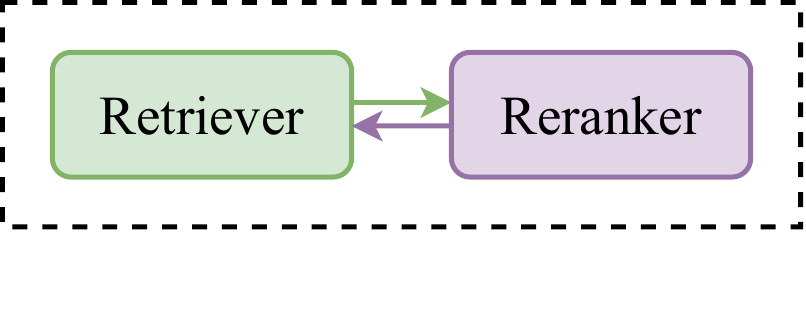}
	}
	\caption{Comparison of the three approaches for  pipeline training of retriever and reranker. }
	\label{fig:reranking-pipeline}
\end{figure*}

\section{Integration with Reranking}\label{sec:rerank}
In a complete retrieval system, it usually consists of first-stage retrievers and subsequent rerankers.  
In what follows, we first briefly introduce the retrieval pipeline,  then present several approaches to optimizing the  retrieval pipeline, and finally extend the discussion to the use of dense retrievers in other application systems. 

\subsection{The Retrieval Pipeline}
This part introduces the retrieval  pipeline and PLM-based rerankers.  

\subsubsection{The General Retrieval Pipeline}
To start our discussion, we consider a simplified  retrieval pipeline, consisting of two major stages, namely the first-stage retrieval and the reranking. To obtain the final ranked list, 
a certain number of possibly relevant documents (\eg several hundreds to thousands) to a given query are retrieved from a corpus by a retriever, such as dense retriever or BM25.
Then, the candidate texts are scored and reranked by a more capable relevance model (called \emph{reranker}).
Finally,   top ranked  texts (\eg several or tens) will be returned as the search results to downstream tasks, such as  question answering and dialog systems. 

Generally, first-stage retrieval and reranking stages have different focuses in a retrieval system~\cite{IR-book4,nogueira2019passage, lin2021pretrained}.
First-stage retrieval aims to efficiently recall relevant candidates from a large text corpus. 
As a result, it is practically infeasible to employ time-consuming models in first-stage retrieval. In contrast,  the goal of reranking is to reorder the candidate results from the proceeding stages, where the number of candidate texts is generally smaller (\eg hundreds or thousands) and more complicated models can be used to implement the rerankers. Therefore, bi-encoder is often used for implementing the retriever, and cross-encoder is often used as the architecture of the reranker.    

Note that a retrieval pipeline usually contains  multiple reranking stages by successively refining a reduced candidate list for producing the final results. Besides, there may also exist multiple first-stage retrievers in a practical retrieval system, where the results from multiple retrievers are aggregated as the input of the rerankers. 
Interested readers can refer to \cite{lin2021pretrained,fan2021pre,IR-book4} for detailed discussions.

\subsubsection{PLM-based Rerankers and Multi-stage Ranking}

In this part, we first discuss the typical reranker models based on PLMs, and then introduce multi-stage ranking mechanism that involves PLM-based rerankers. 

\paratitle{Reranker models}.  To implement the reranker, 
a typical approach is to employ the cross-encoder as the ranking model, showing substantial improvements over the traditional methods~\cite{qiao2019understanding, nogueira2019passage, wang2019multi, yan2019idst}.  Specifically, BERT is widely used to implement the cross-encoder for estimating the relevance score~\cite{nogueira2019passage,qiao2019understanding}, \eg monoBERT~\cite{multistage}.
As the input, a query and a candidate text are  concatenated as a sequence proceeded by the ``\textsc{[CLS]}'' token.
After encoding the query-text sequence, the ``\textsc{[CLS]}'' embedding is adopted as the match representation between the query and text, which will be subsequently fed into a neural network for computing the relevance score  of the text being relevant (Equation~\eqref{eq-prob-rel}). 
The relevance score is computed for each text independently, and the final ranking of texts is obtained by sorting them according to  relevance scores.
For training the rerankers, it usually formulates the ranking problem as a binary classification task~\cite{nogueira2019passage} using the binary cross-entropy~(BCE) loss as shown in Equation~\eqref{eq-L-BCE}. 
To optimize the BCE loss, it also needs to generate negatives for learning, which can be randomly selected or sampled from the top results of the retriever~\cite{Zhou2022TowardsRR,gao2021rethink}.
Furthermore, duoBERT~\cite{multistage} implements the BERT-based text ranking in a pairwise way, where it takes as input a query and a pair of texts to compare (with a concatenation pattern ``\textsc{[CLS]} \emph{query} \textsc{[SEP]} \emph{text}$_1$ \textsc{[SEP]} \emph{text}$_2$"  ). 
The training objective is to reserve the partial order of semantic relevance for a given text pair,  such that it can predict the relevance preference  for ranking the texts. Besides encoder-only models, encoder-decoder based PLMs (\eg T5) have been also utilized to implement the reranker, which takes as input a query-text pair and outputs the relevance label~\cite{nogueira2020document} or ranking score~\cite{T5ranking}.

\paratitle{Multi-stage ranking}. To construct a full system, a commonly used strategy is to arrange the retriever and the reranker(s) in a processing pipeline. Nogueira et al.~\cite{multistage} present a multi-stage ranking architecture in order to construct an effective retrieval pipeline. In this architecture, the first-stage ranker is implemented by the BM25 algorithm, while the second-stage and third-stage rankers are implemented by the monoBERT and duoBERT, respectively.
In this approach, different stages score each candidate separately. 
Recently, a new retriever-reranker integration  architecture has been proposed~\cite{HLATR} for multi-stage ranking optimization~\footnote{A similar approach SetRank~\cite{Pang2020SetRankLA} has been proposed, though it does not use PLMs. }. As the major novelty, the learned relevance information from retriever and reranker are transformed into input embeddings and positional embeddings respectively for another Transformer encoder, which produces the final ranking scores. 
Instead of scoring each candidate individually, the Transformer encoder derives the ranking scores by taking a listwise contrastive loss. 
Thus, the relevance information from the retrieval and reranking stages 
can be effectively utilized and optimized in a listwise way.  
In practice, multi-stage ranking is widely adopted in production systems (\eg Facebook Search~\cite{Huang2020embedding} and Walmart Product Search~\cite{Walmart}),  where more comprehensive factors are considered   and more complicated strategies are designed. 

Besides the above work, there is a large body of studies that utilize PLMs to enhance pre-BERT neural ranking models, \eg CEDR~\cite{macavaney2019cedr}. In these studies, PLMs are used to generate contextual features to enhance the semantic representation or similarity, which do not serve as the ranking backbone. Thus, we omit the discussion of this topic, and interested readers can find a detailed discussion in ~\cite{cai2021semantic,pretrain2020iclr,ma2021pre}.  

\subsection{Retrieval Pipeline Training}
\label{sec:pipeline-training}

The retrieval pipeline forms a  cascaded structure  by stacking the retriever and (one or multiple) reranker(s), and it needs to design suitable optimization algorithms for the entire pipeline.
Next, we introduce three major optimization approaches for the retrieval pipeline as shown in Figure~\ref{fig:reranking-pipeline}. For simplicity, we only discuss the scenario where only a retriever and a reranker are involved in a retrieval pipeline.

\subsubsection{Separate Training}
A straightforward approach is to separately optimize the  retriever and reranker,  considering them as two independent components. 
In this approach, the training of the  retriever and reranker are transparent to each other without information interaction (except the retrieval results) between the two components. Tyically, we can firstly optimize the retriever and then learn the reranker, following the optimization methods in Section 5 and Section 7.1, respectively.

Such an optimization approach cannot sufficiently capture the cascading information correlations between different stages in the retrieval pipeline, thus possibly leading to a suboptimal ranking performance. 
The reasons are twofold. 
First, when the retrieved candidate list improves, the contained negatives also become increasingly difficult to be discriminated by the reranker, which are more likely to be false negatives~\cite{gao2021rethink,rocketqa}.  Second, without considering the first-stage retrieval results, the optimization of reranker cannot be well adjusted to 
the result distribution of the retriever, which is not specially optimized according to  the entire pipeline~\cite{rocketqav2}.
Therefore, it is necessary to incorporate information interaction between the retriever and reranker during training.

\subsubsection{Adaptive Training}

As an improvement approach, adaptive training enables the information interaction  between the retriever and reranker. 
The basic idea is to let the two components adapt to the retrieval results (or intermediate data representations) of each other in order to achieve a better overall retrieval performance. For example,  Qu et al.~\cite{rocketqa} propose to train a cross-encoder by sampling  from top results of the retriever (a dual-encoder) as negatives, which lets the cross-encoder adjust to the distribution of the retrieval results  by the dual-encoder.

Another commonly adopted way is to alternatively  update the 
retriever and reranker during the training process. 
At each iteration, the fixed component can provide necessary relevance information or guidance signals to another being-optimized component. 
Trans-Encoder~\cite{liu2021trans} designs an iterative joint training framework  by alternating between the cross-encoder and bi-encoder. Specifically, during the learning process, one component  will produce pseudo relevance labels for updating the other one.

Furthermore, an adversarial learning approach has been proposed in \cite{zhang2021adversarial}: the roles of the retriever and ranker are to retrieve negative documents and identify ground-truth documents from a mixed list including both positive and negative ones, respectively. The entire optimization process of the retriever and ranker is formulated as a  minimax game.

\subsubsection{Joint Training}
In implementation, retriever and reranker are often optimized in different ways, which makes it difficult for joint optimization. 
Specially, the retriever is usually trained by  maximizing the probabilities of positive passages against a list of sampled negatives.
It is optimized according to the overall ranking for the candidate list of positive and negatives, called \emph{listwise approach}\footnote{Following \cite{rocketqav2},  \emph{listwise} denotes that the optimization is based on an entire candidate list, which is different from that in learning to rank~\cite{cao2007learning}.}.
As a comparison, the training of reranker is modeled  as \emph{pointwise} or \emph{pairwise} optimization, where the model is learned based on a single text or a pair of texts. 
Due to the different optimization ways, it is difficult to jointly train the two components. 

To address the above issue, RocketQAv2~\cite{rocketqav2} proposes a unified listwise learning approach to jointly optimizing a bi-encoder retriever and a cross-encoder reranker.
For both retriever and reranker,  their relevance predictions are modeled as listwise  distributions. 
To unify the learning process, RocketQAv2 designs a \emph{dynamic listwise distillation} mechanism  by distilling the reranker to the retriever. 
Different from previous distillation methods, it adaptively updates the retriever and reranker   by enforcing the interaction of relevance information between the two modules. 
For optimizing the listwise distillation, RocketQAv2 also employs  data augmentation to construct high-quality training data.  

Besides, Sachan et al.~\cite{sachan2021end} present an end-to-end training method to jointly optimize  the retriever and the reader (similar methods can apply to reranker) for 
retrieval-augmented question-answering systems. They propose two training approaches by modeling the retrieved documents,  either \emph{individually} or \emph{jointly}.

\subsection{Integration and Optimization in Other Applications}

In the above subsection, we have discussed how to optimize a retrieval pipeline 
that integrates a dense retriever and reranker(s). 
Besides retrieval systems, dense retrievers have been widely used in a variety of application systems that require external knowledge resources, such as open-domain question answering and entity linking. 
Typically, these application systems contain  two major parts, namely knowledge retriever (the component provides necessary supporting evidences) and task solver (the component reasons over the retrieved evidences for prediction). 
Compared with traditional sparse retrievers, dense retrievers are more flexible to be integrated into downstream tasks. It is conceptually simple, and can be effectively optimized according to specific task goals.  

To optimize a retrieval-augmented application system, a typical approach~\cite{REALM}  is to construct an embedding index for knowledge retriever and then utilize the retrieved text embeddings for optimizing the task solver. 
During the training of the whole system, 
a major difficulty lies in the issue of \emph{index staleness}: at each training step, when we update the knowledge encoder, the indexed embeddings no longer correspond to the latest parameters of the trainer.  
To address this issue, REALM~\cite{REALM} proposes to use  \emph{asynchronous index update}: the index builder is   periodically (\eg each for several hundred training steps) updated,  and the trainer utilizes an existing  index for optimization before it is  reconstructed\footnote{The   asynchronous update strategy can be utilized in other optimization tasks that rely on  indexed embeddings, \eg hard negative retrieval~\cite{ANCE}. }.
Such an approach can be generally applied to optimize downstream systems built on dense retrievers.   
Other related studies~\cite{sachan2021end,Izacard2021LeveragingPR,Lewis2020RetrievalAugmentedGF} also discuss how to optimize retrieval-augmented systems for open-domain question answering.

In general, with dense retrieval techniques, it is  easier  to develop retrieval based  approaches  for downstream tasks. We will discuss the use of dense retrieval techniques for specific applications in Section 9.

%% file: sec/sec-advancedtopics.tex
\section{Advanced Topics}\label{sec:advanced-topics}
In this section, we  discuss several important advanced topics for dense retrieval, including zero-shot dense retrieval, model robustness to query variations, model based retrieval and retrieval-augmented language model.

\subsection{Zero-shot Dense Retrieval}
~\label{sec:advanced-topics-zero}

The success of dense retrievers heavily relies on large-scale relevance judgement data. 
This poses a challenge for a wide range of task scenarios, as it is difficult and expensive to acquire sufficient training corpus when a new domain or task is introduced. Thus, it is important to examine the zero-shot capabilities of dense retrievers, as well as their out-of-distribution performance.

\paratitle{Zero-shot evaluation datasets}.
To our knowledge, BEIR~\cite{thakur2021beir} is the first heterogeneous benchmark for examining the zero-shot capabilities of dense retrieval methods. BEIR includes nine different retrieval tasks (news retrieval, question answering, bio-medical information retrieval, etc.) spanning 18 diverse datasets. 
Based on BEIR, it has been found that a number of bi-encoder dense retrievers that outperform BM25 on in-domain evaluation perform poorly on out-of-distribution evaluation,  while cross-attentional reranking models and late interaction models have better zero-shot performance on BEIR. 
Furthermore, Sciavolino et al.~\cite{Sciavolino2021SimpleEQ} create a dataset  containing simple entity-centric questions, and find that BM25 significantly outperforms dense retrievers on this dataset. Their analysis shows that dense retrievers perform better on common entities than rare entities, and can also generalize to unseen entities to some extent when question patterns exist in the training data. 
Liu et al.~\cite{Liu2021ChallengesIG} further measure three kinds of generalization capacities in different settings based on NQ dataset, including  training set overlap, compositional generalization and novel-entity generalization. They find that dense retrievers perform worse in  the latter two settings than the first setting.

\paratitle{Empirical analysis of the influence factors}. Besides the findings drawn from the overall performance comparison, it is essential to  understand how different factors affect the performance of dense retrievers in a zero-shot setting. 
For this purpose, Ren et al.~\cite{ren-zero} have conducted a   thorough study on the effect of underlying influencing factors on zero-shot retrieval performance based on a total of 12 commonly used retrieval datasets.
This study frames the zero-shot retrieval setting by introducing source domain data (available for training) and target domain data (only available for testing). They mainly focus on examining the influence factors from source training corpus (\ie query set, document set, and data scale), and also discuss other factors such as query type distribution and vocabulary overlap.
They  empirically find that source training dataset has significant influence on the zero-shot retrieval performance of the dense retrieval models, since only source domain data can be utilized for training. Such an effect can be attributed to a number of specific  factors, including vocabulary overlap (\emph{larger is better}),  query type distribution (\emph{more comprehensive is better}), and data scale (\emph{more queries are better but not for documents}). Besides, they also find that  the 
dataset bias of the test set potentially affects the performance comparison between sparse and dense retrieval models:  since some datasets are constructed based on lexical matching models~\cite{tsatsaronis2015overview, suarez2018data}, they tend to be more favorable for sparse retrieval methods due to a larger term overlap.

\paratitle{Exisiting solutions for zero-shot retrieval}. Recently, it has attracted much attention from the research community to improve the zero-shot capabilities of dense retrievers. Following \cite{ren-zero}, we briefly summarize these studies and discuss how they address the issues for zero-shot retrieval in three major aspects. 

$\bullet$ \emph{Augmenting the target training data}. For zero-shot retrieval, the major challenge is that it lacks the training data from the target domain. Considering this issue, a major solution is to generate   large-scale synthetic data for improving the training of 
dense retrievers.  After being trained on  large-scale synthetic training data, the zero-shot capabilities of dense retrievers can be improved to some extent.
Typically, a data generation model is employed for generating large-scale query-text pairs in the target domain~\cite{ma2021zero, liang2020embedding, reddy2021towards, wang2021gpl}. 
Recently, Promptagator~\cite{Dai2022PromptagatorFD} leverages a large language model consisting of 137 billion parameters, \ie FLAN~\cite{Wei2022FinetunedLM}, to generate large-scale query-text pairs by using only a small number of  labeled examples.  It trains a dense retriever with the set of  generated query-text pairs, and shows that the retriever trained on the generated data (with consistency filtering) significantly outperforms ColBERTv2~\cite{santhanam2021colbertv2} that is well-trained on manually annotated dataset (\eg MS MARCO) on BEIR. 
As an alternative approach, knowledge distillation is commonly adopted for tackling the scarcity of training data, {which utilizes a more capable teacher model to improve the student model~\cite{chen2021salient, wang2021gpl, thakur2020augmented} in zero-shot retrieval scenarios.} Besides, several studies~\cite{Izacard2021TowardsUD, xu2022laprador} conduct unsupervised pretraining by leveraging large-scale positive and negative pairs with different data augmentation methods, \eg ICT~\cite{latent2019acl} and SimCSE~\cite{Gao2021SimCSESC}.

$\bullet$ \emph{Enhancing the term matching capability}. Unlike sparse retrievers~(\eg BM25), dense retrievers no longer rely on exact term matching for text retrieval, but instead learn  semantic matching for capturing the text relevance. 
While, empirical studies show that the capability of exact term matching is useful to improve zero-shot retrieval performance~\cite{thakur2021beir, Chen2022OutofDomainST}, since term overlapping is a strong indicator for text relevance.  
Thus, sparse retrievers are easier to adapt to zero-shot settings without training data.  
Inspired by this, several studies propose to enhance the lexical matching capacity of dense retrievers by leveraging sparse retrievers, such as the fusion of rankings~\cite{Chen2022OutofDomainST} or relevance scores~\cite{xu2022laprador} for both sparse and dense retrievers. Besides, we can also employ knowledge distillation
for improving dense retrievers~\cite{chen2021salient}, taking sparse retrievers as the teacher model. 

$\bullet$ \emph{Scaling up the model capacity.} In recent years, scaling law for PLMs has been widely explored for raising the performance bar of various tasks. It has been shown useful to improve the zero-shot performance by increasing the model size of dense retrievers. In \cite{Ni2021LargeDE},  based on a T5-based dense retriever trained on large-scale question-answer pairs, scaling up the model size with multi-stage training can significantly improve the zero-shot retrieval performance on the BEIR  benchmark. Similarly,  performance improvement has been observed when the model size is increased from 0.1 billion to 2.4 billion in ERNIE-Search~\cite{ERNIE-Search}.

Besides, it is also useful to employ multi-task learning
(\eg a joint training with self-supervised tasks, dense retrieval and extractive question answering~\cite{fun2021efficient}) to improve the zero-shot retrieval performance.  {As another related study, Zhan et al.~\cite{Extrapolation} examine how dense retrievers generalize to out-of-domain test queries,  called  \emph{extrapolation capacity} in their paper. They find that cross-encoder can extrapolate well, but not bi-encoder and sparse retriever.  }  This part can be extended to a more general topic \emph{low-resourced  dense retrieval}, and readers can find a comprehensive discussion in a recent  survey~\cite{survey-lowresource}.

\subsection{Improving the Robustness to Query Variations}
Besides the out-of-distribution issue in zero-shot retrieval, it has been shown that dense retrieval models are 
more sensitive to \emph{query variations} than traditional lexical matching methods~\cite{BERT-typo,EvaluatingRobustness,LocalRanking}. 
Generally, query variations exist widely in real search behaviors, \eg unexpected query typos due to spelling errors and diverse query formulations for the same information need. We next discuss the effect of query variations on retrieval performance and introduce several  enhanced training methods. 

\paratitle{Effect of query variations on retrieval}. There are increasing concerns on the robustness of dense retrieval models to query variations.
These studies typically create different types of query variations and examine the performance of dense retrieval models under different query variations.
Zhuang et al.~\cite{BERT-typo} present a study on the impact of query typos based on character operations (insertion, deletion, substitution and swap), showing a significant performance drop for BERT-based  retriever and  reranker. 
Penha et al.~\cite{EvaluatingRobustness} aim to  examine how different types of query variations negatively affect the robustness of the retrieval pipeline.   
In order to generate query variations, they consider four syntax-changing types (misspelling, naturality, ordering and paraphrasing) and two semantics-changing types (gen./specialization, aspect change) for query transformations. 
Experimental results show that retrieval pipelines are sensitive to query variations (especially the misspelling), which lead to an average 20\% drop on performance compared with that evaluated on original queries. 
Similar findings are reported about the performance drop due to query typos in \cite{LocalRanking},  
where they consider eight  types of query typos, including  new transformations such as adding extra punctuation, stopword removal and verb-tense shift.

\paratitle{Enhanced training methods}. In order to improve the robustness, existing studies propose a series of enhanced training methods that take the query variations into consideration. Basically, these methods utilize data augmentation strategies to incorporate  augmented queries in training data, so that dense retrieval models can be  aware of query variations during training. Zhuang et al.~\cite{BERT-typo} propose a typos-aware training approach, which changes the queries (with 50\% chance) in training set according to different typo types. In this way, the retrieval model is trained with a training set consisting of both original queries and augmented queries with different typo types. Given such an augmentation dataset, Sidiropoulos and Kanoulas~\cite{typo++} propose to construct a query-side contrastive loss: modified queries and randomly sampled queries are considered as  positives and negatives, respectively.
Based on the work in  \cite{BERT-typo}, the same authors~\cite{zhuang2022characterbert} further propose a self-teaching approach by incorporating a distillation loss, and it requires the underlying retrieval model to produce similar rankings under original queries and corresponding variations. They characterize such an idea by reducing the KL-divergence loss between the distributions over the same candidate list for original and 
augmented queries.  
Similarly, Chen et al.~\cite{LocalRanking} propose a local ranking alignment method that enforces the original queries and the corresponding variations to produce similar rankings: the probability distributions of the in-batch passages given an original query and its variation should be similar, and the probability distributions over in-batch queries or their variations given a passage should be similar.

While, query variations are not always harmful for retrieval systems.
For example, we can utilize  \emph{query extension} or \emph{query rewriting}~\cite{carpineto2012survey} to derive better search results by reformulating the issued query.  Furthermore, there are several studies that focus on
the vulnerabilities of dense retrieval models~\cite{BERTAdversarial,Order-Disorder}. It has been shown that dense retrieval models are brittle to deliberate attacks and thus adversarial training approaches are often used  for enhancing the model robustness~\cite{BERTAdversarial,Order-Disorder}. 
While, we limit our discussion to query variations, which are more likely to occur in real search scenarios than deliberate attack~\cite{EvaluatingRobustness}.

\subsection{Model based Retrieval}

In recent years, \emph{model based  retrieval}~\cite{metzler2021rethinking,Tay2022TransformerMA} has been proposed as an alternative paradigm for information retrieval. During retrieval,  it no longer examines the  index storing the candidate texts in the collection, but instead directly predicts the identifiers of relevant documents  (\ie docids) based on a parametric model. The underlying model is expected to fully  capture the necessary relevance information, which is the key to this retrieval approach.

\paratitle{The generative scheme}. In essence, model based  retrieval adopts a  generative scheme for predicting relevant texts.  
Indeed, the idea of generative retrieval has been first explored in the task of entity linking~\cite{DeCao2021AutoregressiveER}, where entity identifiers (\eg unique entity names) are generated conditioned on the text context in an autoregressive manner, and then GRLS~\cite{GRLS} extends this idea to long sequence retrieval by introducing multi-step generative retrieval. 
Specially, the perspective paper~\cite{metzler2021rethinking} envisions the  \emph{model-based paradigm}  that simplifies the classic \emph{index-retrieve-then-rank} paradigm  by a unified relevance model. 
Further, a representative work~\cite{Tay2022TransformerMA} called \emph{DSI} develops a generative  retrieval model based on a seq2seq encoder-decoder architecture (\ie T5), where query and docids correspond to input and output, respectively. 
DSI frames the approach by introducing two key procedures: \emph{indexing}~(associating the content of a document with its corresponding docid) and \emph{retrieval}~( autoregressively generating docids given the input query). They are essentially related to the two key issues: (i) how to represent the texts with meaningful docids, and (ii) how to map a query to relevant docids. 
Next, we discuss the two issues in detail. 

\paratitle{Representing docids}. Since, in model based  retrieval, we do not directly use the text content for relevance evaluation, we need to design effective docid representations to reflect the underlying semantics of a text. 
  DSI~\cite{Tay2022TransformerMA} introduces three major docid representation methods, including unstructured atomic identifiers (a unique integer identifier), simple string identifiers (tokenaizable strings) and semantically structured identifiers (clustering-based prefix representation). 
Intuitively,  the second approach is more flexible to use, and in principle one can use various kinds of sequence data to represent a document, \eg titles and URLs; while the third approach seems to be more meaningful at the cost of additional pre-processing (\eg clustering based on BERT embeddings), which can be further effectively leveraged by a prefix-based decoding way.  
DSI compares the three docid assignment methods, and shows that the third method leads to the best performance in its implementation. 
Later extensions almost follow the three major methods for representing docids: 
DynamicRetriever~\cite{Zhou2022DynamicRetrieverAP}~(atomic docid: original docid), SEAL~\cite{Bevilacqua2022AutoregressiveSE}~(simple string docid: full ngram signatures with the efficient support of Ferragina Manzini index), NCI~\cite{NCI} (semantic string docids: hierarchical clustering), DSI-QG~\cite{DSI-QG} (simple string docids: generated queries), Ultron~\cite{Ultron} (simple string docids: URL and titles, semantic string docids: product quantization), GERE~\cite{GERE} (semantic string docids: title and evidence), and CorpusBrain~\cite{CorpusBrain} (semantic string docids: title).

\paratitle{Pretraining for semantic mapping}. The essence of  model based   retrieval is to establish the semantic mapping from queries to docids, without an explicit reference to the text content. Thus, in principle, it is more difficult to train generative retrieval models than previous bi-encoder models. 
For learning query-to-docid mapping, 
DSI employs two major training strategies, namely memorization-based pretraining (content mapping) and retrieval-oriented fine-tuning: the former learns to map the content of a document to its corresponding docid and the latter learns to map queries to relevant docids.
For content mapping, various pretraining tasks are further proposed to enhance the association between the text content and the docids~\cite{Tay2022TransformerMA,Ultron,CorpusBrain}, \eg mapping a sampled content (\eg passage, word sets and ngrams~\cite{Ultron,Zhou2022DynamicRetrieverAP}) or associated content (\eg anchor text~\cite{CorpusBrain}) from a text to its docid.
Such a strategy can be  considered  as \emph{learning from pseudo queries}, where various kinds of sampled contents are considered as pseudo queries.
Besides, since the amount of available relevance judgement data is usually limited, query generation is useful in these generative retrieval models for enhancing the learning from queries to docids~\cite{DSI-QG,Ultron}, where the original content of a document can be employed to generate potential queries.  

\paratitle{Merits}. Compared with traditional retrieval approaches, model based  retrieval potentially leads to a  more unified retrieval paradigm. 
The major merits of this approach are threefold. Firstly, it   simplifies the classic index-retrieve-then-rank paradigm by \emph{a model-based paradigm}, where a more unified solution to information retrieval tasks could be developed (as envisioned in \cite{metzler2021rethinking}). 
Due to the flexibility, there are increasing studies that utilize model based retrieval for solving different knowledge-intensive tasks~\cite{CorpusBrain,GERE}. 
It can be potentially  extended to index and retrieve various Web resources, \eg figures and videos. 
Secondly,
since such an approach  basically takes an encoder-decoder architecture, it is more easier to be optimized in an end-to-end way. 
We could potentially get rid of the tedious pipeline training in existing  retrieval systems. 
Thirdly, it directly removes  the use of elaborate inverted-index or  large-sized embedding index. During retrieval, we do not need to manage a complicated  index structure for retrieval, which can largely simplify the data structure for supporting  the retrieval system. 

Despite these merits (unified paradigm,  training flexibility and simplified structure), this emerging retrieval paradigm is  still under exploration. So far, most of model based retrieval studies only demonstrate the effectiveness of their approaches on MS MARCO subsets or task-specific datasets, while evaluation at a larger scale (\eg the full MS MARCO dataset) is still challenging for model based  retrieval. 

\subsection{Retrieval-Augmented Language Model}
\label{sec:application-retrieval-augmented-lm}
In recent years,  neural language models~(LM), especially PLMs, have proven to be powerful~\cite{Bengio2000ANP,Mikolov2013DistributedRO,Peters2018DeepCW,Radford2018ImprovingLU} in a variety of tasks. 
It is shown that PLMs can encode a large amount of semantic knowledge~\cite{Petroni2019LanguageMA,brown2020language} into  large-scale model parameters. In order to enhance the representation capacity, a number of studies  explore the scaling law of PLMs~\cite{brown2020language,Raffel2020ExploringTL} by increasing the model size, which makes the training and use of PLMs prohibitively expensive in a resource-limited setting.

In order to alleviate this issue, retrieval-augmented LMs have been proposed by allowing the model to explicitly access external data~\cite{Khandelwal2020GeneralizationTM,He2021EfficientNN,Yogatama2021AdaptiveSL,REALM,Borgeaud2021ImprovingLM}. 
Retrieval-augmented models tend to be more parameter efficient, since they retrieve the knowledge rather than fully storing it in model parameters.  
Khandelwal et al.~\cite{Khandelwal2020GeneralizationTM} present $k$NN-LM, an approach that linearly interpolates between the LM's next word distribution and the next word distribution of $k$-nearest neighbor model.
The experiments demonstrate that $k$NN-LM works particularly well when predicting rare patterns, since it allows explicit retrieval of rare patterns. 
Furthermore, Yogatama et al.~\cite{Yogatama2021AdaptiveSL} incorporate a gating mechanism to replace the fixed 
interpolation parameter in $k$NN-LM, which adaptively combines short-term memory with long-term memory (\ie knowledge retrieved from external memory) for prediction conditioned on the context.

Since external knowledge resource is very important to retrieval-augmented LMs, 
Borgeaud et al.~\cite{Borgeaud2021ImprovingLM} further explore scaling up the retrieval corpus to trillions of tokens for large LMs ranging from 150M to 7B parameters, which leads to an improved capacity of LMs. 
In these studies, LMs are frozen pretrained retrievers. As a comparison, Guu et al.~\cite{REALM} propose REALM to jointly optimize the knowledge  retriever and the knowledge-augmented encoder (detailed in  Section~\ref{sec:retrieval-augmented-pretraining}).

%% file: sec/sec-application.tex
\section{Applications}\label{sec:application}

As an important information seeking technique, dense retrieval has been widely applied in a variety of applications from different fields. 
In this section, we review the applications of dense retrieval technique in three aspects, namely the applications to information retrieval tasks (\emph{extensions to different retrieval settings}), the applications to natural language processing tasks (\emph{benefiting downstream tasks}) and industry practice (\emph{industry-level use and adaptation}).  

\subsection{Information Retrieval  Applications}

In this part, we describe several specific retrieval settings and corresponding dense retrieval approaches. 

\subsubsection{Temporal Retrieval}

Most of existing works conduct the study of dense retrieval  on synchronic document collections (\eg Wikipedia). 
However, there is a large proportion of temporal-dependent questions in real application scenarios. To advance the research on temporal retrieval, Zhang et al.~\cite{zhang2021situatedqa} create an open-retrieval question answering dataset called \emph{SituatedQA}, where the systems are required to answer questions given the tempo-spatial  contexts. Their experiments  show that  existing dense retrievers cannot produce satisfying answers that are frequently updated. 
Besides, Wang et al.~\cite{wang2021archivalqa}  create another large question answering dataset corpus called \emph{ArchivalQA}, consisting of 1,067,056 question-answer pairs, which is specially constructed for temporal news question answering. 
These datasets can be used to develop time-sensitive dense retrievers. 
\subsubsection{Structured Data Retrieval}

Previous sections mainly review the progress of dense retrieval on unstructured text. Recently, dense retrieval has also been applied to structured data, \eg lists, tables and knowledge bases. 
Herzig et al.~\cite{herzig2021open} apply a bi-encoder based dense retriever on table retrieval, which is a Transformer-based language model pretrained on millions of tables with the awareness of tabular structure. As shown in \cite{herzig2021open}, the proposed approach significantly outperforms BM25  on the NQ-TABLES dataset. 
Oguz et al.~\cite{oguz2020unified} further study the problem of unifying open-domain question answering with  both structured and unstructured data. They first linearize tables and knowledge bases into texts, and then train a bi-encoder based dense retriever on multiple datasets consisting of texts, lists, tables and knowledge bases. The experimental results show that the proposed approach performs well on both entity-oriented and text-oriented benchmarks. 
Kostic et al.~\cite{kostic2021multi} also explore the unified retrieval by proposing  a tri-encoder, with one unique encoder each  for question, text and table, respectively. Their experimental results show that the tri-encoder performs better than the bi-encoder with one encoder for the question, the other one for both text and tables.

\subsubsection{Multilingual Retrieval} 
In the literature, monolingual retrieval (mainly in English)  is most commonly studied, since available  large-scale labeled datasets are mainly in English (\eg MS MARCO and NQ). 
To  facilitate the research on other languages rather than English, Bonifacio et al.~\cite{bonifacio2021mmarco} create a multilingual version of the MS MARCO passage ranking dataset using machine translation, called \emph{mMARCO},  consisting of 13 languages.  
The experimental results show that the models fine-tuned on multilingual datasets perform better than that fine-tuned only on the original English dataset.  
 {Zhang et al.~\cite{mrTyDi} present the  multi-lingual benchmark dataset called \emph{Mr. TYDI}, consisting of monolingual retrieval corpus in eleven langauges. They show that BM25 performs  better than  a multi-lingual adaptation of DPR  on this benchmark dataset. }

The above multilingual retrieval settings aim to find  the answer from the corpus in the same language as the question. 
However, many languages face the scarcity of text resources, where there are few reference texts written in a specific language.
Hence, Asai et al.~\cite{asai2021xor} propose a cross-lingual task, called Cross-lingual Open Retrieval Question Answering (XOR QA), that requires answering questions in one language based on a text in another language. 
They create a large-scale dataset consisting of 40K questions across seven non-English languages for this proposed task.
The proposed approach relies on machine translation modules for question translation and answer translation.  Since the multi-step machine translations (\eg query translation and answer translation) in previous approaches may lead to error propagation,  {Asai et al.~\cite{asai2021one} further propose a multilingual dense passage retriever that extends DPR to retrieve candidate passages  from multilingual document collections in a cross-lingual way. }
It adopts an iterative training approach to fine-tuning a  multilingual PLM (\eg mBERT~\cite{pires2019multilingual}), and achieves performance improvement in a number of languages.

\subsubsection{Other Retrieval Tasks} 

So far,  dense retrieval has been widely applied to various domain-specific  retrieval tasks, such as mathematical IR~\cite{math},  code retrieval~\cite{code} and biomedical IR~\cite{Biomedical}. When considering a specific retrieval task, we need to adapt the dense retrievers to effectively fit domain-specific text data (continual pretraining is often needed), deal with special content features or formats (\eg formulas or code) and resolve other possible training issues (\eg lack of training data).

Besides, another important topic is cross-modal retrieval~\cite{CMR-survey}, where we consider a retrieval setting across different modalities (\eg text-to-image retrieval  and text-to-video retrieval). The key to these cross-modal tasks is to establish the fusion or mapping between different modalities. With the development of vision-language pretraining models exemplified by CLIP~\cite{CLIP}, the capacity of learning shared representations across modalities has been largely enhanced, which can improve the downstream cross-modal retrieval tasks.  
While, this survey is mainly focused on text resources for retrieval, and interested readers can refer to \cite{CMR-survey} for a detailed discussion. 

\subsection{Natural Language Processing Applications}

Many NLP tasks need to retrieve relevant supporting evidence or knowledge from external sources, \eg text corpus,  knowledge graph and tables.  
In this section, we will discuss the applications of  dense retrieval to three typical NLP tasks, including question answering, entity linking and dialog system.

\subsubsection{Question Answering}

Open-domain question answering~(QA) is an information seeking  task that  aims to find accurate answers to a question instead of relevant texts.  A QA system typically conducts a 
two-stage approach~\cite{chen17openqa}, where a text retriever  is firstly employed to find relevant documents (or passages) from a large text corpus and then another component (called \emph{reader}) derives  the exact answer from these documents. 

Traditionally,  the text retriever in QA systems is often implemented by sparse retrieval methods such as  BM25 and tf-idf. 
 As an important progress, ORQA~\cite{latent2019acl} (for the first time) demonstrates that dense retrieval models can outperform BM25 on multiple open-domain QA datasets. 
It pretrains the retriever with an  Inverse Cloze Task in a self-supervised manner (as discussed in Section~\ref{sec:training-pretrain}), and then jointly fine-tunes the retriever and reader. However, ORQA needs computationally intensive pretraining. 
By fine-tuning the BERT model only on annotated question-passage pairs,
DPR~\cite{dpr2020} no longer requires  
additional pretraining, while the experiments demonstrate that DPR performs even better than ORQA on multiple QA datasets. 
Moreover, RocketQA~\cite{rocketqa} proposes an optimized training approach to dense retrieval,  leading to a better performance on QA tasks. These studies rely on extractive readers to obtain answers from retrieved documents. Different from these studies, RAG~\cite{Lewis2020RetrievalAugmentedGF} and Fusion-in-Decoder~\cite{Izacard2021LeveragingPR} use seq2seq based generative models  for predicting the answer, showing a  better performance on open-domain question answering than extractive readers.

The above studies focus on the simple  setting where the answers can be derived from a single supporting document, \ie the single-hop setting.
While, in a more complicated setting, the QA system needs to collect sufficient evidence from multiple documents for finding the answer, which requires the capacity of multi-hop reasoning (\eg HotpotQA~\cite{yang2018hotpotqa} and HoVer~\cite{Jiang2020HoVerAD}). In early studies~\cite{Asai2020LearningTR},   hyperlink structure in the corpus has been utilized to thread relevant documents for locating the answer. More recently, 
without using hyperlinks, 
\emph{multi-hop dense retrieval}~\cite{Khattab2021BaleenRM,mh-dr}  has been proposed to enhance  the performance of complex  QA tasks~\cite{yang2018hotpotqa}. 
Specifically, it iteratively encodes the combination of the question and the retrieved documents as a new question embedding, and then retrieves the documents relevant to the new query for the next hop. 
Experiment results show that these approaches work well for two-hop tasks.
However, the many-hop retrieval task remains challenging, as each hop is likely to propagate errors.
To better deal with many-hop (more than two hops) tasks, Khattab et  al.~\cite{Khattab2021BaleenRM} further propose the \emph{condensed retrieval} architecture,  which enhances the query informativeness by extracting the facts from each hop as part of the query for the next hop. 

More recently, the survey paper~\cite{QASurvey} extensively discusses the recent progress on  open domain question answering based on dense retrieval. The readers can refer to this survey for a detailed introduction.

\subsubsection{Entity Linking}

Entity linking is a semantic disambiguation task that relates specific contextual references to real-world entities. 
Given a text context containing a specific mention, the entity linking system aims to link it to a corresponding  entity in a knowledge base. 
Since knowledge bases are usually large (\eg English Wikipedia contains more than six million articles~\footnote{As of 25 November 2022, there are 6,579,781 articles in the English Wikipedia.}), a major challenge of entity linking is to efficiently retrieve a small number of candidate entities for linkage. 
Traditional approaches mainly use a high-coverage alias table or sparse retrieval methods (\eg TF-IDF and BM25) for efficient candidate retrieval, and further  employ neural models for effective candidate ranking~\cite{FrancisLandau2016CapturingSS,Gupta2017EntityLV,Ganea2017DeepJE}. 

Based on dense retrieval, researchers use PLMs for candidate recall or reranking in entity linking. 
Logeswaran et al.~\cite{Logeswaran2019ZeroShotEL} explore BERT-based rankers with different architectures (\eg bi-encoder and cross-encoder) for candidate ranking but rely on traditional sparse retrieval for candidate recall. 
As neural two-tower architecture (pre-BERT models) is shown effective for candidate recall~\cite{entitylinking}, Wu et al.~\cite{wu2020scalable} further propose a BERT-based two-stage approach for entity linking, namely BLINK. In its first stage, BLINK uses a BERT-based bi-encoder to retrieve candidates. The retrieved candidates are then reranked by a BERT-based cross-encoder. 
In BLINK, when training the bi-encoder,  hard negative mining and  knowledge distillation  are also used to improve its retrieval performance.

\subsubsection{Dialog System}

Recent studies  (\eg Meena~\cite{Adiwardana2020TowardsAH}, BlenderBot~\cite{Roller2021RecipesFB}, and PLATO-2~\cite{Bao2021PLATO2TB}) that pretrain dialog models on large-scale corpora have demonstrated their ability to converse like a human to some extent. 
However, these models are still prone to suffering from factual incorrectness and knowledge hallucination in an open-domain setting~\cite{Thoppilan2022LaMDALM}.

Inspired by the success of Retrieval-Augmented Generation (RAG) in the task of  open-domain question answering~\cite{Lewis2020RetrievalAugmentedGF}, 
 the RAG architecture has been applied to open-domain knowledge-grounded dialog to address this issue~\cite{Shuster2021RetrievalAR,Huang2021PLATOKAGUK}. 
In this approach, a retriever  trained in an end-to-end manner is used to access external knowledge sources based on dialog contexts. 
The retrieved knowledge is then used to enhance dialog generation. 
Experiment results show that these approaches can alleviate the hallucination issue without sacrificing the conversational ability~\cite{Shuster2021RetrievalAR}. 

Furthermore, several studies~\cite{Komeili2021InternetAugmentedDG,Thoppilan2022LaMDALM} propose to generate a pseudo search query with generative models according to the dialog contexts. 
The generated query is then used to retrieve knowledge from a search engine or a pre-built information retrieval system. 
The retrieved knowledge and conversation history will be subsequently used to generate a knowledge-grounded response.

\subsection{Industrial Practice}

Besides research-oriented studies,  considerable  efforts have attempted to deploy dense retrieval techniques in real systems\footnote{Due to the privacy of companies, we mainly survey the public industrial reports from the literature or blogging articles.}, including 
Web search~\cite{Liu2021PretrainedLM}, sponsor search~\cite{fan2019mobius,Zhang2022UniRetrieverTL}, product search~\cite{li2021embedding,Liu2021Que2SearchFA} and personalized search~\cite{Huang2020embedding}.

In Web search, the term mismatch between search queries and Web pages is one of the major challenges in practice. 
In order to enhance the semantic matching capacity, 
Liu et al.~\cite{Liu2021PretrainedLM} propose a multi-stage training approach to optimizing dense retriever by leveraging both large-scale weakly supervised training data (\eg search logs) and small-scale human labeled training data.
Such a multi-stage training approach leads to significant improvement compared to single-stage training.  Besides, 
since dense retrieval is less capable in lexical matching (as we discussed in Section 4.3), an integration of dense retrieval and sparse retrieval techniques has been employed in Baidu search~\cite{Liu2021PretrainedLM} and Spotify search\footnote{\url{https://engineering.atspotify.com/2022/03/introducing-natural-language-search-for-podcast-episodes/}}.

In sponsor search~\cite{SponsoredSearch}, the systems should optimize multiple objectives simultaneously, \eg  query-advertisement~(ad) relevance and click-through rate (CTR). 
Under a traditional retrieval pipeline, the retriever first retrieves the relevant ads by only considering the relevance between queries and ads, and then the reranker reorders the retrieved ads according to the CTR criterion. Fan et al.~\cite{fan2019mobius} argue that the separation of  relevance learning and  CTR prediction tends to produce  suboptimal performance, and they propose to train a retriever that jointly considers both relevance and CTR. Specifically, they train the retriever from the synthetic clicked logs augmented by a  teacher model. The proposed system is called \emph{Mobius} and has been deployed into Baidu's sponsor search. 
As another solution, Zhang et al.~\cite{Zhang2022UniRetrieverTL} propose to unify knowledge distillation and contrastive learning for training a more capable retriever, namely \emph{Uni-Retriever}. Specifically, the retriever learns to retrieve high-relevance ads by distilling the knowledge from a relevance teacher, while learning to retrieve high-CTR ads by contrastive learning.  It has been reported that Uni-Retriever has been included as one of the major multi-path retrievers in Bing’s sponsor search~\cite{Zhang2022UniRetrieverTL}. 
Besides, Dodla et al.~\cite{HEARTS} propose to integrate dense retrievers and non-autoregressive generators in a multi-task setting (sharing the same encoder), showing a significant performance gain by combining the two techniques in bid keyword retrieval.

In personalized search and product search, 
the personalized features and product features are important to consider in the retrievers. Huang et el.~\cite{Huang2020embedding} incorporate the embeddings of search context (social relation, location, etc.) as the input of dense retriever. 
Liu et al.~\cite{Liu2021Que2SearchFA} incorporate the embeddings of product images as the input of dense retriever. Besides, Huang et al.~\cite{Huang2020embedding} and Li et al.~\cite{li2021embedding} also find that hard negatives are important for training dense retrievers (as we discussed in Section 5.2). Further, 
Magnani et al.~\cite{Walmart} design a  dense-sparse hybrid system for e-commerce search deployed at Walmart, and they select negatives based on three strategies including product type matching, query token matching and monoBERT-based scoring.